\documentclass{article}

\PassOptionsToPackage{numbers, compress}{natbib}
\usepackage[main, preprint]{neurips_2025}

\makeatletter
\renewcommand{\@noticestring}{%
  Accepted for publication in \textit{Information Fusion}. DOI: 10.1016/j.inffus.2026.104286%
}
\makeatother

\usepackage[utf8]{inputenc} 
\usepackage[T1]{fontenc}    
\usepackage{hyperref}       
\usepackage{url}            
\usepackage{booktabs}       
\usepackage{amsfonts}       
\usepackage{nicefrac}       
\usepackage{microtype}      
\usepackage{xcolor}         

\usepackage{mathtools}
\usepackage{hyperref,amsmath,amsthm, cleveref, scalerel, stackengine, pgfplots, dsfont, booktabs, multirow, multicol}
\usepackage{caption, subcaption, xcolor}
\usepackage{graphicx}
\usepackage[ruled,vlined]{algorithm2e}
\usepackage{color}
\usepackage{kotex}
\usepackage{longtable}

\usepackage{amssymb}
\usepackage{algpseudocode}
\usepackage{bm}

\usepackage[figuresright]{rotating}

\numberwithin{equation}{section}

\title{\Large{Diffolio: A Diffusion Model \\ for Multivariate Probabilistic Financial Time-Series Forecasting and Portfolio Construction}}

\author{%
    So-Yoon Cho\thanks{These authors contributed equally to this work.} \\
    Division of Finance \& AI \\
    Hankuk University of Foreign Studies \\
    Gyeonggi-do 17035 \\
    Republic of Korea \\
    \And
    Jin-Young Kim$^*$ \\
    Independent Researcher \\
    Seoul \\
    Republic of Korea \\
    \And
    Kayoung Ban \\
    School of Physics \\
    Korea Institute for Advanced Study \\
    Seoul 02455 \\
    Republic of Korea \\
    \And
    Hyeng Keun Koo \\
    Dept. of Financial Engineering \\
    Ajou University \\
    Gyeonggi-do 16499 \\
    Republic of Korea \\
    \And
    Hyun-Gyoon Kim\thanks{Corresponding author} \\
    Dept. of Financial Engineering \& \\
    Dept. of Artificial Intelligence\\
    Ajou University \\
    Gyeonggi-do 16499 \\
    Republic of Korea \\
}

\begin{document}

\maketitle

\begin{abstract}
Probabilistic forecasting is crucial in multivariate financial time-series for constructing efficient portfolios that account for complex cross-sectional dependencies. In this paper, we propose Diffolio, a \textcolor{black}{novel} diffusion\textcolor{black}{-based} model \textcolor{black}{specifically} designed for multivariate financial time-series forecasting and portfolio construction. 
\textcolor{black}{Unlike existing diffusion-based financial time-series models that typically process financial covariates without structural distinction,}
Diffolio employs a denoising network with a hierarchical attention architecture, comprising both asset-level and market-level layers.
This structure effectively extracts salient features not only from historical returns but also from asset-specific and systematic covariates \textcolor{black}{in a manner inspired by asset pricing theory}, and significantly enhances the performance of forecasts and portfolios.  
Furthermore, to \textcolor{black}{rigorously} reflect cross-sectional correlations, we introduce a correlation-guided regularizer informed by a stable estimate of the target correlation matrix. 
\textcolor{black}{By explicitly aligning the attention map with this target correlation, our approach enforces that the learned dependency structures are well-conditioned and economically meaningful.}
Experimental results on the daily excess returns of 12 industry portfolios show that Diffolio outperforms various probabilistic forecasting baselines in \textcolor{black}{both statistical} multivariate forecasting accuracy and portfolio performance. Moreover, in portfolio experiments, portfolios constructed from Diffolio's forecasts show consistently robust performance, thereby outperforming those from benchmarks by achieving \textcolor{black}{remarkably} higher Sharpe ratios for the mean-variance tangency portfolio and higher certainty equivalents for the growth-optimal portfolio. These results \textcolor{black}{underscore} the superiority of our proposed Diffolio in terms of not only statistical accuracy but also economic significance. 
\end{abstract}

\section{Introduction}

Multivariate time-series forecasting is a crucial problem across various domains such as traffic, energy, and weather, and it is particularly critical in finance for the construction of multi-asset portfolios. As initiated by Markowitz \cite{markowitz1952portfolio}, modern portfolio theory is founded on the principle of strategically allocating assets to diversify risk and maximize risk-adjusted returns by taking the cross-sectional dependencies into account. Consequently, accurately predicting both individual asset returns and their complex joint dynamics is crucial for constructing efficient investment portfolios.

However, relying solely on point estimates of these multivariate returns is insufficient. Widely-used portfolio optimization strategies, such as the mean-variance efficient portfolio or utility maximization portfolio, require estimates of the expected returns, the covariance matrix, and often the entire joint distribution of returns to determine optimal portfolio weights. This necessitates density estimation rather than point estimation, emphasizing the need for probabilistic forecasting to characterize the distribution of future returns. 

Probabilistic forecasting, in contrast to deterministic forecasting, is an approach that predicts the probability distribution of future outcomes rather than a single point value. 
While traditional methods for this task include parametric modeling, bootstrapped residuals, and quantile regression, more recent deep learning approaches have predominantly utilized generative models such as generative adversarial networks (GANs) \cite{goodfellow2014generative}, variational autoencoders (VAEs) \cite{kingma2013auto}, diffusion models \cite{sohl2015deep, ho2020denoising}, and flow matching \cite{lipman2023flow}. In this work, we employ a diffusion model, which is renowned for its ability to generate high-fidelity and diverse samples, to accurately forecast financial time-series. These approaches are discussed further in Section \ref{sec:related_works}.

The necessity for probabilistic forecasting in financial time-series also stems from its inherent uncertainty of the future. Deterministic forecasting can be viewed as predicting the expectation marginalized over this uncertainty (e.g., \cite{gneiting2014probabilistic}), resulting in a loss of valuable information. In finance, this uncertainty manifests as \textit{risk}, a central component in portfolio management, risk hedging, and derivative pricing. By adopting a probabilistic approach, we aim to implicitly characterize this uncertainty. However, this characterization is inherently challenging; it is infeasible to know all the factors or covariates that influence future returns, let alone the distribution of the residual uncertainty that remains after accounting for them. Nevertheless, extensive research has identified several covariates that exhibit predictability for asset returns (e.g., \cite{welch2008comprehensive, gu2020empirical}). Therefore, designing a model architecture capable of effectively incorporating these covariates is paramount to refine the predictive distribution.

Many time-series forecasting models (e.g., \cite{kim2021explainable,tashiro2021csdi,alcaraz2022diffusion,kim2023predicting,kollovieh2023predict, fan2024mg}) generate predictions conditioned solely on the historical sample. This approach can be considered to be based on the stationary assumption that temporal patterns observed in the past will recur. However, financial time-series exhibit time-varying characteristics conditioned on dynamic asset-specific and systematic environments \cite{campbell1988dividend, welch2008comprehensive, gu2020empirical}, rather than merely reproducing fixed temporal patterns. Consequently, relying only on historical data often fails to yield consistent forecasting results for financial time-series, and it is more appropriate to incorporate the effects of the time-varying economic environment through covariates. 
{While approaches that model stock returns using these covariates have been explored 
\cite{kelly2019characteristics, gu2020empirical, gu2021autoencoder}, 
studies that leverage such information for \emph{probabilistic forecasting} of return distributions 
and its integration into portfolio construction remain scarce, highlighting the need for further research.}

{\color{black} To effectively incorporate two distinct types of covariates---asset-specific and systematic---an architecture is required that mirrors the structural dependency of financial markets. Classic and modern asset pricing theories provide the inspiration. While traditional factor models \cite{sharpe1964capital, ross1976arbitrage} posit that asset returns are driven by exposures to systematic factors, contemporary paradigms such as instrumented principal component analysis (IPCA) \cite{kelly2019characteristics} demonstrate that these exposures are not static but are instrumented by asset-specific characteristics. 

This logic---distilling salient information at the individual asset level and subsequently modeling the dependency structure with systematic factors at the market level---necessitates a hierarchical modeling framework. 
{By decoupling highly noisy individual information from the modeling of cross-asset dependencies, this hierarchy acts as a structural regularizer, distinguishing it from non-hierarchical architectures that typically process features without structural distinction and often suffer from signal interference or information leakage.}
We implement this logic via a hierarchical attention architecture \cite{yang2016hierarchical}---a structure widely used in language models to process information at word and sentence levels separately. We adapt this hierarchical structure to financial time-series forecasting by operating on two distinct levels, asset-level and market-level, enabling the model to elaborately capture complex cross-sectional dependencies and significantly enhance predictability.
}

{Furthermore, to better capture} the cross-sectional dependency that may not be fully explained by the covariates (e.g., due to cascading effects arising from supply chain relationships), we introduce a correlation-guided regularizer. This regularizer is applied during the market-level self-attention stage and is designed to guide the attention probabilities between asset-level latent tensors to align with a stable estimate of the correlation matrix among assets. This mechanism enhances the model's ability to capture cross-sectional dependencies and potentially improves portfolio performance.

{\color{black}
Our proposed model presents notable conceptual and architectural advancements that distinguish it from existing diffusion-based financial time-series models \cite{wang2024financial, huang2024generative, takahashi2025generation, tanaka2025cofindiff}. Most recent works have applied diffusion models to univariate pattern replication, data denoising, or prediction for individual assets. Although some of these models extend to  multiple assets, they typically treat them as independent sequences and simply aggregate their outputs, thereby overlooking the complex joint dynamics essential for portfolio management. In contrast, our model recognizes multivariate financial time-series forecasting as a joint density estimation problem in which the modeling of cross-sectional correlation is paramount.
By characterizing time-varying cross-sectional dependencies with the aid of the correlation-guided regularizer, our model yields the predicted dependency structures that are both statistically robust and economically reliable. 
{Moreover,} although some existing models (e.g., \cite{tanaka2025cofindiff}) incorporate basic statistical conditions such as trend and volatility, they lack the structural flexibility to integrate the various covariates grounded in recent asset pricing theory. Our hierarchical attention architecture facilitates the effective use of these informative covariates, and our model thus captures a more comprehensive representation of the time-varying economic environment.
}

\textcolor{black}{Ultimately, the primary goal of this paper is to propose a multivariate financial time-series forecasting model that transcends mere statistical precision to deliver economic reliability. 
As Gu et al. \cite{gu2020empirical} highlighted, accurate predictions for individual assets do not necessarily translate into reliable portfolio forecasts, largely because portfolio performance is critically dependent on the complex dependence structure among assets.
To address the challenge of generating forecasts that are both statistically accurate and economically significant in multivariate settings, we propose \textit{Diffolio}.
Accordingly, Diffolio is evaluated from two corresponding perspectives: statistical forecasting accuracy and portfolio performance.  }
As demonstrated in Section \ref{sec:experiments}, Diffolio exhibits strong performance in both aspects. Notably, in terms of portfolio performance, our model significantly outperforms all baseline models considered and consistently achieves performance superior or comparable to the market benchmark. These results demonstrate that our proposed model architecture allows for the construction of stable and efficient portfolios.

The remainder of this paper is organized as follows. We begin by briefly reviewing related literature in Section \ref{sec:related_works} and providing the necessary background on diffusion models and probabilistic time-series forecasting in Section \ref{sec:backgrounds}. We then detail the architecture of our proposed model in Section \ref{sec:diffolio}. Section \ref{sec:experiments} presents the experimental setup and analyzes the empirical results of our proposed model against several baselines. Section \ref{sec:conclusion} concludes the paper.        

\section{Related Works} \label{sec:related_works}

The application of deep generative models, including GANs, VAEs, and diffusion models, has significantly advanced the field of probabilistic time-series forecasting. 
GAN-based approaches, with TimeGAN \cite{yoon2019time} being a representative example, employ an adversarial training process to learn temporal dynamics and generate realistic time-series samples. 
Models applied in the financial domain (e.g., FIN-GAN \cite{takahashi2019modeling} and QuantGAN \cite{wiese2020quant}) focus on generating univariate financial data to reproduce stylized facts.
For multivariate conditional generation, Fin-GAN \cite{vuletic2024fin} introduces an economics-driven loss function for forecasting, and SigCWGAN \cite{liao2024sig} integrates Wasserstein GANs with path signature transforms. 
\textcolor{black}{More recently, MarketGAN \cite{huh2026marketgans} proposes a factor-model-based generative framework that incorporates an explicit asset-pricing structure to preserve high-dimensional cross-sectional dependencies, primarily focusing on data augmentation under scarcity.
However, GAN-based models often suffer from training instability and mode collapse, which limits their ability to fully capture the heavy-tailed distributions characteristic of asset returns. Furthermore, they typically lack an explicit likelihood formulation, making it difficult to quantify forecast uncertainty accurately.} 

On the other hand, VAE-based models are developed to learn probabilistic latent representations of sequential data. An example application in finance is FactorVAE \cite{duan2022factorvae}, which integrates a VAE with a dynamic factor model to learn latent factors for predicting stock returns. \textcolor{black}{While VAEs offer explicit density estimation, they often struggle to generate  realistic, fine-grained samples compared to diffusion models due to the restrictive assumptions on the approximate posterior.}

Diffusion models \cite{sohl2015deep, ho2020denoising} have emerged as a powerful approach for generative tasks, renowned for their ability to produce high-fidelity samples; this strength has also been shown to be effective in the time-series domain.  DiffWave \cite{kong2020diffwave}, originally for audio synthesis, utilizes 1-dimensional dilated convolutions in its denoising network $\epsilon_{\theta}$. TSDiff \cite{kollovieh2023predict} proposes a flexible approach for conditional forecasting by training an unconditional model and applying self-guidance at inference time. Work on multivariate conditional diffusion models has also been extensive. TimeGrad \cite{rasul2021autoregressive} adopts an autoregressive recurrent neural network (RNN) to encode historical information. Models such as CSDI \cite{tashiro2021csdi} and SSSD \cite{alcaraz2022diffusion} utilize different backbones---transformer and structured state space models, respectively---to focus on imputation and forecasting. MG-TSD \cite{fan2024mg} is proposed to enhance training stability by using multi-granularity data as targets. Diffusion-TS \cite{yuan2024diffusion} trains an unconditional transformer-based model to predict the clean sample $x_0$ directly, and then employs classifier-free guidance at inference time to generate conditional forecasts. 

{\color{black}
Recently, several studies have adapted diffusion models to the financial domain, although their primary objectives often lie in synthetic data generation or data purification rather than multivariate time-series forecasting. Huang et al. \cite{huang2024generative} introduced FTS-Diffusion, which deconstructs financial time-series into irregular and scale-invariant patterns for univariate augmentation and prediction. While effective for capturing localized temporal dynamics, its focus remains restricted to univariate modeling and does not account for external systematic covariates. In a different approach, Takahashi and Mizuno \cite{takahashi2025generation} utilized wavelet transformations to represent market features as multi-channel images for synthetic data generation and emphasized data synthesis over forecasting. However, this work models the multivariate structure of features for a single asset rather than the cross-sectional dependencies across multiple assets.

Beyond generation, diffusion models have been explored for data refinement and controllable simulation. Wang and Ventre \cite{wang2024financial} proposed a conditional diffusion denoiser to enhance the signal-to-noise ratio of individual stock prices for downstream tasks. Similarly, Tanaka et al. \cite{tanaka2025cofindiff} presented CoFinDiff, a controllable model designed to generate price trajectories according to conditions such as trend and volatility. Despite their contributions, these purification and simulation models are not inherently designed for the multivariate time-series forecasting of multiple assets. They typically treat assets as independent sequences, aggregate them post-hoc, and fail to address the joint density estimation problem essential for capturing time-varying cross-sectional dependencies. Furthermore, as with the aforementioned models, they lack the structural flexibility to integrate the various covariates required to accurately represent the complex dynamics of financial markets.

While these finance-specific models and general time-series models demonstrate strong performance, their architectures are often either generic or limited in scope.} As mentioned above, research on diffusion models for multivariate financial time-series forecasting remains particularly scarce. Moreover, both the generic models and those tailored for finance are often not explicitly designed to incorporate the two distinct types of covariates---asset-specific and systematic---crucial for financial forecasting \cite{welch2008comprehensive, gu2020empirical}, thereby limiting their ability to fully exploit this significant predictive information. Furthermore, while these models implicitly capture cross-sectional dependencies, they often lack explicit mechanisms designed to model and refine the intricate relationships paramount for efficient portfolio construction. Diffolio addresses these gaps as a multivariate conditional diffusion model tailored for financial time-series forecasting, utilizing a hierarchical attention architecture and a correlation-guided regularizer.

{A related strand of research also leverages attention mechanisms to jointly model temporal and cross-sectional dependencies for portfolio construction. In particular, Cong et al. \cite{cong2021alphaportfolio} propose AlphaPortfolio, which uses a historical attention module at the individual asset level and a cross-asset attention network to directly learn portfolio weights through deep reinforcement learning. 
{While both approaches share the architectural intuition of employing distinct attentions on individual assets and across assets, their detailed backbone architectures and handling of covariates diverge. Unlike AlphaPortfolio's network, which considers asset-specific covariates only, Diffolio's architecture is designed to hierarchically incorporate both asset-specific and systematic covariates. Beyond these architectural differences, their objectives also fundamentally differ.}
AlphaPortfolio optimizes investment performance by maximizing a reward function over portfolio returns, effectively learning a policy for asset allocation. In contrast, Diffolio focuses on probabilistic multivariate forecasting---that is, modeling the conditional joint distribution of future returns---to provide calibrated uncertainty estimates and to enable risk-aware portfolio construction. Thus, whereas AlphaPortfolio represents a reinforcement learning approach to direct portfolio optimization, our model constitutes a generative approach that produces distributional forecasts applicable to a wide range of decision frameworks.}

{\color{black}{
\section{Backgrounds} \label{sec:backgrounds}
This section provides an overview of the diffusion models that form the foundation of Diffolio. 
We denote the diffusion steps using the index $\tau \in \{1, 2, \cdots, T\}$ to distinguish them from the real time index $t$. For a comprehensive list of symbols and their definitions used in this paper, we refer to ~\ref{appendix: Nomenclature}.

\subsection{Denoising Diffusion Probabilistic Models (DDPM)}
Denoising diffusion probabilistic models (DDPMs) \cite{ho2020denoising, sohl2015deep} consist of a forward process that gradually adds Gaussian noise to a data sample $\mathbf{x}^0$ and a reverse process that learns to recover the original data sample.
A key property of the forward process is that any noisy sample $\mathbf{x}^{\tau}$ can be expressed directly in terms of $\mathbf{x}^0$ using the reparameterization trick
\begin{equation} \label{eqn:reparameterization}
\mathbf{x}^{\tau} = \sqrt{\bar{\alpha}_{\tau}}\mathbf{x}^{0} + \sqrt{1-\bar{\alpha}_{\tau}}\boldsymbol{\epsilon}, \quad \text{where } \boldsymbol{\epsilon} \sim \mathcal{N}(\mathbf{0}, \mathbf{I}),
\end{equation}
where $\alpha_\tau :=1-\beta_\tau$, ${\bar{\alpha}}_\tau:=\prod_{i=1}^\tau \alpha_i$, and $\{\beta_\tau \in (0,1)\}_{\tau=1}^T$ is a predefined variance schedule. 
The reverse process recovers the original data $\mathbf{x}^0$ by modeling the transitions $p_\theta(\mathbf{x}^{\tau-1}|\mathbf{x}^{\tau})$ that are parameterized as Gaussian distributions $\mathcal{N}(\mathbf{x}^{\tau-1}; \boldsymbol{\mu}_\theta (\mathbf{x}^\tau, \tau), \bm{\Sigma}_\theta (\mathbf{x}^\tau, \tau))$. As shown by Ho et al. \cite{ho2020denoising}, this optimization can be simplified by reparameterizing the mean $\boldsymbol{\mu}_{\theta}$ to predict the noise $\boldsymbol{\epsilon}$ from \eqref{eqn:reparameterization} using a neural network $\boldsymbol{\epsilon}_\theta (\mathbf{x}^\tau, \tau)$.
The simplified training objective is
\begin{equation} \label{eqn:ddpm_loss}
\mathcal{L}_{\text{DDPM}}(\theta) = \mathbb{E}_{\tau, \mathbf{x}^0, \boldsymbol{\epsilon}} \left[ \lVert \boldsymbol{\epsilon} - \boldsymbol{\epsilon}_{\theta}(\mathbf{x}^\tau, \tau) \rVert^2 \right].
\end{equation}

\subsection{Denoising Diffusion Implicit Models (DDIM)} \label{sec:ddim}
While DDPMs yield high-quality samples, the reverse process requires $T$ sequential denoising steps, which can be computationally intensive. Denoising diffusion implicit models (DDIM) \cite{song2020denoising} accelerate the sampling process by defining a non-Markovian generative process that shares the same training objective as DDPM but allows for sampling using fewer steps.
During inference, DDIM first estimates the original data $\mathbf{x}^0$ from the noisy input $\mathbf{x}^\tau$ using the trained denoising network $\boldsymbol{\epsilon}_{\theta}$ as
\begin{equation*}
\hat{\mathbf{x}}^0 = \frac{\mathbf{x}^{\tau} - \sqrt{1-\bar{\alpha}_{\tau}}\boldsymbol{\epsilon}_{\theta}(\mathbf{x}^{\tau}, \tau)}{\sqrt{\bar{\alpha}_{\tau}}}.
\end{equation*}
Then, DDIM generates the sample at a preceding step $\tau'$ (where $\tau' < \tau$) using the following generalized update rule,
\begin{equation} \label{eqn:ddim_sampling}
\mathbf{x}^{\tau'} = \sqrt{\bar{\alpha}_{\tau'}} \hat{\mathbf{x}}^0 + \sqrt{1 - \bar{\alpha}_{\tau'} - (\sigma_{\tau'})^2} \cdot\boldsymbol{\epsilon}_{\theta}(\mathbf{x}^{\tau}, \tau) + \sigma_{\tau'} \boldsymbol{\epsilon}',
\end{equation}
where $\boldsymbol{\epsilon}'\sim \mathcal{N} (\mathbf{0}, \mathbf{I})$ and $\sigma_{\tau'} = \eta \sqrt{(1 - \bar{\alpha}_{\tau'}) / (1 - \bar{\alpha}_{\tau})} \sqrt{1 - \bar{\alpha}_{\tau} / \bar{\alpha}_{\tau'}}$ for $\eta\in[0, 1]$. 
We employ the deterministic DDIM sampling procedure ($\eta = 0$) during inference, enabling faster generation by using a reduced number of steps while maintaining high sample quality.

\subsection{Conditional Diffusion Models for Multivariate Time-Series Forecasting}
In the context of this paper, our objective is the probabilistic forecasting of $N$-dimensional asset excess returns conditioned on covariates.
Diffolio models the conditional distribution $p(\mathbf{x}^0 | \mathbf{c})$, where the data sample at the zero diffusion step $\mathbf{x}^0_{t+1}$ corresponds to the $N$-dimensional asset excess return vector at the next time step, $\mathbf{r}_{t+1}\in\mathbb{R}^N$, with $\mathbf{c}$ denoting the conditioning information.
Specifically, $\mathbf{c}$ consists of two components: the historical excess returns over a lookback window of $M$ steps, $\mathbf{r}_{t-(M-1):t}$, and a set of asset-specific and systematic covariates.
Our approach characterizes the conditional predictive distribution $p(\mathbf{r}_{t+1}|\mathbf{r}_{t-(M-1):t}, \mathbf{c}_{t-(M-1):t})$ by generating a large set of plausible future return trajectories through a diffusion model. 
The resulting distribution is then utilized to optimize investment portfolios, a task for which accurately modeling the joint distribution of asset returns is paramount.

The denoising network $\boldsymbol{\epsilon}_\theta (\mathbf{x}^\tau, \tau, \mathbf{c})$ is trained to predict the noise $\boldsymbol{\epsilon}$ by minimizing the conditional objective:\begin{equation}\label{eqn:conditional_ddpm_loss}
\mathcal{L}_{\text{DDPM-cond}}(\theta) = \mathbb{E}_{\tau, \mathbf{x}^0, \boldsymbol{\epsilon}, \mathbf{c}} \left[ \lVert \boldsymbol{\epsilon} - \boldsymbol{\epsilon}_{\theta}(\mathbf{x}^{\tau}, \tau, \mathbf{c}) \rVert^2 \right].
\end{equation}
In Diffolio, the architecture $\boldsymbol{\epsilon}_{\theta}$ utilizes attention structures to extract salient features from the conditioning information $\mathbf{c}$, enhancing its predictive capability for conditional generation. 
}}

\section{Diffolio} \label{sec:diffolio}

{\color{black}
We propose Diffolio, a conditional diffusion model specifically tailored for multivariate financial time-series forecasting and portfolio construction. The core of our model lies in the denoising network, $\boldsymbol{\epsilon}_\theta$, which is engineered with a hierarchical attention architecture to effectively capture the complex joint dynamics of asset returns by utilizing historical data alongside asset-specific and systematic covariates. {Our design is inspired by established asset pricing theories. While traditional factor models \cite{sharpe1964capital, ross1976arbitrage} posit that asset returns are driven by static exposures to systematic factors, contemporary paradigms like IPCA \cite{kelly2019characteristics} demonstrate that these exposures are dynamically instrumented by asset-specific characteristics.} Diffolio reflects this sequential progression from individualized distillation to systematic integration by organizing the information flow into a two-stage hierarchy.


{This hierarchy functions as a structural regularizer by decoupling the processing of high-noise individual asset information from the modeling of cross-asset dependencies, thereby allowing for the effective capture of predictive signals rather than being submerged by noise.}
Specifically, the model operates in two stages.
{First, at the lower, asset-level hierarchy, we process each asset individually using cross-attention to infuse information from its historical returns and asset-specific covariates---which together form the context vector---into individualized latent tensors. This stage prevents individual information from leaking across assets while extracting the latent state of each asset. Subsequently, at the higher, market-level hierarchy, these asset-level latent tensors interact with one another and with systematic covariates via a self-attention mechanism. This enables the model to elaborately capture complex cross-sectional dependencies and significantly enhance predictability. }
Furthermore, to capture dependency structures that may not be fully accounted for by covariates, we introduce a correlation-guided regularizer in the market-level stage to align attention scores with a stable estimate of the asset correlation matrix.}


\subsection{Hierarchical Attention Denoising Network}
The denoising network $\boldsymbol{\epsilon}_\theta$ takes a noisy data sample $\mathbf{x}_{t+1}^\tau$ at diffusion step $\tau$ and conditioning information $\mathbf{c}$ to predict the original noise $\boldsymbol{\epsilon}$ that was added to the clean data $\mathbf{x}_{t+1}^0$. The conditioning information $\mathbf{c}$ is composed of data from a lookback window of $M$ steps, including: (1) historical excess returns for $N$ assets, $\mathbf{r}_{t-(M-1):t} \in \mathbb{R}^{M\times N}$; (2) a set of sequences of asset-specific covariates for each asset, $\{\mathbf{z}_{i, t-(M-1):t}\}_{i=1}^N$; and (3) a sequence of systematic covariates, $\mathbf{y}_{t-(M-1):t}$. The network processes this information through a two-stage hierarchical attention mechanism.

\textbf{Stage 1: Asset-Specific Feature Infusion via Cross-Attention.}
The first stage performs \textit{asset-level attention}, where the model extracts relevant features for each asset from its own historical data and asset-specific covariates. For each asset $i$, we form a \textit{query} vector $\mathbf{q}_i \in \mathbb{R}^D$ by embedding the concatenation of its noisy future return $\mathbf{x}_{i, t+1}^\tau$ and the diffusion step embedding. The corresponding \textit{key} and \textit{value} tensors, $\mathbf{K}_i, \mathbf{V}_i \in \mathbb{R}^{M\times D}$, are formed by embedding the concatenated sequence of the asset's historical returns $\mathbf{r}_{i, t-(M-1):t}$ and the corresponding asset-specific covariates $\mathbf{z}_{i, t-(M-1):t}$. A cross-attention layer then infuses these inputs into an asset-specific latent vector $\mathbf{h}_{i}\in \mathbb{R}^D$.

For the specific architecture of this layer, we adopt the attention structure proposed by Bolya et al. \cite{bolya2025perception} and define our CrossAttentionBlock as follows: 
\begin{equation} 
\begin{split}
\mathbf{h}_i &\coloneqq \text{CrossAttentionBlock}(\mathbf{q}_i, \mathbf{K}_i, \mathbf{V}_i) \\
&= \text{CA} (\mathbf{q}_i, \mathbf{K}_i, \mathbf{V}_i) + \text{MLP}(\text{LayerNorm}(\text{CA} (\mathbf{q}_i, \mathbf{K}_i, \mathbf{V}_i))), 
\end{split}
\end{equation}
where the intermediate representation CA is computed using the standard scaled dot-product cross-attention,
\begin{equation*}
\begin{split}
\text{CA}(\mathbf{q}_i, \mathbf{K}_i, \mathbf{V}_i) &= \text{softmax}\left(\frac{\tilde{\mathbf{q}}_i \tilde{\mathbf{K}}_i^\top}{\sqrt{D}}\right)\tilde{\mathbf{V}}_i, \\
\text{where } \tilde{\mathbf{q}}_i = \mathbf{q}_i\mathbf{W}_q, &\ \tilde{\mathbf{K}}_i = \mathbf{K}_i\mathbf{W}_K, \text{ and } \tilde{\mathbf{V}}_i = \mathbf{V}_i\mathbf{W}_V.
\end{split}
\end{equation*}
Here, the learnable weight matrices $
\mathbf{W}_q, 
\mathbf{W}_K, 
\mathbf{W}_V \in \mathbb{R}^{D\times D}$ perform a linear transformation to project the input tensors.
This CrossAttentionBlock applies a layer normalization (LayerNorm) and a multi-layer perceptron (MLP) to the intermediate representation with a residual connection. The MLP is composed of two layers with a Gaussian error linear unit (GELU) activation function. This attention block allows the model to learn how each asset's future return is influenced by its individual historical patterns and characteristics. Note that the embedding layers for returns and asset-specific covariates and the cross-attention layer share parameters across all $N$ assets, promoting generalization and model efficiency.

\textbf{Stage 2: Cross-Asset and Systematic Feature Integration via Self-Attention.}
The second stage performs \textit{market-level attention} to model the cross-sectional dependencies among all assets and their exposure to market-level systematic features.
The asset-specific latent vectors from the first stage, $\{\mathbf{h}_{i}\}_{i=1}^N$, are stacked into a single tensor. 
This tensor is then concatenated with $\{\mathbf{h}_j\}_{j=N+1}^{N+N_y}$, the embeddings of the systematic covariates $\mathbf{y}_{t-(M-1):t}$, to form an integrated tensor $\mathbf{h}\in\mathbb{R}^{(N+N_y)\times D}$. Here, $N_y$ is the number of systematic covariates.

To facilitate interaction between these features, we apply a self-attention mechanism. We define a SelfAttentionBlock analogous to the one in the first stage as
\begin{equation}
\begin{aligned}
\mathbf{h}' \coloneqq& \text{SelfAttentionBlock} (\mathbf{h}) \\ =& \text{SA}(\mathbf{h}) + \text{MLP}(\text{LayerNorm}(\text{SA}(\mathbf{h}))),
\end{aligned}
\end{equation}
where the intermediate representation SA is computed using scaled dot-product self-attention,
\begin{equation*}
\begin{split}
\text{SA}(\mathbf{h}) &= \text{softmax}\left(\frac{\mathbf{Q} \mathbf{K}^\top}{\sqrt{D}}\right)\mathbf{V}, \\
\text{where } \mathbf{Q} &= \mathbf{h}\mathbf{U}_Q, \mathbf{K} = \mathbf{h}\mathbf{U}_K, \text{ and } \mathbf{V} = \mathbf{h}\mathbf{U}_V.
\end{split}
\end{equation*}
and the MLP block is composed of two layers with a GELU activation function.
Here, $
\mathbf{U}_Q, 
\mathbf{U}_K, 
\mathbf{U}_V \in \mathbb{R}^{D\times D}$ are the learnable weight matrices to perform a linear transformation to project the input tensor $\mathbf{h}$ for the query, key, and value roles. 
This self-attention structure allows for an all-to-all interaction between asset-specific representations and systematic covariates.
We then slice the first $N$ vectors from the output tensor $\mathbf{h}'$, each vector corresponding to each asset.
These vectors are then passed through a decoding layer, which shares parameters across all assets, to produce the final noise prediction $\hat{\boldsymbol{\epsilon}}=\{\hat{\boldsymbol{\epsilon}}_i\}_{i=1}^{N}$. 

In summary, the first stage infuses the information of historical returns and asset-specific covariates for each asset into the corresponding latent tensor through the CrossAttentionBlock. As each latent tensor is formed individually, this stage prevents asset-specific information from leaking across assets. The second stage employs the SelfAttentionBlock to model the interdependencies among assets in the market-level hierarchy. This self-attention allows for capturing the intricate cross-asset dependencies while simultaneously accounting for their exposure to systematic covariates. This hierarchical approach, which effectively incorporates both asset-level and market-level features, makes the model particularly well-suited for financial time-series forecasting.

\subsection{Correlation-Guided Attention Training}

In financial applications, particularly for asset return data, the correlation structure among assets is of paramount importance. To guide the model to learn the correlation structure more precisely, we introduce a correlation-guided regularization, $\mathcal{L}_{\text{corr}}$, which acts on the attention probabilities in the SelfAttentionBlock of the market-level hierarchy. The total training objective is then a weighted sum of the standard conditional denoising loss and the correlation-guided regularizer defined as
\begin{equation}
\mathcal{L}_{\text{Diffolio}} = \mathcal{L}_{\text{DDPM-cond}} + \lambda_{\text{corr}} \mathcal{L}_{\text{corr}},
\end{equation}
where $\mathcal{L}_{\text{DDPM-cond}}$ is the mean squared error between the true and predicted noise as defined in \eqref{eqn:conditional_ddpm_loss}, and $\lambda_{\text{corr}}$ is a hyperparameter balancing the two terms.

The correlation-guided regularizer $\mathcal{L}_{\text{corr}}$ is designed to align the asset-to-asset submatrix of the attention probability matrix, say $\mathbf{A}\in \mathbb{R}^{N\times N}$, with a robustly estimated target correlation matrix, $\bm{\Sigma}_t^{\text{target}} \in \mathbb{R}^{N\times N}$. 
The alignment is achieved by minimizing the average of the negative cosine similarities between the corresponding row vectors of the two matrices,
\begin{equation}\label{eqn:cossim}
\mathcal{L}_{\text{corr}} = -\text{CosineSimilarity}(\mathbf{A}, \bm{\Sigma}_{t}^{\text{target}}):=-\frac{1}{N} \sum_{i=1}^N \frac{\mathbf{A}_i \cdot \bm{\Sigma}_{i, t}^{\text{target}}}{ \lVert \mathbf{A}_i\rVert_2 \lVert \bm{\Sigma}_{i, t}^{\text{target}}\rVert_2},
\end{equation}
where $\mathbf{A}_i$ and $\bm{\Sigma}_{i, t}^{\text{target}}$ are the $i$-th row vectors of matrices $\mathbf{A}$ and $\bm{\Sigma}_{t}^{\text{target}}$, respectively.

An important consideration is the stable estimation of the time-varying target correlation matrix $\bm{\Sigma}_t^{\text{target}}$. 
While it is crucial to use the most recent information to reflect the time-varying asset correlations, estimating the sample correlation matrix from only the recent input window ($M$ steps) is prone to be unstable and can result in near-singular matrices, especially when $M$ is not much larger than $N$.
To mitigate this, we employ the Ledoit-Wolf shrinkage estimator \cite{ledoit2003improved}. This method provides a robust estimate of covariance matrix by shrinking an unstable sample covariance matrix towards a stable one. 
In our experiments, the sample covariance is calculated from the recent window of $\mathbf{r}_{t-(M-1):t}$, and we use the covariance matrix pre-computed from the entire training period, $\bm{\Sigma}^{\text{train}}$, as the stable one.
From this robust covariance estimate, we derive our well-conditioned and economically meaningful target correlation matrix $\bm{\Sigma}_t^{\text{target}}$, which provides a stable guide for the self-attention.

\begin{figure*}[!t] 
    \centering
    \includegraphics[width=0.9\textwidth]{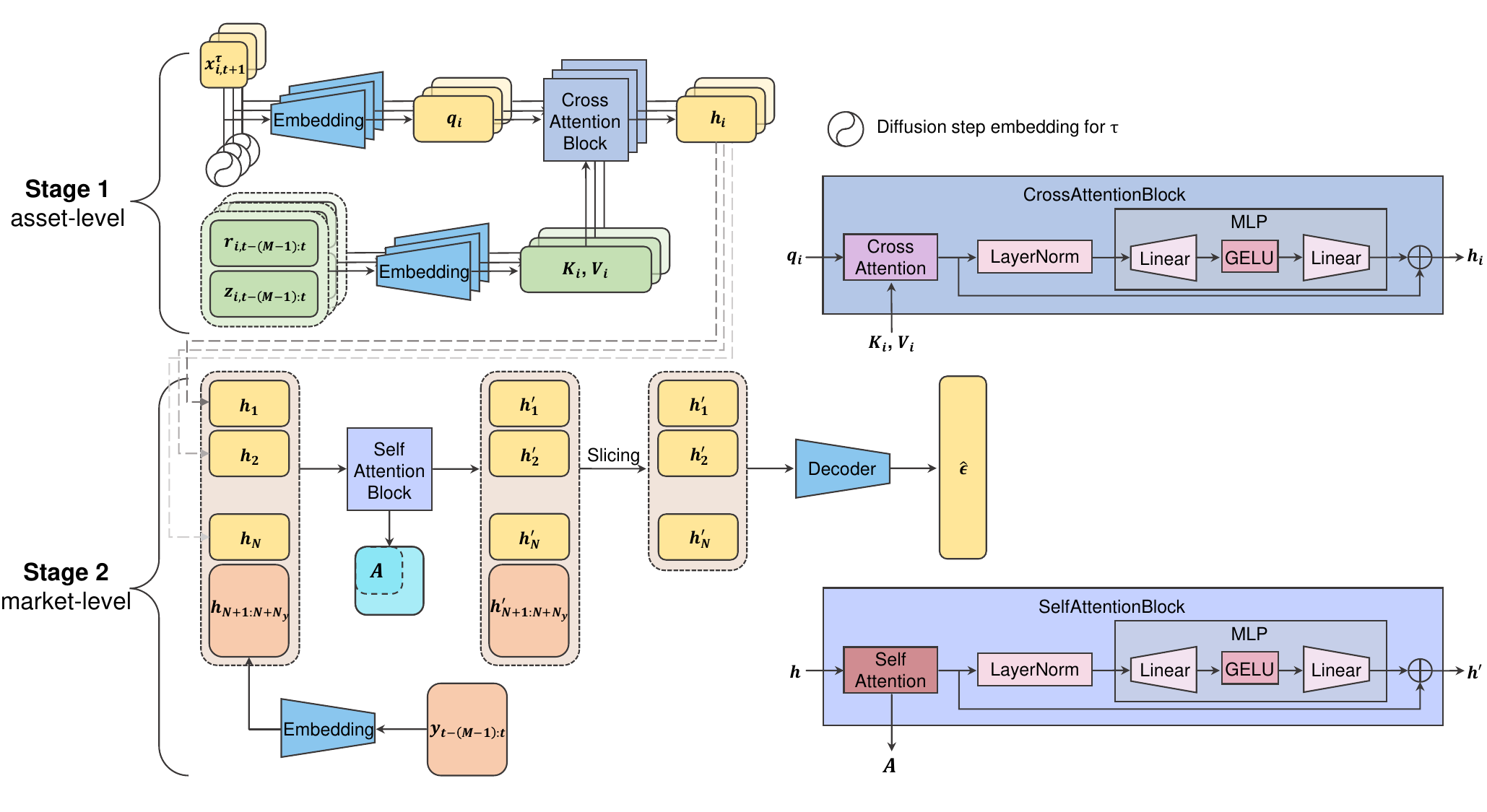} 
    \caption{The structure of the hierarchical attention denoising network of Diffolio.}
    \label{fig:diffolio_structure}
\end{figure*}

The overall structure of the hierarchical attention denoising network and its training process are presented in Figure \ref{fig:diffolio_structure}, and Algorithm \ref{alg:diffolio_training}, respectively. For inference, we employ the deterministic DDIM sampling procedure described in Section \ref{sec:ddim}, which allows for efficient generation of return samples in a reduced number of steps.

{ \linespread{1.2}
\begin{algorithm}[!tb]
\caption{Diffolio Training Process}
\label{alg:diffolio_training}
\SetAlgoNoLine
\LinesNumbered

\For{\textnormal{step from 1 to TrainingSteps}}{

$\mathbf{x}^0$ $\leftarrow$ $\mathbf{r}_{t+1}$; \ \ 
$\mathbf{c}\leftarrow \{\mathbf{r}_{t-(M-1):t}, \{\mathbf{z}_{i, t-(M-1):t}\}_{i=1}^N, \mathbf{y}_{t-(M-1):t}\}$

$\tau\sim \text{Uniform}(\{1, 2, \cdots, T\})$

$\boldsymbol{\epsilon}\sim \mathcal{N}(\mathbf{0}, \mathbf{I})$

$\mathbf{x}^\tau \leftarrow \sqrt{\bar{\alpha}_{\tau}}\mathbf{x}^0 + \sqrt{1-\bar{\alpha}_\tau} \boldsymbol{\epsilon}$

$\hat{\boldsymbol{\epsilon}}, \mathbf{A}\leftarrow\boldsymbol{\epsilon}_\theta (\mathbf{x}^\tau, \tau, \mathbf{c})$  \hfill \textit{$\triangleright$ Hierarchical Attention Denoising Network}

$\boldsymbol{\Sigma}_t^{\text{target}} \leftarrow \text{Ledoit-Wolf Shrinkage}(\mathbf{r}_{t-(M-1):t}, \bm{\Sigma}^{\text{train}})$ 

$\mathcal{L}_{\text{DDPM-cond}} \leftarrow \lVert \boldsymbol{\epsilon} - \hat{\boldsymbol{\epsilon}} \rVert^2$

$\mathcal{L}_{\text{corr}} \leftarrow -\text{CosineSimilarity}(\mathbf{A}, \bm{\Sigma}_t^{\text{target}})$

$\mathcal{L}_{\text{Diffolio}} \leftarrow \mathcal{L}_{\text{DDPM-cond}} + \lambda_{\text{corr}}\mathcal{L}_{\text{corr}}$  \hfill \textit{$\triangleright$ Calculate losses}

Take a gradient descent step on $\nabla_\theta \mathcal{L}_{\text{Diffolio}}$
}
\end{algorithm}
}

\section{Experiments} \label{sec:experiments}
\subsection{Experimental Setup} \label{sec:experimental_setup}

\textbf{Datasets.}
In this study, we forecast one-day-ahead daily excess returns of 12 industry portfolios ($N=12$), which serve as our target assets\footnote{The data is sourced from Kenneth R. French's data library: https://mba.tuck.dartmouth.edu/pages/faculty/ken.french/data\_library.html.}. \textcolor{black}{These portfolios---comprising Consumer Non-durables (NoDur), Consumer Durables (Durbl), Manufacturing (Manuf), Energy (Enrgy), Chemicals (Chems), Business Equipment (BusEq), Telecommunications (Telcm), Utilities (Utils), Shops, Healthcare (Hlth), Money, and Other---represent a broad cross-section of the U.S. equity market according to the Fama-French 12-industry classification \cite{fama1997industry}.} The use of industry portfolios allows for a purer signal by diversifying away asset-specific idiosyncratic risks. 
The total dataset spans from 1958 to 2023 (16,613 trading days). We partition this data into a training set (1958-1999), a validation set (2000-2004), and a test set (2005-2023). 
The constituents of these industry portfolios are determined from stocks listed on the NYSE, AMEX, and NASDAQ based on their four-digit SIC codes. 
Detailed descriptions of these industries are provided in \ref{appendix: data description}.

For the model's covariates, we employ asset characteristics as the asset-specific variables and macroeconomic variables as the systematic variables, both of which are widely used in machine learning-based financial applications (see, e.g., \cite{gu2020empirical}).
Specifically, we use a set of ten asset characteristics. 
To enhance the reproducibility and accessibility of our methodology, we select characteristics that can be constructed from the historical return and market data, such as momentum and volatility, while excluding those that rely on data that are not freely available such as from accounting databases\footnote{\color{black} The proposed architecture is designed to accommodate diverse data sources as most variables naturally map to our asset- or market-level hierarchies (e.g., individual news sentiments or market-wide liquidity). While omitted variables like accounting metrics or news sentiment may offer incremental predictive gains, they are not incorporated in the present study to prioritize data accessibility. We expect Diffolio to remain robust or even improve with the inclusion of such variables by leveraging the attention mechanism’s capacity to prioritize salient signals, and we leave the rigorous analysis of these alternative sources for future work.}. 
This choice is also supported by the empirical literature, as notable studies \cite{moritz2016tree, gu2020empirical, chen2024deep} have documented that return-based predictors are often more significant in forecasting asset returns than accounting-based variables. 
These characteristics include \texttt{mom1m}, \texttt{mom6m}, \texttt{mom12m}, \texttt{mom36m}, \texttt{chmom}, \texttt{retvol}, \texttt{maxret}, \texttt{beta}, \texttt{betasq}, and \texttt{idiovol}.

For the systematic covariates, we employ eight macroeconomic variables known for their predictive power for stock returns as identified by \cite{welch2008comprehensive}\footnote{The data is obtained from Amit Goyal's website: https://sites.google.com/view/agoyal145.}. These variables are \texttt{tbl}, \texttt{d/p}, \texttt{e/p}, \texttt{b/m}, \texttt{tms}, \texttt{dfy}, \texttt{ntis}, and \texttt{svar}. Since these are provided at a monthly frequency, we fill the daily values using the most recently available monthly observation. All covariates are normalized using the mean and standard deviation of the entire training set. The detailed definitions and descriptions for all covariates are given in \ref{appendix: data description}.

\textbf{Evaluation Metrics.}
To reflect the two primary objectives of Diffolio---achieving accurate probabilistic forecasts and ensuring strong economic performance---we assess the performance of Diffolio across two perspectives: 
statistical forecasting accuracy, and portfolio performance. 
For all metrics requiring an empirical estimation of the predictive distribution, we generate 100 forecast samples from the model and then use them for computing metrics.

First, to measure the accuracy of the forecasted distributions, we use two proper scoring rules \cite{gneiting2007strictly}: the \textit{continuous ranked probability score (CRPS)} and the \textit{energy score (ES)}. The CRPS is a widely adopted metric that measures the discrepancy between the forecasted distribution and the ground truth observation \cite{gneiting2014probabilistic}. As the CRPS is defined for univariate distributions, we apply it to each asset's marginal distribution to assess their individual forecast quality. To evaluate the multivariate joint distribution across all assets, we utilize the ES, which generalizes the CRPS to multi-dimensions. As a multivariate metric, the ES can somewhat capture the dependency structure among assets. For both metrics, lower values indicate a more accurate forecast.

Second, we assess the economic performance of the forecasts through portfolio experiments based on the forecast samples generated by Diffolio. We construct daily rebalanced long-only portfolios using two optimization strategies: the \textit{mean-variance tangency portfolio (MVP)}, which relies on the first and second moments to maximize the Sharpe ratio (SR), and the \textit{growth-optimal portfolio (GOP)}\footnote{Here we consider the no-leverage GOP, that is, without short-selling the risk-free asset.
},  which maximizes the expected logarithmic 
utility, thereby implicitly accounting for the entire predictive distribution. By rebalancing the portfolios daily, the accuracy of the distributional forecast at each time step is directly compounded into the investment results, allowing the overall portfolio performance to serve as an integrated measure of the model's ability to forecast time-varying distributions. To evaluate the portfolios, we focus on the annualized SR as the primary metric for the MVP and the annualized certainty equivalent (CE) for the GOP. While the GOP maximizes expected logarithmic utility, this value lacks a direct financial interpretation. The CE translates this abstract utility into an intuitive financial metric; it represents the return on a certain (risk-free) investment that would yield the same level of utility as the uncertain portfolio investment. 

These portfolio-based evaluations not only offer a crucial economic perspective but also implicitly measure the consistency of a model's performance unlike the static metrics such as CRPS and ES. If a model's forecasting performance is inconsistent over time, the resulting portfolio returns will also be erratic, leading to higher volatility and, consequently, lower SR and CE values\footnote{SR decreases with higher volatility by definition. For CE, due to the concavity of the logarithmic utility function, its expected value decreases as the volatility of portfolio returns increases.}. Therefore, SR and CE are important because they inherently measure the consistency of a model's performance over time, while also depending on both forecast accuracy and the modeling of cross-sectional correlations. Higher values indicate better performance for SR and CE. For a comprehensive view, we also report other summary statistics: annualized return (Ret), annualized volatility (Vol), maximum drawdown (MDD), \textcolor{black}{and turnover.}

{\color{black} These evaluation metrics described so far are centered on the core objectives of Diffolio, achieving accurate probabilistic forecasts and strong economic performance. Separately from these primary concerns, we use the \textit{correlation score (CorrScore)} and the \textit{log-determinant divergence (LogDet)}, also known as Stein’s loss, as supplementary diagnostic tools to assess how well the model reproduces the dependency structure among assets. Since these scores measure neither predictive accuracy nor economic significance, they serve a supplementary role in verifying the cross-sectional correspondence of the generated synthetic data. Specifically, CorrScore is calculated as the Frobenius norm of the matrix difference to quantify entry-wise Euclidean distances, and LogDet evaluates the discrepancy based on eigenvalues to assess structural alignment. Both scores are estimated between the correlation matrices of the real data and the mean of the synthetic data, and smaller values indicate closer correspondence.

Furthermore, to evaluate the probabilistic reliability of our forecasts, we examine {forecast calibration} using the \textit{prediction interval coverage probability (PICP)} and the \textit{average coverage error (ACE)}. PICP measures the proportion of actual observations that fall within a prediction interval, calculated as an average across all assets and time points in the test period. We evaluate these metrics across multiple confidence levels---50\%, 80\%, 90\%, 95\%, and 99\%---where ACE represents the discrepancy between the observed PICP and the target confidence. Additionally, we utilize \textit{reliability diagrams} to visually inspect the alignment between theoretical quantiles and empirical coverage, ensuring that the generated distributions are well-calibrated and reliably represent uncertainty.
Detailed descriptions of these evaluation metrics, summary statistics, and supplementary scores are provided in \ref{appendix: evaluation metrics}. }

\textbf{Baselines.}
\textcolor{black}{For a comprehensive evaluation, we compare Diffolio against seven models for multivariate probabilistic time-series forecasting. We first include the dynamic conditional correlation GARCH (DCC-GARCH) model \cite{engle2002dynamic} as a classic econometric benchmark extensively used for multivariate correlation modeling. In addition, we consider several deep learning-based models}: TimeGAN \cite{yoon2019time}, TimeGrad \cite{rasul2021autoregressive}, CSDI \cite{tashiro2021csdi}, Diffusion-TS \cite{yuan2024diffusion}, MG-TSD \cite{fan2024mg} , and SigCWGAN \cite{liao2024sig}. This selection comprises a widely-recognized time-series generative model (TimeGAN), several prominent diffusion-based models (TimeGrad, CSDI, Diffusion-TS, MG-TSD), and a model tailored for asset return forecasting (SigCWGAN). {\color{black}Following standard convention, DCC-GARCH parameters are estimated via rolling windows during the test period, whereas deep learning parameters remain fixed after being optimized during training.}

These models are employed for forecasting through conditional generation where future time-series distributions are predicted based on historical observations. 
A distinctive feature of our proposed Diffolio is its architecture, particularly designed to incorporate both asset-specific and systematic covariates.
Since the baseline models were not originally designed to accommodate such covariates, we implement and evaluate two versions of each to ensure a fair comparison.
We denote the original version using only historical return data as (H), and the modified version that includes the additional covariates as (H+C) (e.g., TimeGAN (H) and TimeGAN (H+C)).

{\color{black}The method of incorporating covariates is tailored to each model. In the case of DCC-GARCH, covariates are integrated into the mean equation; specifically, while DCC-GARCH (H) employs a vector autoregression (VAR) to model mean returns based on historical return data, DCC-GARCH (H+C) leverages Ridge regression\footnote{While OLS is one of the most widely-used traditional approach for covariate-based predictive regressions (e.g., \cite{welch2008comprehensive}), we observe that it yields significantly unstable results and faces over-parameterization issues in our high-dimensional setting as shown in Gu et al. \cite{gu2020empirical}. Thus, we employ Ridge regression to ensure more robust estimation of the mean returns.} that includes lagged returns alongside systematic and asset-specific covariates. For both versions, the resulting residuals are modeled using univariate GARCH(1,1) for volatility and DCC framework for the dynamic correlation structure. Regarding the deep learning baselines, the covariates are integrated into the architectural components responsible for predictive output. The models utilizing latent space embeddings---TimeGAN, TimeGrad, and MG-TSD---concatenate the covariate vector with their latent embeddings.} For Diffusion-TS, covariates were concatenated directly to the historical input data. In SigCWGAN, they were appended to the signature of the historical path, and for CSDI, they were integrated as side information following its prescribed mechanism for auxiliary features. 
\textcolor{black}{While these modifications were designed to facilitate a fair comparison by providing all models with the same set of covariates, their inherent structural designs may not be as suited for incorporating two distinct types of covariates as Diffolio's. Consequently, while these adaptations follow standard practices for each architecture, these fundamental structural differences should be kept in mind when comparing the performance results.}

\textbf{Implementation Details.} 
We select the optimal hyperparameters based on the configuration that achieves the lowest ES on the validation set.
This metric was chosen because, as a multivariate generalization of the CRPS, it provides a balanced evaluation of probabilistic accuracy while also somewhat reflecting cross-sectional dependencies.
To be specific, we set the input window size to 63, which corresponds to one quarter of trading days, to provide the model with sufficient historical context for capturing asset dynamics. 
The model architecture is further composed of a hidden dimension of 128 and 4 heads for all attention blocks, and the MLP within each attention block has a hidden dimension of 512.
All embedding and decoding layers are implemented as linear layers.
We set the total diffusion time steps to $T=1,000$ with a linear noise schedule ranging from $10^{-4}$ to 0.02 as in~\cite{ho2020denoising}. A 32-dimensional sinusoidal positional embedding is used to encode each step. 
We train the model for a total of 100,000 steps using the AdamW optimizer and a batch size of 1024. 
The learning rate is set to a maximum value of $10^{-4}$ with a linear warmup for the first 1,000 steps followed by a cosine decay schedule for the remaining steps.
The coefficient for the correlation-guided regularizer, $\lambda_{corr}$, is set to 0.05. 
During inference, we use 50 steps for the DDIM sampling process.

We keep the experimental setup for the baseline models consistent with Diffolio wherever applicable. 
We employ the same AdamW optimizer, a batch size of 1024, and the learning rate schedule. 
{\color{black} The input window size of 63 is consistently applied across all models; specifically, for DCC-GARCH, this window is utilized for the rolling-window estimation of the conditional volatility and dynamic correlation.}
For all diffusion-based baselines, we utilize the same 32-dimensional sinusoidal embedding for the diffusion time steps. 
Consistent with the hyperparameter selection for Diffolio, the optimal values for key model-specific hyperparameters of each baseline are chosen from their respective search spaces to minimize the ES on the validation set.
{\color{black} Detailed descriptions of Diffolio's hyperparameter search spaces, along with a sensitivity analysis regarding its diffusion and DDIM sampling steps, and the configurations of the baseline models are provided in \ref{appendix: hyperparam}.}

\subsection{Empirical Results and Analysis}

\begin{figure*}[!t]
    \centering
    \includegraphics[width=\linewidth]{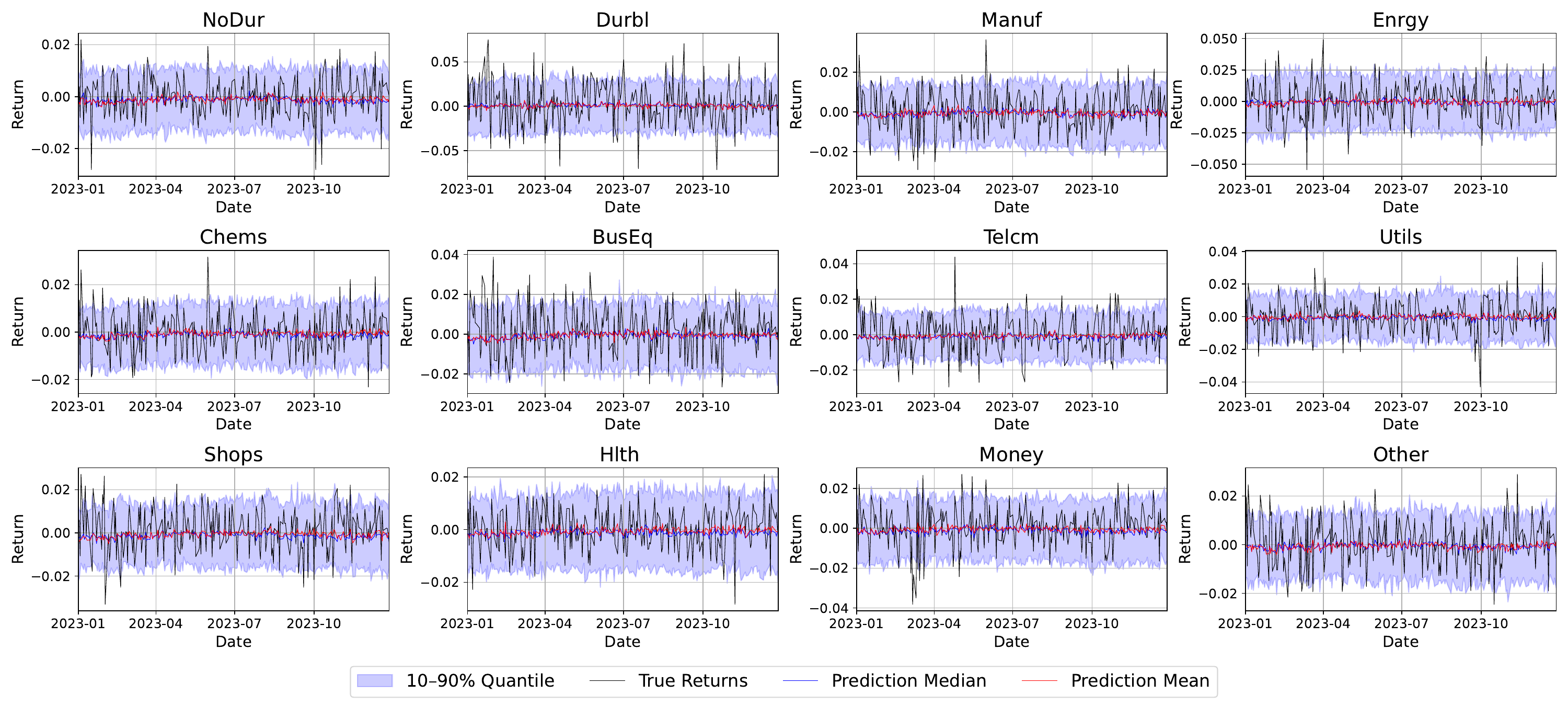}
    \caption{Example of predicted excess returns for 12 assets in 2023 generated by Diffolio. The plot shows the mean, median, and the 10\%-90\% quantile range from 100 generated sample paths, along with the ground truth.
    }
    \label{fig:2023}
\end{figure*}

We begin with a qualitative figure of the generated forecast samples. As an illustrative example, Figure \ref{fig:2023} shows the predicted excess returns for 12 assets during 2023, the most recent year of the test period. For each asset, the figure displays the mean, median, and the 10\%-90\% quantile range derived from 100 generated sample paths, plotted alongside the ground truth. Visually, the variation in the forecasts generated by Diffolio appears consistent with that of the actual data. Furthermore, the mean and median of the predictions, while exhibiting less volatility than the true values, seem to capture the general directional movements of the market.

{\color{black}
\subsubsection{Statistical Accuracy and Cross-sectional Dependency}}
For a quantitative analysis, we summarize the performance of Diffolio and the baseline models across the entire test period in Table \ref{tab:performance_comparison_detailed}. \textcolor{black}{The table presents four primary metrics: CRPS for marginal accuracy, ES for multivariate accuracy, and MVP-SR and GOP-CE for portfolio performance. Additionally, we include CorrScore and LogDet as supplementary diagnostics to provide a more comprehensive assessment of how accurately the models reproduce the cross-sectional dependency structures.}

\begin{table*}[!t]
\centering
\caption{Comparison of evaluation metrics and supplementary diagnostics for Diffolio and the baseline models. CRPS is reported as the mean ($\pm$ standard deviation) calculated marginally for each of the 12 assets. Lower values are better for CRPS, ES, CorrScore, \textcolor{black}{and LogDet}, while higher values are better for MVP-SR and GOP-CE. For each evaluation metric, the best result is shown in \textbf{bold} and the second-best is \underline{underlined}.}
\label{tab:performance_comparison_detailed}
\resizebox{\textwidth}{!}{
\begin{tabular}{lcccc ccc}
\toprule
& \multicolumn{4}{c}{Evaluation Metrics} & & \multicolumn{2}{c}{Supplementary Diagnostics} \\
\cmidrule(lr){2-5} \cmidrule(lr){7-8}
\textbf{Model} & \textbf{CRPS} & \textbf{ES} & \textbf{MVP-SR} & \textbf{GOP-CE} & & \textbf{CorrScore} & \textbf{LogDet} \\
\midrule
DCC-GARCH (H)$^\dagger$ & 0.007974 ($\pm$ 0.001777) & 0.032154 & 0.3752 & 0.0553 & & \textbf{0.2444} & \textbf{0.1188} \\
DCC-GARCH (H+C)$^\dagger$ & 0.007340 ($\pm$ 0.001695) & 0.029957 & 0.4062 & 0.0975 & & 2.5825 & 2.8481 \\
TimeGAN (H) & 0.009421 ($\pm$ 0.002538) & 0.038188 & 0.4080 & 0.0729 & & 3.0991 & 894506.8 \\
TimeGAN (H+C) & 0.009544 ($\pm$ 0.002004) & 0.038601 & 0.3732 & 0.0527 & & 3.1025 & 34849.7 \\
TimeGrad (H) & 0.007329 ($\pm$ 0.001770) & 0.029622 & 0.4312 & 0.0450 & & 6.6886 & 9.0429 \\
TimeGrad (H+C) & 0.007363 ($\pm$ 0.001754) & 0.029721 & 0.4721 & 0.0527 & & 6.7158 & 8.7694 \\
CSDI (H) & 0.007377 ($\pm$ 0.001712) & 0.029701 & \underline{0.4886} & 0.0606 & & 5.6606 & 7.2736 \\
CSDI (H+C) & 0.007409 ($\pm$ 0.001667) & 0.029903 & 0.3694 & 0.0367 & & 7.2040 & 8.5027 \\
Diffusion-TS (H) & \textbf{0.007159 ($\pm$ 0.001616)} & 0.029034 & 0.4565 & \underline{0.1124} & & 2.0170 & 84.5137 \\
Diffusion-TS (H+C) & 0.007343 ($\pm$ 0.001739) & 0.029854 & 0.2932 & 0.0533 & & 2.3362 & 390.5657 \\
MG-TSD (H) & 0.007214 ($\pm$ 0.001640) & \underline{0.029027} & 0.4155 & 0.1011 & & 2.7654 & 3.6756 \\
MG-TSD (H+C) & 0.007244 ($\pm$ 0.001627) & 0.029157 & 0.4805 & 0.0826 & & 6.0587 & 8.0304 \\
SigCWGAN (H) & 0.007213 ($\pm$ 0.001624) & 0.029132 & 0.4196 & 0.0720 & & 1.5586 & 150.0877 \\
SigCWGAN (H+C) & 0.007338 ($\pm$ 0.001686) & 0.029534 & 0.3782 & 0.0993 & & 1.4706 & 43.8911 \\
\midrule
\textbf{Diffolio (Ours)} & \underline{0.007169 ($\pm$ 0.001601)} & \textbf{0.028960} & \textbf{0.7206} & \textbf{0.1611} & & \underline{1.0808} & \underline{2.1982} \\
\bottomrule
\multicolumn{8}{l}{\footnotesize $^\dagger$ Parameters of these models are updated via rolling-window estimation over the test period.}
\end{tabular}
}
\end{table*}

The results show that Diffolio achieves superior performance across all evaluation metrics with the sole exception of CRPS, where it ranks as the second-best. Notably, the difference in CRPS between Diffolio and the top-performing model, Diffusion-TS (H), is slight, especially when compared to the gap between Diffolio and the third-best model, SigCWGAN (H). It is also worth noting that Diffolio exhibits the lowest standard deviation in CRPS across the 12 assets, which suggests a more robust forecasting capability across individual assets.



In contrast, for the metrics evaluating multivariate accuracy, Diffolio demonstrates a clear advantage. It achieves the best ES, indicating its effectiveness in capturing the joint distribution of multi-asset excess returns. This superior performance in modeling cross-sectional dependencies is further corroborated by the CorrScore {\color{black} and LogDet. Diffolio significantly outperforms all baselines except DCC-GARCH (H)\footnote{\color{black} While the traditional DCC-GARCH (H) yields the lowest CorrScore and LogDet, this is due to its use of rolling windows estimation. This approach possesses a methodological advantage by updating the correlation structure with data up to the immediate past, whereas deep learning models utilize fixed parameters. }, successfully capturing robust cross-sectional dependencies while ensuring well-conditioned correlation structures. On the contrary, while deep learning baselines such as SigCWGAN or Diffusion-TS may achieve relatively low CorrScores, their disproportionately high LogDet values indicate that their generated correlation matrices are near-singular.
}

\begin{figure}[!t]
    \centering
    \includegraphics[width=\linewidth]{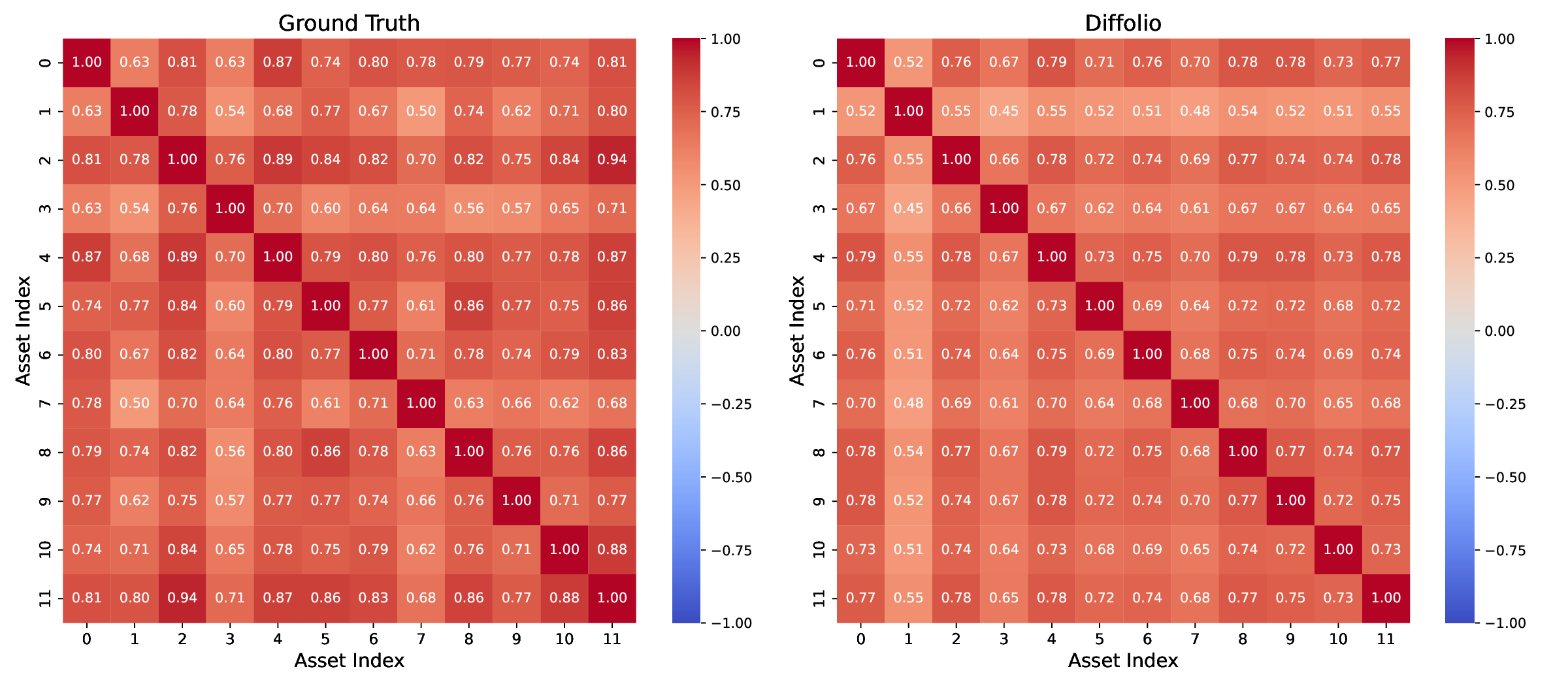}
    \caption{Comparison of correlation matrices estimated via samples from real data (left) and the mean sample path of Diffolio's synthetic data (right).
    }
    \label{fig:corrmatrix}
\end{figure}

We attribute Diffolio's ability to capture the dependencies to three key design choices: (1) the hierarchical design for effectively incorporating covariates known to reflect cross-sectional dynamics; (2) the correlation-guided regularizer, $\mathcal{L}_{\text{corr}}$ in \eqref{eqn:cossim}, that guides the model to better learn the cross-asset dependency using a stable correlation matrix estimated via the Ledoit-Wolf shrinkage method; and (3) the self-attention mechanism adept at capturing cross-asset relationships. 
\textcolor{black}{To visually substantiate the model’s ability to capture the dependency structure, Figure \ref{fig:corrmatrix} displays the unconditional correlation matrices for the real data and Diffolio's mean synthetic data estimated over the test period, demonstrating that the complex dependency structure among assets is accurately preserved.
A more detailed analysis of the impacts of the covariates and the correlation-guided regularizer is provided in the subsequent ablation study.}

{\color{black} The significance of these architectural choices is further underscored by the performance of baseline models, which often fail to benefit from the inclusion of covariates. In traditional frameworks like DCC-GARCH (H+C), incorporating covariates improves predictive accuracy but is accompanied by a degradation in diagnostic scores; as the mean equation explains a larger portion of return variance, the resulting noisier residuals complicate the estimation of dynamic correlation structures. Furthermore, many deep learning baselines exhibit an even more comprehensive performance degradation, where both predictive accuracy and diagnostic scores decline upon the introduction of covariates. This points to a fundamental architectural incompatibility and stems from several factors. First, instead of isolating the asset-specific covariates for each asset, these models allow features to leak across assets. This leads to signal interference that obscures distinct, meaningful signals. Second, they lack a refinement stage to mitigate the low signal-to-noise ratio inherent in individual returns and characteristics, thereby utilizing noisy data directly for prediction. Finally, these models are not equipped to organize and integrate asset-level signals within a market-wide context while accounting for systematic covariates. Consequently, they often become preoccupied with uninformative noise rather than effectively leveraging the predictive information in the additional covariates.
Diffolio addresses these structural hurdles by employing a hierarchical architecture specifically designed to isolate and distill individual asset-level signals before integrating them with broader market dynamics.
}


{\color{black}
\subsubsection{Forecast Calibration and Probabilistic Reliability}

In probabilistic forecasting, especially for financial risk management, the reliability of the predicted distributions is as critical as their accuracy. A well-calibrated model ensures that the predicted uncertainty (e.g., the width of a 95\% prediction interval) accurately reflects the frequency of observed outcomes. To quantify this reliability, we employ the PICP, which represents the actual frequency of observations falling within the predicted intervals, and the ACE, which measures the deviation of this empirical coverage from the target confidence level. The results for Diffolio and the baseline models are summarized in Table \ref{tab:calibration_summary}.
}

\begin{table*}[!t]
\centering
\caption{Summary of PICP and ACE (in {parenthesis}) at various target confidence levels. Better calibration is indicated by PICP values closer to the target coverage and ACE values closer to zero. Best results are in \textbf{bold} and second-best are \underline{underlined}.}
\label{tab:calibration_summary}
\resizebox{\textwidth}{!}{
\begin{tabular}{lccccc}
\toprule
& \multicolumn{5}{c}{\textbf{Target Coverage (Prediction Level)}} \\
\cmidrule(lr){2-6}
\textbf{Model} & \textbf{50\%} & \textbf{80\%} & \textbf{90\%} & \textbf{95\%} & \textbf{99\%} \\
\midrule
DCC-GARCH (H) & 0.4080 ($-$0.0920) & 0.6703 ($-$0.1297) & 0.7746 ($-$0.1254) & 0.8378 ($-$0.1122) & 0.9030 ($-$0.0870) \\
DCC-GARCH (H+C) & 0.4500 ($-$0.0500) & 0.7176 ($-$0.0824) & 0.8146 ($-$0.0854) & 0.8717 ($-$0.0783) & 0.9256 ($-$0.0644) \\
TimeGAN (H) & 0.0018 ($-$0.4982) & 0.0035 ($-$0.7965) & 0.0047 ($-$0.8953) & 0.0056 ($-$0.9444) & 0.0068 ($-$0.9832) \\
TimeGAN (H+C) & 0.0030 ($-$0.4970) & 0.0059 ($-$0.7941) & 0.0076 ($-$0.8924) & 0.0089 ($-$0.9411) & 0.0110 ($-$0.9789) \\
TimeGrad (H) & 0.3697 ($-$0.1303) & 0.6182 ($-$0.1818) & 0.7227 ($-$0.1773) & 0.7878 ($-$0.1622) & 0.8628 ($-$0.1272) \\
TimeGrad (H+C) & 0.3603 ($-$0.1397) & 0.6056 ($-$0.1944) & 0.7093 ($-$0.1907) & 0.7758 ($-$0.1742) & 0.8530 ($-$0.1370) \\
CSDI (H) & 0.3457 ($-$0.1543) & 0.5869 ($-$0.2131) & 0.6945 ($-$0.2055) & 0.7635 ($-$0.1865) & 0.8491 ($-$0.1409) \\
CSDI (H+C) & 0.3857 ($-$0.1143) & 0.6367 ($-$0.1633) & 0.7357 ($-$0.1643) & 0.7967 ($-$0.1533) & 0.8696 ($-$0.1204) \\
Diffusion-TS (H) & \textbf{0.5350 (+0.0350)} & \underline{0.7858 ($-$0.0142)} & \underline{0.8705 ($-$0.0295)} & \underline{0.9164 ($-$0.0336)} & \underline{0.9659 ($-$0.0241)} \\
Diffusion-TS (H+C) & 0.3507 ($-$0.1493) & 0.6231 ($-$0.1769) & 0.7475 ($-$0.1525) & 0.8249 ($-$0.1251) & 0.9160 ($-$0.0740) \\
MG-TSD (H) & 0.4527 ($-$0.0473) & 0.7293 ($-$0.0707) & 0.8312 ($-$0.0688) & 0.8884 ($-$0.0616) & 0.9457 ($-$0.0443) \\
MG-TSD (H+C) & 0.4247 ($-$0.0753) & 0.7052 ($-$0.0948) & 0.8150 ($-$0.0850) & 0.8798 ($-$0.0702) & 0.9446 ($-$0.0454) \\
SigCWGAN (H) & 0.5396 (+0.0396) & \textbf{0.7913 ($-$0.0087)} & 0.8668 ($-$0.0332) & 0.9083 ($-$0.0417) & 0.9501 ($-$0.0399) \\
SigCWGAN (H+C) & 0.3385 ($-$0.1615) & 0.6287 ($-$0.1713) & 0.7586 ($-$0.1414) & 0.8400 ($-$0.1100) & 0.9221 ($-$0.0679) \\
\midrule
\textbf{Diffolio (Ours)} & \underline{0.5386 (+0.0386)} & 0.8163 (+0.0163) & \textbf{0.9030 (+0.0030)} & \textbf{0.9466 ($-$0.0034)} & \textbf{0.9812 ($-$0.0088)} \\
\bottomrule
\end{tabular}
}
\end{table*}

{\color{black} The results in Table \ref{tab:calibration_summary} highlight Diffolio's remarkable ability to provide reliable uncertainty estimates. While several baselines exhibit significant negative ACE values---indicating that intervals are too narrow to cover actual returns---Diffolio maintains low calibration errors across the entire range of target coverage levels. Specifically, Diffolio ranks as the second-best model at the 50\% level and third-best at the 80\% level, with marginal differences from the top performers.

Crucially, at the higher prediction levels (90\%, 95\%, and 99\%), which are paramount for tail-risk management, Diffolio consistently achieves the best performance. For instance, at the 95\% target level, Diffolio yields a PICP of 0.9466 with an ACE of only -0.0034, remarkably outperforming all baselines which tend to underestimate tail risks significantly. This indicates that the predicted distributions from Diffolio offer a highly faithful representation of the tail density, ensuring that the model’s probabilistic characterization of extreme events remains statistically aligned with actual market realizations.
}

\begin{figure}[!ht] \centering \includegraphics[width=0.8\linewidth]{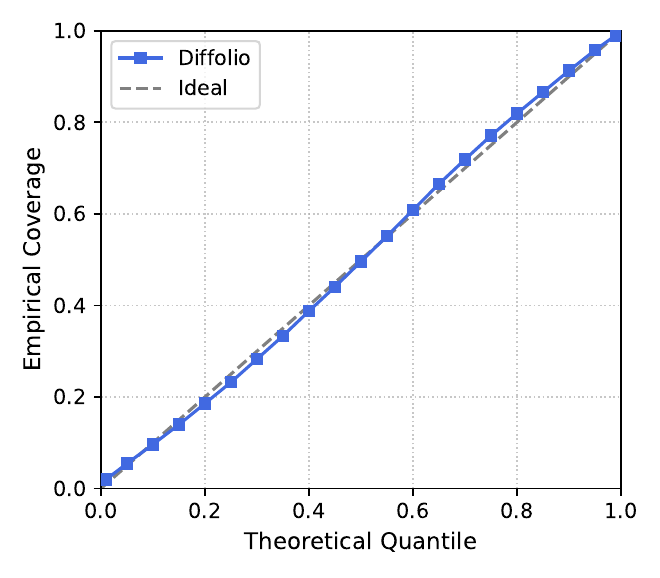} \caption{Reliability diagram of Diffolio. The theoretical quantiles are plotted against the empirical coverage calculated over the entire test period. The dashed diagonal line represents the ideal calibration.} \label{fig:reliability_diffolio} \end{figure}

{\color{black} The calibration quality is further substantiated by the reliability diagram in Figure \ref{fig:reliability_diffolio}. The curve for Diffolio aligns closely with the ideal diagonal line, indicating that the predicted quantiles are highly consistent with the empirical frequencies observed in the data. While most baseline models show substantial deviations from the ideal line due to their inability to effectively handle financial noise and complex dependencies, Diffolio remains robust. For a comparative view, the reliability diagrams for all baseline models are provided in \ref{appendix: reliability diagram}. }

\begin{figure*}[!t]
    \centering
    \includegraphics[width=0.92\linewidth]{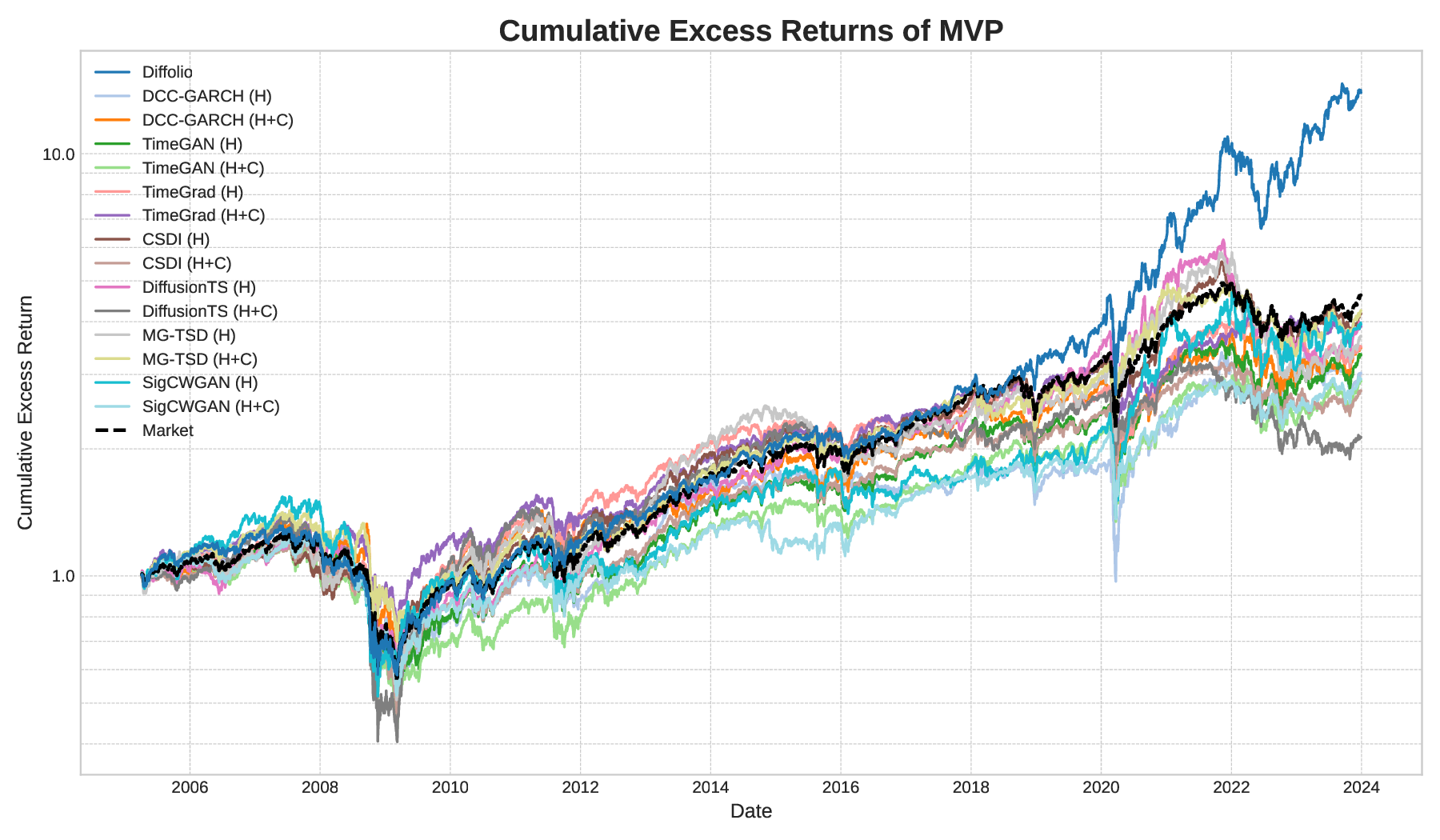}
    \caption{Cumulative excess returns of the MVP for Diffolio, baseline models, and the market benchmark. The returns are plotted on a logarithmic scale.
    }
    \label{fig:mvp_cumulative}
\end{figure*}

\begin{figure*}[!t]
    \centering
    \includegraphics[width=0.92\linewidth]{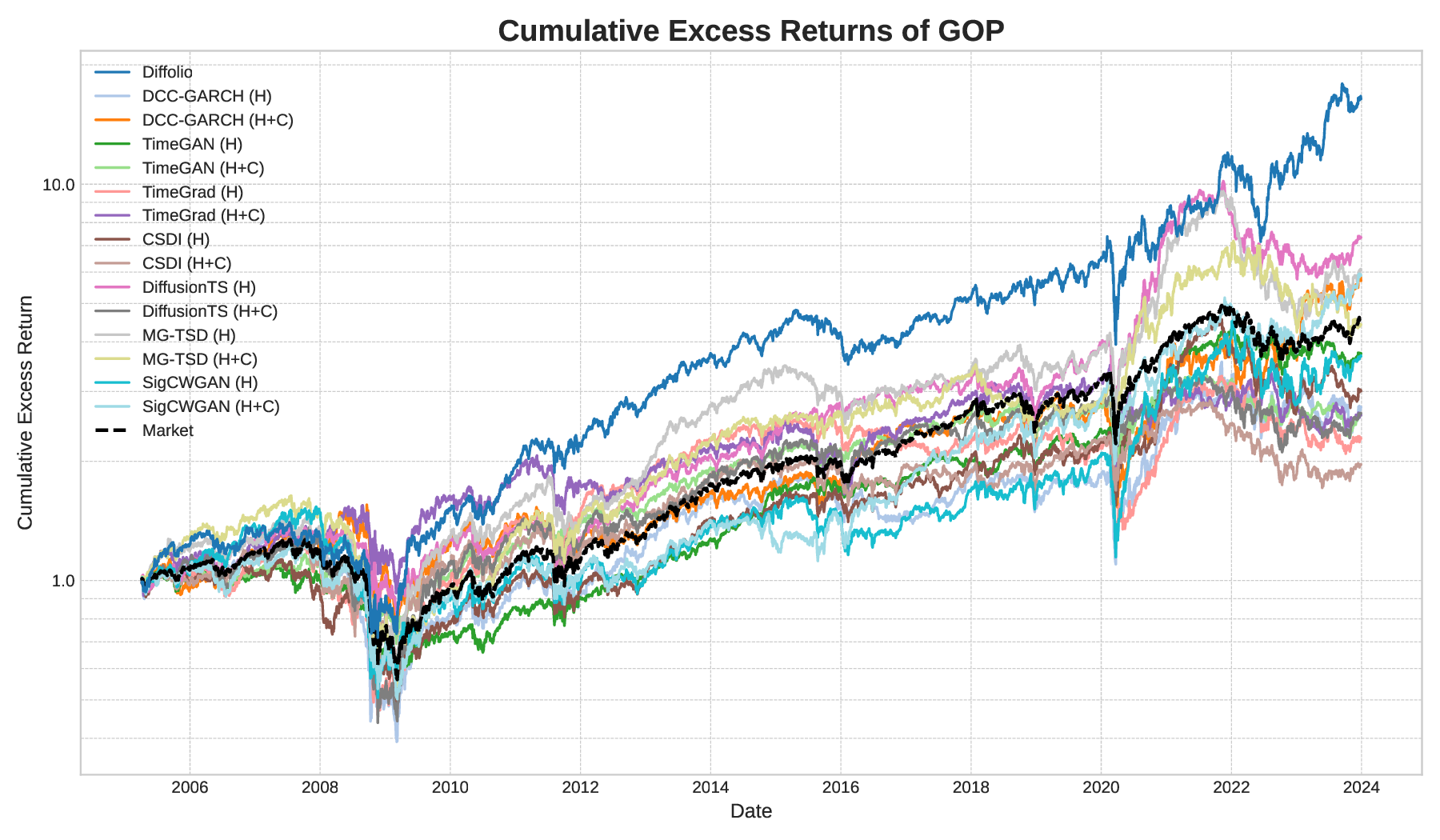}
    \caption{Cumulative excess returns of the GOP for Diffolio, baseline models, and the market benchmark. The returns are plotted on a logarithmic scale.}
    \label{fig:gop_cumulative}
\end{figure*}

{\color{black}
\subsubsection{Economic Significance and Portfolio Performance}}
Beyond statistical accuracy, we now evaluate the economic significance of the forecasts. 
Table \ref{tab:portfolio_performance} provides a comprehensive summary of the MVP and GOP strategies, including SR, Ret, Vol, MDD, \textcolor{black}{turnover,} and CE. For a chronological perspective, Figures \ref{fig:mvp_cumulative} and \ref{fig:gop_cumulative} illustrate the cumulative returns of these portfolios, respectively, benchmarked against baseline models and a market portfolio. The market portfolio, proxied by the value-weighted returns on all U.S. stocks listed on the NYSE, AMEX, or NASDAQ in our experiment, serves as a formidable benchmark. Grounded in modern portfolio theory, it represents the fully diversified portfolio, which in principle cancels out the idiosyncratic risks of individual assets and thus contains only systematic risk. As such, consistently outperforming this highly efficient benchmark is known to be challenging.

\begin{table*}[!t]
\centering
\caption{Comprehensive portfolio performance for the MVP and GOP strategies. All reported evaluation metrics and summary statistics are annualized, except for Turnover which is reported as an average monthly one-way value.  For each primary evaluation metric, MVP-SR and GOP-CE, higher values indicate better performance, and the best result is shown in \textbf{bold} with the second-best \underline{underlined}. As the remaining summary statistics serve as supplements to the analysis, they are not highlighted.}
\label{tab:portfolio_performance}
\resizebox{\textwidth}{!}{%
\begin{tabular}{l ccccc cccccc}
\toprule
& \multicolumn{5}{c}{\textbf{MVP}} & \multicolumn{6}{c}{\textbf{GOP}} \\
\cmidrule(lr){2-6} \cmidrule(lr){7-12}
\textbf{Model} & \textbf{SR} & \textbf{Ret} & \textbf{Vol} & \textbf{MDD} & \textbf{Turnover} & \textbf{SR} & \textbf{Ret} & \textbf{Vol} & \textbf{MDD} & \textbf{Turnover} & \textbf{CE} \\
\midrule
DCC-GARCH (H) & 0.3752 & 0.0847 & 0.2258 & -0.6533 & 17.7108 & 0.3349 & 0.0900 & 0.2688 & -0.7335 & 17.5621 & 0.0553 \\
DCC-GARCH (H+C) & 0.4062 & 0.0925 & 0.2276 & -0.5566 & 10.8140 & 0.4825 & 0.1286 & 0.2665 & -0.5382 & 10.4605 & 0.0975 \\
TimeGAN (H) & 0.4080 & 0.0872 & 0.2138 & -0.6409 & 0.1456 & 0.4747 & 0.0873 & 0.1840 & -0.5264 & 17.1944 & 0.0729 \\
TimeGAN (H+C) & 0.3732 & 0.0793 & 0.2124 & -0.6401 & 2.2591 & 0.3839 & 0.0663 & 0.1728 & -0.4062 & 8.4461 & 0.0527 \\
TimeGrad (H) & 0.4312 & 0.0872 & 0.2023 & -0.5705 & 16.0926 & 0.3075 & 0.0704 & 0.2290 & -0.6443 & 18.5055 & 0.0450 \\
TimeGrad (H+C) & 0.4721 & 0.0915 & 0.1938 & -0.4648 & 9.9261 & 0.3349 & 0.0799 & 0.2385 & -0.4623 & 12.9288 & 0.0527 \\
CSDI (H) & 0.4886 & 0.0955 & 0.1955 & -0.5086 & 16.0174 & 0.3724 & 0.0848 & 0.2278 & -0.5701 & 17.8579 & 0.0606 \\
CSDI (H+C) & 0.3694 & 0.0740 & 0.2003 & -0.6083 & 7.6319 & 0.2692 & 0.0657 & 0.2440 & -0.5993 & 8.6712 & 0.0367 \\
Diffusion-TS (H) & 0.4565 & 0.0950 & 0.2082 & -0.4974 & 11.4517 & 0.5960 & 0.1305 & 0.2189 & -0.4344 & 12.8257 & \underline{0.1124} \\
Diffusion-TS (H+C) & 0.2932 & 0.0654 & 0.2229 & -0.7072 & 1.7436 & 0.3433 & 0.0773 & 0.2252 & -0.6783 & 2.3442 & 0.0533 \\
MG-TSD (H) & 0.4155 & 0.0969 & 0.2331 & -0.5965 & 18.6626 & 0.5239 & 0.1246 & 0.2378 & -0.6139 & 18.6124 & 0.1011 \\
MG-TSD (H+C) & 0.4805 & 0.0983 & 0.2045 & -0.5113 & 17.6219 & 0.4494 & 0.1088 & 0.2422 & -0.6696 & 17.2904 & 0.0826 \\
SigCWGAN (H) & 0.4196 & 0.1040 & 0.2479 & -0.6636 & 16.8614 & 0.4012 & 0.1018 & 0.2537 & -0.6703 & 16.6566 & 0.0720 \\
SigCWGAN (H+C) & 0.3782 & 0.0794 & 0.2100 & -0.5817 & 12.6414 & 0.5163 & 0.1235 & 0.2391 & -0.6041 & 11.6213 & 0.0993 \\
\midrule
Market & \underline{0.5119} & 0.1012 & 0.1977 & -0.5565 & - & 0.5119 & 0.1012 & 0.1977 & -0.5565 & - & 0.0850 \\
\midrule
\textbf{Diffolio (Ours)} & \textbf{0.7206} & 0.1680 & 0.2332 & -0.5577 & 16.3672 & 0.7042 & 0.1833 & 0.2602 & -0.4882 & 17.2513 & \textbf{0.1611} \\
\bottomrule
\end{tabular}%
}
\end{table*}

As shown in the MVP cumulative return plot in Figure \ref{fig:mvp_cumulative}, Diffolio's performance is comparable to the market until 2018, after which it begins to significantly outperform all other models and the market.
Notably, from 2022 onwards, Diffolio is the only model to outperform the market benchmark in terms of cumulative return.
The GOP cumulative return presented in Figure \ref{fig:gop_cumulative} shows an even more pronounced result, with Diffolio consistently and substantially outperforming all other methods and the market from mid-2010 onwards.

While cumulative return is an important indicator, a higher return often comes with increased risk. A more informative metric is the SR, which measures the excess return earned per unit of risk. In both MVP and GOP experiments, Diffolio achieves the highest SR, indicating that its forecasting ability allows for the construction of portfolios with superior risk-adjusted performance. Similarly, for the GOP strategy, which maximizes an investor's expected logarithmic utility, we focus on the CE. The CE translates the abstract concept of utility into an intuitive financial value: the guaranteed risk-free return an investor would consider equivalent to the uncertain portfolio outcome. 
Diffolio has the highest CE value of 0.1611, meaning that for a growth-oriented risk-averse investor, this uncertain portfolio strategy yields a utility equivalent to an investment that guarantees a risk-free annual return of 16.11\%.

{\color{black}Note that, compared with the results in Table \ref{tab:performance_comparison_detailed} and Table \ref{tab:portfolio_performance}, a discernible discrepancy often exists between statistical metrics and portfolio performance. This discrepancy arises because statistical measures like CRPS and ES are typically calculated as averages over the entire test period, where local forecasting errors can be mitigated by high accuracy in other periods. In contrast, portfolio-based metrics such as MVP-SR and GOP-CE represent risk-adjusted measures, making them highly sensitive to both temporal consistency and precise tail-risk modeling.
The observed performance discrepancy for some baselines such as SigCWGAN (H) suggests that they may struggle to consistently track the time-varying joint density even while maintaining competitive average precision. Since portfolio-based metrics penalize volatility and large drawdowns, such intermittent but substantial failures in modeling the dynamic joint density lead to disproportionate risk, making it difficult to recover overall performance even with subsequent accurate forecasts. 
The fact that Diffolio delivers strong results in statistical metrics and portfolio performance, along with forecast calibration at the tails as in Table \ref{tab:calibration_summary}, indicates its capability to consistently capture the time-varying joint density across the entire test period and its economic reliability. }

{\color{black}
Another notable observation from Table \ref{tab:portfolio_performance} is the relatively high turnover exhibited by most models. This is a typical attribute of daily-rebalanced MVP and GOP portfolios, which frequently adjust weights to exploit time-varying market signals \citep{gu2020empirical}. While elevated turnover is typically unfavorable in practical settings due to transaction costs, it is not considered detrimental in this frictionless experimental environment. Since the portfolio optimization process is conducted without transaction cost constraints, the models focus exclusively on the predicted joint distribution to determine optimal weights at each time step. In this context, high turnover is incidental to the primary objectives of maximizing MVP-SR or GOP-CE; it serves as a measure of the model's responsiveness, reflecting the frequency and magnitude of strategic weight adjustments in response to evolving market conditions. Conversely, the lower turnover observed in models such as TimeGAN and Diffusion-TS (H+C) under the MVP strategy suggests a lack of dynamic responsiveness to daily variations in the multivariate distribution.
}

{\color{black}
Following the portfolio results in the frictionless environment, we assess the practical viability of our proposed model under more realistic investment conditions. To reflect practical considerations, {\color{black}we explicitly integrate a transaction cost of 10 basis points (bps) as a penalty term directly into the portfolio optimization objective function and adjust the rebalancing frequency to a five-day interval (weekly)\footnote{While monthly or quarterly rebalancing could be adopted, we selected a weekly interval to reconcile the investment horizon with our one-day forecasting horizon.}, and its detailed formulations are provided in \ref{appendix: evaluation metrics}. This ensures that the resulting portfolio allocation is cost-aware, as the optimization process accounts for trading frictions when determining the optimal weights.} 
In the financial literature, transaction costs are often viewed as a regularizer for portfolio rebalancing, as these frictions impose a hurdle that discourages frequent reallocations unless the expected gains from weight adjustments are sufficient to offset the transaction costs. Viewed from this perspective, an exceptionally low turnover under such frictions suggests that the temporal shifts in the predicted joint return distribution are not substantial enough to overcome the cost penalty, reflecting a tendency to produce nearly invariant forecasts across all time steps or a lack of predictive conviction regarding time-varying market dynamics.

\begin{table*}[!t]
\centering
\caption{Comprehensive portfolio performance for the MVP and GOP strategies under a 10 bp transaction cost and weekly rebalancing. All reported evaluation metrics and summary statistics are annualized, except for Turnover which is reported as an average monthly one-way value. For each primary evaluation metric, MVP-SR and GOP-CE, the best result is shown in \textbf{bold} and the second-best is \underline{underlined}.}
\label{tab:performance_weekly_final}
\resizebox{\textwidth}{!}{
\begin{tabular}{l ccccc cccccc}
\toprule
& \multicolumn{5}{c}{\textbf{MVP}} & \multicolumn{6}{c}{\textbf{GOP}} \\
\cmidrule(lr){2-6} \cmidrule(lr){7-12}
\textbf{Model} & \textbf{SR} & \textbf{Ret} & \textbf{Vol} & \textbf{MDD} & \textbf{Turnover} & \textbf{SR} & \textbf{Ret} & \textbf{Vol} & \textbf{MDD} & \textbf{Turnover} & \textbf{CE} \\
\midrule
DCC-GARCH (H) & 0.1056 & 0.0242 & 0.2295 & -0.6250 & 2.6851 & 0.2353 & 0.0617 & 0.2622 & -0.5658 & 2.3992 & 0.0273 \\
DCC-GARCH (H+C) & 0.2276 & 0.0553 & 0.2429 & -0.6339 & 2.0956 & 0.1768 & 0.0483 & 0.2731 & -0.6802 & 1.8543 & 0.0106 \\
TimeGAN (H) & 0.5532$^\ddagger$ & 0.0858 & 0.1552 & -0.3912 & 0.0000 & 0.5840 & 0.1014 & 0.1736 & -0.4775 & 0.0255 & 0.0889$^\ddagger$ \\
TimeGAN (H+C) & \textbf{0.6114}$^\ddagger$ & 0.1156 & 0.1891 & -0.4633 & 0.0000 & 0.5170 & 0.0847 & 0.1638 & -0.4024 & 0.0372 & 0.0737$^\ddagger$ \\
TimeGrad (H) & 0.3726$^\dagger$ & 0.0777 & 0.2087 & -0.6934 & 0.0763 & 0.2697 & 0.0562 & 0.2084 & -0.5617 & 0.2514 & 0.0347 \\
TimeGrad (H+C) & 0.3170$^\dagger$ & 0.0623 & 0.1966 & -0.5218 & 0.0611 & 0.1765 & 0.0512 & 0.2903 & -0.7667 & 0.0548 & 0.0089$^\dagger$ \\
CSDI (H) & 0.5131$^\dagger$ & 0.0937 & 0.1827 & -0.5306 & 0.0683 & 0.5163 & 0.1000 & 0.1937 & -0.5788 & 0.0475 & 0.0846$^\ddagger$ \\
CSDI (H+C) & 0.3992$^\dagger$ & 0.0799 & 0.2002 & -0.6045 & 0.0851 & 0.3774 & 0.0747 & 0.1980 & -0.5788 & 0.0043 & 0.0567$^\ddagger$ \\
Diffusion-TS (H) & \underline{0.5721}$^\ddagger$ & 0.0992 & 0.1734 & -0.3712 & 0.0000 & 0.6462 & 0.1163 & 0.1800 & -0.4652 & 0.0261 & 0.1045$^\ddagger$ \\
Diffusion-TS (H+C) & 0.3922$^\ddagger$ & 0.0772 & 0.1968 & -0.5764 & 0.0000 & 0.6431 & 0.1447 & 0.2249 & -0.5473 & 0.0008 & \underline{0.1267}$^\ddagger$ \\
MG-TSD (H) & 0.5177 & 0.1349 & 0.2606 & -0.5615 & 0.2165 & 0.4596 & 0.1015 & 0.2208 & -0.4980 & 0.4912 & 0.0798 \\
MG-TSD (H+C) & 0.3149 & 0.0762 & 0.2421 & -0.7460 & 0.7016 & 0.4292 & 0.1047 & 0.2439 & -0.6544 & 1.1362 & 0.0776 \\
SigCWGAN (H) & 0.5553$^\ddagger$ & 0.1185 & 0.2133 & -0.5473 & 0.0170 & 0.5942 & 0.1380 & 0.2322 & -0.5700 & 0.0987 & 0.1167$^\dagger$ \\
SigCWGAN (H+C) & 0.4362 & 0.0932 & 0.2136 & -0.5605 & 0.1117 & 0.5189 & 0.1201 & 0.2315 & -0.5473 & 0.0751 & 0.0977$^\dagger$ \\
\midrule
\textbf{Diffolio (Ours)} & 0.4971 & 0.1313 & 0.2642 & -0.5759 & 0.7307 & 0.5787 & 0.1625 & 0.2807 & -0.5583 & 0.8687 & \textbf{0.1307} \\
\bottomrule
\multicolumn{12}{l}{\footnotesize $^\dagger$ 0.05 $\le$ Turnover < 0.10, $^\ddagger$ Turnover < 0.05.
}
\end{tabular}
}
\end{table*}

Table \ref{tab:performance_weekly_final} presents the portfolio results under these practical constraints.
While Diffolio continues to excel in GOP-CE, its performance in the MVP strategy may seem less pronounced upon initial inspection.
This relative standing, however, is largely attributable to the high risk-adjusted returns reported by models with near-zero turnover, such as TimeGAN and Diffusion-TS. These models effectively adopt a complete buy-and-hold stance. While such results might be interpreted as successful long-term asset selection, this claim is undermined by their mediocre performance in GOP-CE and their lack of consistency in the frictionless environment as shown in Table \ref{tab:performance_comparison_detailed}. If these models truly possessed high-conviction asset selection abilities, they should have demonstrated superior performance even without the transaction cost constraints. Their extreme passivity implies that their predicted dynamics are too weak to surpass the threshold imposed by transaction costs; in other words, their signals are buried under the regularizing effect of frictions. The fact that these models exhibit low turnover across both MVP and GOP strategies further confirms that their underlying distribution dynamics are insufficient to trigger active rebalancing. 

Given that these are fundamentally time-series forecasting models, a severe lack of turnover indicates a failure to capture evolving asset dynamics. Since such models offer little economic value for active management, a more meaningful comparison should focus on models that actively predict such dynamics. In this context, if we filter for portfolios with a turnover exceeding 10\%, Diffolio ranks as the second-best performer for the MVP-SR, following MG-TSD (H) by a narrow margin. In this regard, Diffolio excels by demonstrating active adaptation to time-varying market dynamics. This outcome underscores that while frictionless results reflect the sophisticated precision of distributional forecasting, performance under costs hinges on the ability to track market dynamics with sufficient conviction to overcome frictions---demonstrating that Diffolio excels in both respects.

}

{\color{black}
\subsubsection{Robustness across Market Regimes}}

\textcolor{black}{A closer look at the cumulative return plots in Figures \ref{fig:mvp_cumulative} and \ref{fig:gop_cumulative} reveals the robustness of our model's performance over time.} Unlike several baseline models that exhibit volatile performance---fluctuating between outperforming and underperforming the market---Diffolio consistently maintains a strong performance relative to the market benchmark without experiencing significant periods of underperformance.
Diffolio's robust performance contrasts with that of models like Diffusion-TS (H) and MG-TSD (H), which, despite their strong results on static metrics like CRPS and ES, show erratic portfolio performance. This inconsistency amplifies their volatility and thus leads to lower risk-adjusted metrics like SR and CE.

\begin{figure*}[!tb]
    \centering
    \includegraphics[width=\linewidth]{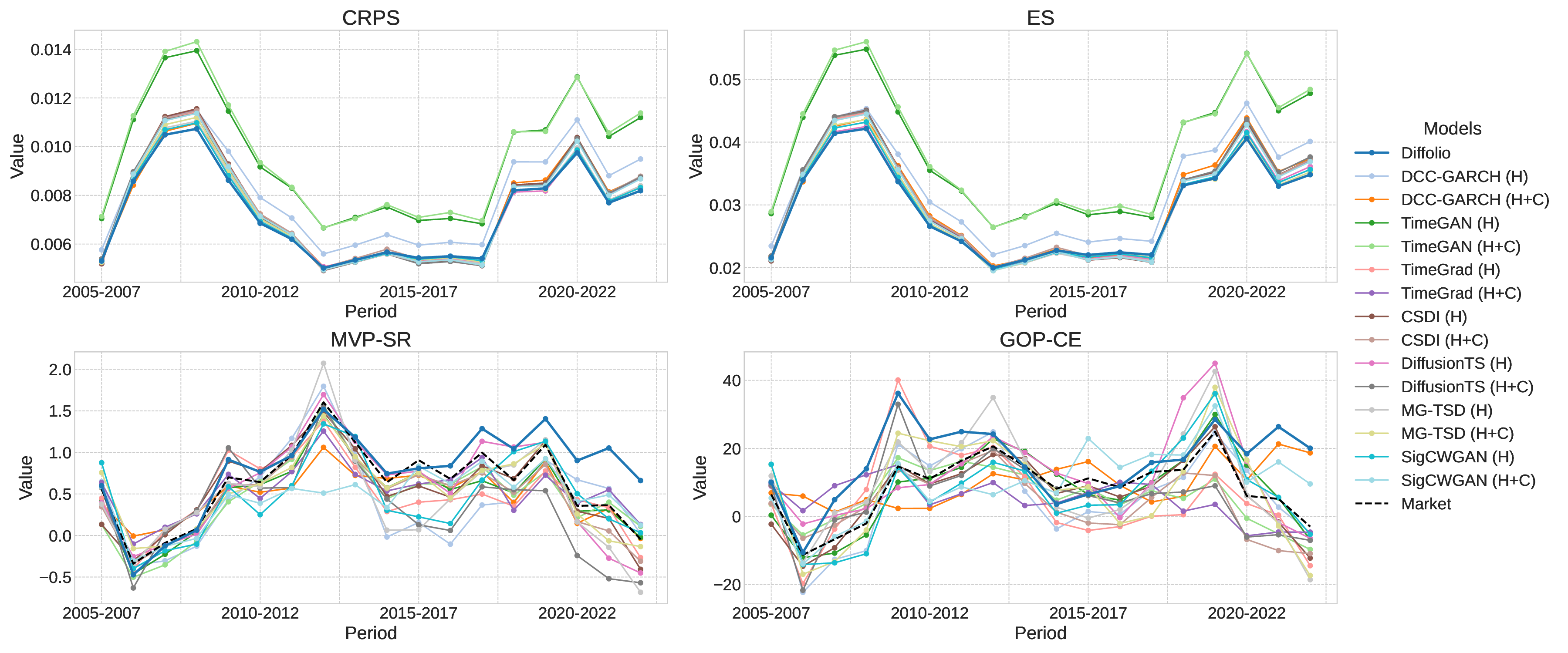}
    \caption{The evolution of evaluation metrics over time for Diffolio and baseline models. For a stable assessment, metrics are calculated each year based on a 3-year rolling window. Lower values are better for CRPS, and ES, while higher values are better for MVP-SR and GOP-CE. 
    }
    \label{fig:metrics_over_time}
\end{figure*}

To further investigate this robustness over time \textcolor{black}{and across different market regimes}, we present the evolution of the evaluation metrics in Figure \ref{fig:metrics_over_time}. Each metric is calculated each year on a 3-year rolling window to ensure a stable assessment, as portfolio performance can be distorted when evaluated over too short a period. 
While Diffolio's statistical forecasting accuracy, CRPS and ES, shows a brief period of weaker performance in the mid-to-late 2010s \textcolor{black}{which is characterized by low volatility and stable growth}, it remains strong overall. 
Diffolio’s robustness appears to be evident particularly during financial crises, where it demonstrates superior performance, such as during the 2008 global financial crisis and the 2020 COVID-19 pandemic. 
In terms of portfolio-based metrics, both MVP-SR and GOP-CE consistently remain comparable to or outperform the market. The MVP-SR shows particularly outstanding performance from the late 2010s onwards, and the GOP-CE exhibits significant outperformance around the 2008 global financial crisis and again in the post-pandemic era. This consistent strength contrasts sharply with most baseline models, which exhibit significant fluctuations \textcolor{black}{depending on the market conditions}. For instance, MG-TSD (H) temporarily led in MVP-SR during 2012-2014, but it subsequently underperformed the market in the next period and delivered the worst performance in the period that followed. Similarly, while several baselines showed strong GOP-CE results in 2019-2021, their performance was followed by a sharp decline in the subsequent period, falling below Diffolio's. In the following periods, most of these models significantly underperformed Diffolio and the market. This is a critical distinction from the perspective of risk-averse portfolio management: in contrast with the fluctuating baselines, Diffolio's consistency in delivering robust results \textcolor{black}{across diverse market regimes}  demonstrates reliable and outstanding efficiency.


In summary, Diffolio's superior ability to consistently capture the time-varying dynamics allows for the construction of more stable and efficient portfolios. This leads to an outstanding risk-adjusted performance that is \textcolor{black}{both economically meaningful and practically reliable}.

{\color{black}
\subsection{Variable Importance Analysis} \label{sec:variable_importance_analysis}

{In this section, we investigate the relative importance of input covariates by analyzing the attention scores produced during return forecasting over the test period. Specifically, we interpret the attention-weight structure in Diffolio’s hierarchical architecture as an informative proxy for how the model allocates predictive emphasis across both asset-specific characteristics and market-level systematic covariates.

This analysis helps clarify how different inputs contribute to the construction of the model’s joint return distribution. Within Diffolio’s hierarchical design, the cross-attention layer primarily captures the influence of asset-specific covariates, while the self-attention layer reflects the role of systematic market-level signals shared across assets. Together, these mechanisms provide transparency into how Diffolio aggregates information across multiple economic dimensions when forming return forecasts.}
The average attention scores recorded throughout the synthetic data generation process are illustrated in Figure \ref{fig:total_attention_scores}.


    

\begin{figure}[!t] 
    \centering
    \begin{subfigure}{\linewidth} 
        \centering
        \includegraphics[width=\linewidth]{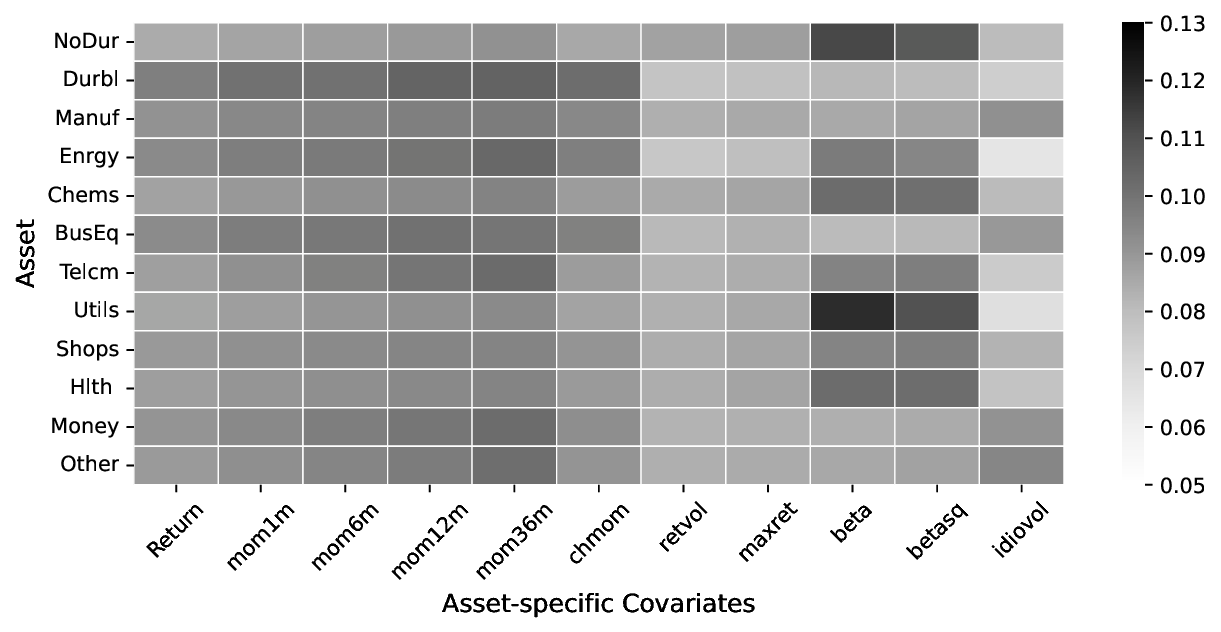}
        \caption{Asset-level cross-attention scores}
        \label{fig:asset_char_cross_attention}
    \end{subfigure}

    \vspace{1em}
    
    \begin{subfigure}{\linewidth}
        \centering
        \includegraphics[width=\linewidth]{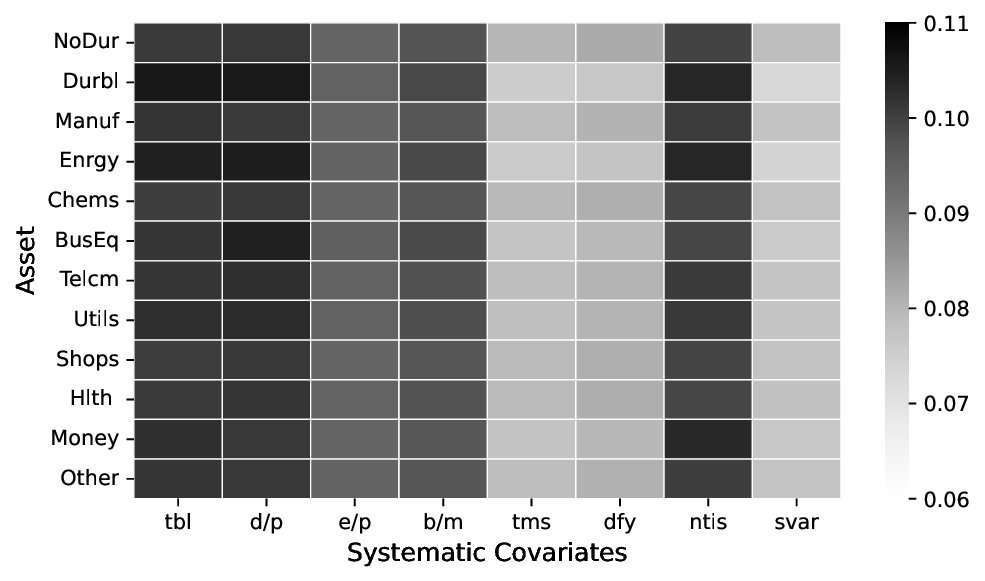}
        \caption{Market-level self-attention scores}
        \label{fig:systematic_self_attention}
    \end{subfigure}

    \caption{Relative importance of covariates via attention maps.}
    \label{fig:total_attention_scores}
\end{figure}

We first examine the relative importance of covariates across the entire investment universe to identify general tendencies. Regarding asset-specific covariates, momentum-related features are observed to exhibit overall high importance across all assets, aligning with the empirical findings of Gu et al. \cite{gu2020empirical}. However, while Gu et al. \cite{gu2020empirical} identify short- or mid-term momentum as a primary driver for individual stocks at a monthly frequency, Diffolio assigns a more pronounced weight to long-term momentum ($\texttt{mom36m}$) for industry portfolios at a daily frequency. This distinction likely reflects the stronger persistence of industry-level structural trends compared to idiosyncratic individual stock signals. Furthermore, in the context of high-frequency daily forecasting, $\texttt{mom36m}$ may serve as a robust proxy for long-term industrial cycles, providing a more stable informational signal amidst the inherent noise of daily returns.

The consistency with established literature is even more pronounced in the systematic covariate analysis. Diffolio exhibits a configuration of importance weights substantially analogous to the findings of Gu et al. \cite{gu2020empirical} as it assigns high predictive weight to $\texttt{tbl}$, $\texttt{b/m}$, and $\texttt{d/p}$, moderate importance to $\texttt{e/p}$, and relatively marginal roles to $\texttt{svar}$ and $\texttt{dfy}$. The lower relevance of $\texttt{tms}$ represents the only distinction but remains inherently reasonable; given its well-documented characterization as a long-term leading indicator that typically influences markets with a significant lag, its immediate informational value for daily return distributions can be lower. This alignment with fundamental economic structures reinforces the reliability and economic grounding of Diffolio’s results.

Next, we examine the relative importance of covariates within individual sectors to evaluate how Diffolio identifies the dominant predictors of return distributions according to each sector's economic context.
For defensive industries---defined here by their low realized betas---specifically Utils, NoDur, Chems, and Hlth\footnote{Detailed realized betas for each industry sector during the test period are provided in \ref{appendix: average beta}.}, the model assigns high importance scores to $\texttt{beta}$ and $\texttt{betasq}$. This observation aligns with the Betting Against Beta (BAB) anomaly \cite{frazzini2014betting}, which documents that low-beta assets frequently earn higher risk-adjusted returns than predicted by standard theory. By prioritizing these characteristics, Diffolio recognizes the low-beta effect as a crucial informational signal for these defensive assets.

On the other hand, the model assigns elevated scores for residual volatility ($\texttt{idiovol}$) to Manuf, BusEq, Money, and Other. This is consistent with the established understanding that these assets are often driven by intra-sectoral dynamics\footnote{These dynamics include operational production cycles in Manuf, technological shifts in BusEq, and credit cycles in Money.} rather than common systematic signals and tend to decouple from broader market flows. In contrast, the Enrgy sector shows a significantly lower reliance on $\texttt{idiovol}$, as its return distributions are predominantly synchronized with macroeconomic common factors such as global commodity prices\footnote{Since Enrgy returns are heavily dictated by external systematic factors, the industry-specific residual---excluding market systematic risk---is measured as minimal, making $\texttt{idiovol}$ a less critical informational signal.}. These alignments with known economic structures further validate Diffolio’s ability to capture structural market dynamics.

In addition to intra-sectoral dynamics, Diffolio's prioritization of informational signals encompasses macroeconomic variables. For example, within cyclical industries such as Durbl and Enrgy, $\texttt{tbl}$ and $\texttt{d/p}$ emerge as dominant predictors. This reflects their documented capital intensity and vulnerability to business cycles\footnote{The prominence of $\texttt{tbl}$ reflects the impact of borrowing costs on both industrial production and consumer demand, while $\texttt{d/p}$ captures the time-varying risk premiums that characterize cyclical assets.}. Similarly, the high predictive importance assigned to $\texttt{tbl}$ and $\texttt{ntis}$ in the Money sector reinforces the validity of Diffolio's predictive mechanism, as it aligns with the sector’s inherent exposure to monetary policy and capital issuance cycles\footnote{Financial institutions are sensitive to interest rate spreads (captured by $\texttt{tbl}$) and rely on equity expansion cycles ($\texttt{ntis}$) as primary signals for predicting returns, providing more critical signals than broad market trajectories.}. Collectively, these prioritization results demonstrate that Diffolio is effectively informed via covariates by the underlying economic context and the structural relationships governing different market segments.
}

\subsection{Ablation Study \textcolor{black}{and Sensitivity Analysis}}

We conduct an ablation study to investigate the individual contributions of Diffolio's methodological components \textcolor{black}{and the interdependencies between them}. We evaluate several configurations where specific components are individually \textcolor{black}{or jointly} removed or altered while holding all other hyperparameters unchanged: (1) without the correlation-guided regularizer $\mathcal{L}_{\text{corr}}$, where the regularizer is removed (i.e., $\lambda_{\text{corr}}=0$); (2) without Ledoit-Wolf Shrinkage, where the regularizer's target correlation matrix is computed using a simple sample-based estimate over a 63-day window instead of the Ledoit-Wolf shrinkage estimator; (3) without asset-specific covariates, where all asset characteristics are replaced with zeros; (4) without systematic covariates, where all macroeconomic variables are replaced with zeros; and (5) without both covariates, where both types of covariates are replaced with zeros. \textcolor{black}{The comprehensive results, ranging from single-component removals to joint ablations involving multiple components to analyze their synergistic interdependencies, are summarized in Table \ref{tab:ablation_study}.}

\begin{table*}[!t]
\centering
\caption{Ablation study of the Diffolio model. Each row shows the evaluation metrics and supplementary diagnostics after removing or altering specific components. For CRPS, ES, CorrScore, \textcolor{black}{and LogDet,} lower values are better. For MVP-SR and GOP-CE, higher values are better. The top row shows the performance of the full Diffolio model for reference.}
\label{tab:ablation_study}
\resizebox{\linewidth}{!}{%
\begin{tabular}{lcccc cc}
\toprule
& \multicolumn{4}{c}{Evaluation Metrics} & \multicolumn{2}{c}{Supplementary Diagnostic} \\
\cmidrule(lr){2-5} \cmidrule(lr){6-7}
\textbf{Model Configuration} & \textbf{CRPS} & \textbf{ES} & \textbf{MVP-SR} & \textbf{GOP-CE} & \textbf{CorrScore} & \textbf{LogDet} \\
\midrule
Diffolio & 0.007169 ($\pm$ 0.001601) & 0.028960 & 0.7206 & 0.1611 & 1.0808 & 2.1982 \\
\midrule
w/o Correlation-Guided Regularizer $\mathcal{L}_{\text{corr}}$ & 0.007128 ($\pm$ 0.001617) & 0.028852 & 0.6232 & 0.1391 & 3.9200 & 5.1065 \\
w/o Ledoit-Wolf Shrinkage for $\mathcal{L}_{\text{corr}}$ & 0.007179 ($\pm$ 0.001639) & 0.029108 & 0.5678 & 0.1155 & 5.5998 & 7.8329 \\
\midrule
w/o Asset-specific Covariates & 0.007191 ($\pm$ 0.001658) & 0.029891 & 0.5374 & 0.0593 & 6.3558 & 8.7254 \\
w/o Systematic Covariates & 0.007167 ($\pm$ 0.001666) & 0.029827 & 0.5970 & 0.0465 & 8.1487 & 12.8705 \\
w/o Both Covariates & 0.007245 ($\pm$ 0.001682) & 0.030153 & 0.4991 & 0.0691 & 7.4334 & 10.6339 \\
\midrule
w/o $\mathcal{L}_{\text{corr}}$ \&  Asset-specific Covariates & 0.007171 ($\pm$ 0.001657) & 0.029460 & 0.5694 & 0.0083 & 6.1784 & 8.4620 \\
w/o $\mathcal{L}_{\text{corr}}$ \& Systematic Covariates & 0.007187 ($\pm$ 0.001674) & 0.029742 & 0.5681 & 0.0786 & 7.7834 & 11.5094 \\
w/o $\mathcal{L}_{\text{corr}}$ \&  Both Covariates & 0.007163 ($\pm$ 0.001653) & 0.029464 & 0.6612 & 0.0824 & 7.9840 & 12.2114 \\
\bottomrule
\end{tabular}%
}
\end{table*}

Interestingly, removing the correlation-guided regularizer $\mathcal{L}_{\text{corr}}$  leads to an improvement in the metrics for the statistical predictive accuracy, CRPS and ES. However, this comes at the cost of a marked deterioration in the CorrScore \textcolor{black}{and LogDet} and a noticeable drop in both portfolio performance metrics, MVP-SR and GOP-CE. This result appears to suggest a trade-off: the use of the correlation-guided regularizer involves trading a modest loss in statistical predictive accuracy for a considerable gain in capturing cross-sectional dependencies, which is essential for constructing efficient multi-asset portfolios. The resulting improvement in portfolio performance underscores the economic significance of accurately modeling these dependencies in financial time-series.

The importance of a stable target for the regularizer is also evident. When the Ledoit-Wolf estimator is replaced with a sample-based correlation estimate, performance degrades across all metrics---even more so than when removing the regularizer entirely. This suggests that guiding the model with a noisy and unstable target correlation matrix can be rather detrimental to learning cross-sectional dependencies, emphasizing the need for careful selection of the target matrix.

The ablation study on the covariates elucidates their respective effects. Zeroing the influence of asset characteristics leads to a broad degradation across all metrics, demonstrating their importance in modeling both individual asset dynamics and their cross-sectional dependencies. On the other hand, zeroing macroeconomic variables shows an asymmetric effect; while the impact on the marginal predictive accuracy, CRPS, seems almost negligible, the performance on metrics sensitive to cross-sectional dependency, such as ES, MVP-SR, GOP-CE, CorrScore, \textcolor{black}{and LogDet} worsens considerably. This suggests that macroeconomic variables play a crucial role in capturing the dependencies among assets. 
{\color{black}
Notably, removing the influence of both types of covariates results in a more significant deterioration in forecasting accuracy than removing either one individually. These observations demonstrate that both asset-specific and systematic covariates contain essential information about the time-varying joint distribution of returns.

The joint ablation results reveal the interdependencies within the architecture of Diffolio. First, the interaction between asset-specific covariates and the correlation-guided regularizer $\mathcal{L}_{\text{corr}}$ is revealed by comparing configurations where asset-specific covariates are omitted. Since these covariates capture the time-varying characteristics of individual assets, their absence causes the latent representations derived from the asset-level hierarchy to 
become rigid and less informative. Applying $\mathcal{L}_{\text{corr}}$ on top of such impoverished features acts as an over-constraint, as the model is forced to align a dependency structure without sufficient asset-level context. This explains why removing the regularizer in this context leads to slight improvements in CRPS, ES, and MVP-SR by relieving this over-constraint, yet simultaneously causes a catastrophic decline in GOP-CE. Such a discrepancy indicates that without structural safeguards, the model may produce unstable estimates of the joint dynamics necessary for risk-aware allocation.

The interaction between systematic covariates and the regularizer shows divergent outcomes across metrics. Removing $\mathcal{L}_{\text{corr}}$ alongside systematic information results in improved GOP-CE and ES, but leads to worsened CRPS and MVP-SR compared to the configuration missing only systematic context. 
This indicates that in the absence of macroeconomic context---provided by time-varying macroeconomic variables reflecting current market states---$\mathcal{L}_{\text{corr}}$ lacks the necessary grounding to function effectively. Even if the internal attention mechanisms maintain their relative scores, the absence of market-level dynamic information prevents the model from contextualizing these scores within prevailing conditions, leaving the joint distribution alignment without an economically meaningful basis.
Furthermore, the fact that configurations lacking both $\mathcal{L}_{\text{corr}}$ and either type of covariate perform consistently worse across all metrics than the model without the regularizer alone confirms that covariates play an essential role in predicting financial time-series. 

Lastly, the configuration excluding all covariates and the regularizer achieves a CRPS even superior to the full Diffolio model. This observation suggests that a simplified model can achieve high statistical stability for individual assets by focusing exclusively on recurring historical patterns without the potential noise or optimization complexity introduced by complex covariates. Nevertheless, its failure to capture cross-sectional dependencies---the very essence of multivariate financial time-series forecasting---results in significantly lower GOP-CE compared to the full Diffolio. This re-emphasizes the necessity of both informative covariates and structural regularizers. In conclusion, these findings confirm that the components of Diffolio are complementary and interdependent, interacting within a hierarchical attention network to balance statistical precision with economic reliability.
}

\begin{figure*}[!t]
    \centering
    \includegraphics[width=\linewidth]{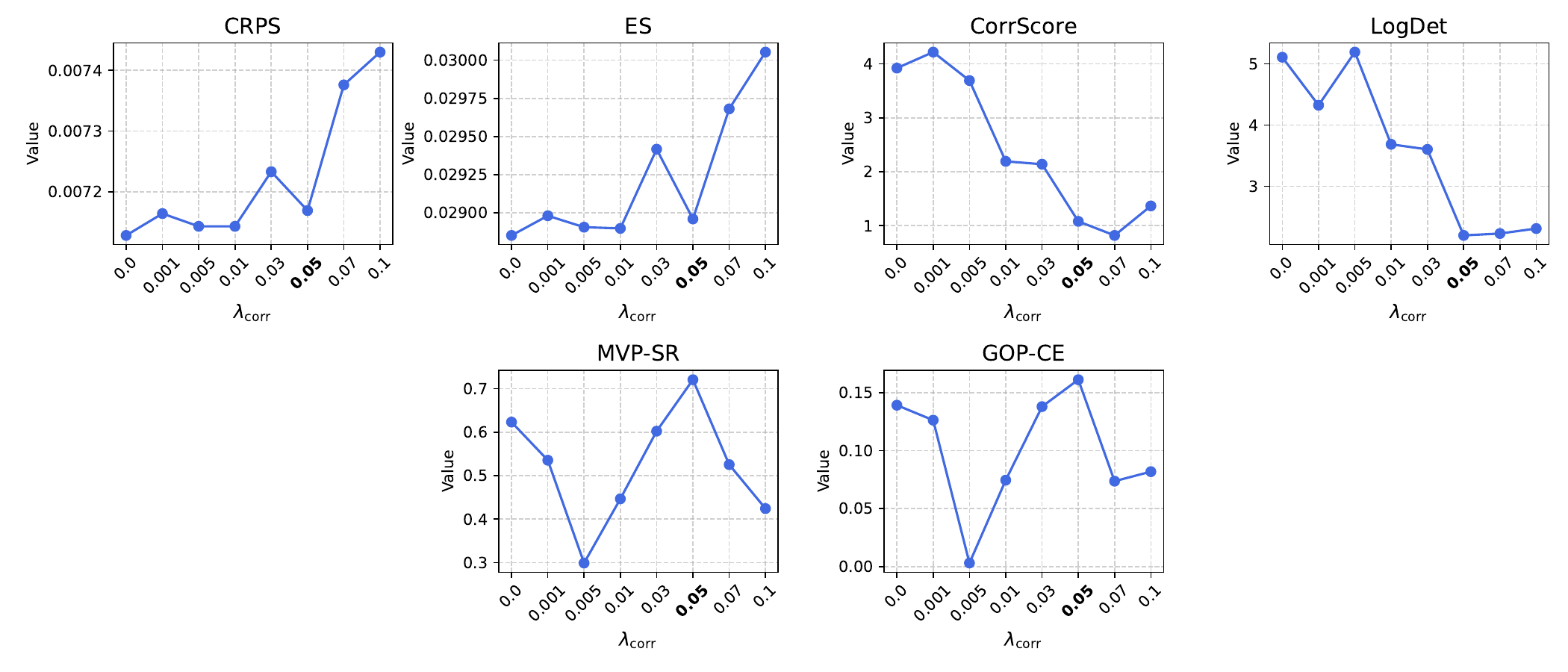}
    \caption{Sensitivity analysis of the correlation-guided regularizer coefficient, $\lambda_{\text{corr}}$. Lower values are better for CRPS, ES,  CorrScore, \textcolor{black}{and LogDet}, whereas higher values are better for MVP-SR and GOP-CE. The dotted lines represent the best performance among the baseline models for each metric. Our final model corresponds to the case of $\lambda_{\text{corr}}=0.05$, which achieved the lowest ES on the validation set. 
    }
    \label{fig:lambda_corr_sensitivity}
\end{figure*}

{\color{black}
To further analyze the effect of the correlation-guided regularizer, we conduct a sensitivity analysis on its coefficient, $\lambda_{\text{corr}}$. We train Diffolio with varying values of $\lambda_{\text{corr}}$ from 0 to 0.1, and the results for the evaluation metrics as well as the CorrScore and LogDet are presented in Figure \ref{fig:lambda_corr_sensitivity}.
The result illustrates a trade-off induced by the regularizer: as $\lambda_{\text{corr}}$ increases, the modeling of the correlation structure (CorrScore and LogDet) tends to improve, whereas the statistical forecasting precision (CRPS and ES) generally degrades.
Notably, we observe that portfolio-based metrics show their highest performance when these two respective aspects are well-balanced.
In our experiments, based on the validation result for the ES which takes both the marginal and cross-sectional aspects into account, we find that $\lambda_{\text{corr}}=0.05$ provides a delicate balance.
The strong performance in MVP-SR and GOP-CE demonstrates that this well-balanced setting leads to economically meaningful outcomes. 
}


{\color{black}
Collectively, the results of the ablation study and sensitivity analysis suggest three generalizable insights for multivariate financial time-series modeling that transcend the specific performance of Diffolio. First, both asset-specific and systematic covariates are essential for capturing dependencies within the joint distribution; consequently, this underscores the necessity of an architecture capable of effectively integrating these distinct information sources to model complex cross-sectional relationships. Second, given that financial data is inherently noisy, injecting domain knowledge as an inductive bias---such as stable shrinkage targets---is crucial for maintaining structural stability. Finally, the results demonstrate that statistical predictive accuracy guarantees neither the quality of dependency modeling nor the model's economic utility, as improvements in marginal precision do not inherently translate to superior portfolio outcomes.
}


{\color{black}
\subsection{Computation time}

Finally, we compare the computational efficiency of Diffolio and baseline models to assess their practical feasibility. All experiments were conducted using  dual Intel Xeon Silver 4510 processors (2.40 GHz, 24 cores in total) and an NVIDIA RTX A6000 GPU with a batch size of 4,096 for all models. For a fair comparison, all diffusion-based models, including Diffolio, used 50 sampling steps during inference. The inference time represents the time required to generate 100 sample paths for the entire test period (4,781 days).Table \ref{tab:computation_cost} presents the total training time and the total inference time across all models.

\begin{table}[!t]
\centering
\caption{Comparison of training time and inference time. The inference time represents the time required to generate 100 sample paths for the entire test period (4,781 days).}
\label{tab:computation_cost}
\begin{tabular}{lcc}
\toprule
\textbf{Model} & \textbf{Training Time} & \textbf{Inference Time} \\
\midrule
DCC-GARCH (H) & N/A & 11m 12s \\
DCC-GARCH (H+C) & N/A & 10m 57s \\
TimeGAN (H) & 4m 1s & 11s \\
TimeGAN (H+C) & 4m 38s & 12s \\
TimeGrad (H) & 3h 43m 4s & 33s \\
TimeGrad (H+C) & 3h 22m 46s & 51s \\
CSDI (H) & 10h 55m 41s& 3h 6m 15s \\
CSDI (H+C) & 21h 57m 56s & 3h 6m 30s \\
Diffusion-TS (H) & 9h 29m 7s& 4h 32m 55s \\
Diffusion-TS (H+C) & 20h 3m 15s & 5h 4m 38s \\
MG-TSD (H) & 3h 39m 36s & 44s \\
MG-TSD (H+C) & 3h 40m 25s & 40s \\
SigCWGAN (H) & {2h 4m 3s} & {3s} \\
SigCWGAN (H+C) & 2h 7m 20s& {2s} \\
\midrule
\textbf{Diffolio (Ours)} & 3h 18m 35s & 2m 37s \\
\bottomrule
\end{tabular}
\end{table}



Regarding training times, Diffolio requires approximately 3 hours and 18 minutes. For DCC-GARCH, training time is not reported as parameters are repeatedly re-estimated through rolling windows during the test period. Among GAN-based models, TimeGAN has a very short training time due to its relatively simple architecture, and SigCWGAN also requires less training time than Diffolio.
On the other hand, diffusion-based baseline models require similar or significantly more training time compared to Diffolio. Notably, Diffusion-TS (H) and CSDI (H) take approximately three times longer than Diffolio despite being trained for only half the number of epochs. Furthermore, Diffusion-TS (H+C) and CSDI (H+C) were trained for the same number of epochs as Diffolio but required nearly six times as much time, illustrating the high computational intensity of their architectures.

Turning to inference efficiency, Diffolio takes about 2 minutes and 37 seconds to generate 100 sample paths for nearly 20 years of the test period. \textcolor{black}{Notably, Table \ref{tab:stage_inference_time} reports the inference latency for Diffolio across its various model stages. The results show that the latency per diffusion step is approximately 23.81 ms. This translates to a total inference time of roughly 1.2 seconds to generate 100 samples for a single time step. Such rapid inference underscores the model's practical feasibility, enabling portfolio rebalancing every two seconds and suggesting potential applicability in higher frequency trading scenarios.} For DCC-GARCH, the inference process takes longer as it requires parameter re-estimation at each time step $t$, and GAN-based models show fast inference results. Among diffusion-based baselines, although Diffolio shows slower inference compared to TimeGrad or MG-TSD, it is markedly faster than CSDI and Diffusion-TS, which require approximately 3 and 5 hours for inference, respectively. This translates to Diffolio being roughly 70 and 100 times faster than these models \footnote{The high latency of CSDI is primarily due to its imputation-based prediction, which requires generating a full window of length $M$ to predict each subsequent time step. For Diffusion-TS, the latency arises from its reconstruction-guided sampling during inference, an optimization process that necessitates multiple backpropagation steps.}.
{While Diffusion-TS (H) showed competitive results in several metrics earlier, its considerable computational costs limit its utility in real-world deployment. In contrast, Diffolio demonstrates superior practical feasibility by maintaining high predictive accuracy with significantly lower computational overhead.}

\begin{table}[!tb]
\centering
\caption{Inference latency analysis by model stage of Diffolio. Values represent the average latency per diffusion step.}
\label{tab:stage_inference_time}
\resizebox{\columnwidth}{!}{\begin{tabular}{lcc}
\toprule
\textbf{Stage} & \textbf{Average Latency (ms)} & \textbf{Ratio (\%)} \\
\midrule
Stage 1 (Asset-Level) & 19.31 & 81.1\% \\
Stage 2 (Market-Level) & 4.50 & 18.9\% \\
\midrule
{Total (Model Inference)} & {23.81} & {100.0\%} \\
\bottomrule
\end{tabular}}
\end{table}

Based on these results, we further discuss the scalability and practical feasibility of Diffolio for higher-dimensional asset universes. To investigate whether the $O(N^2)$ complexity of the market-level self-attention poses a practical bottleneck, \textcolor{black}{we focus on the latency results presented in Table \ref{tab:stage_inference_time}.}  Stage 2 currently accounts for only 18.9\% of the total inference latency---approximately one-fourth of the computational load of Stage 1.  While Stage 2 possesses $O(N^2)$ complexity, Stage 1 scales linearly ($O(N)$) as it operates independently on individual assets. This modest initial proportion for the quadratic stage suggests substantial computational headroom, providing a highly favorable condition for expanding Diffolio to higher-dimensional asset universes.

To assess the model's viability in larger-scale portfolios, we gauge the computational requirements for a universe of $N=100$ assets by roughly scaling our empirical results based on the Big-O complexity of each stage. Under this simplified calculation,\footnote{These estimates assume a scenario without additional overhead, such as data loading or memory transfer. We also note that since Stage 1 operates independently for each asset, it is highly parallelizable, which could reduce the practical burden below $O(N)$; however, we utilized $O(N)$ for a conservative estimation.} the total inference time for $N=100$ is expected to be 52 minutes (consisting of 18 minutes for Stage 1 and 34 minutes for Stage 2), {with single time-step generation taking approximately 24 seconds}. Under similar assumptions, the training time is expected to be roughly 65 hours. While increased memory constraints may necessitate smaller batch sizes---potentially extending actual training times---an inference time of under one hour remains highly feasible for real-world daily or weekly rebalancing tasks.

For even larger deployments involving hundreds or thousands of assets, a shift toward structural sparsity or modular substitution becomes inevitable. Specifically, the potential $O(N^2)$ bottleneck in Stage 2 can be mitigated by replacing the standard self-attention block with linear-complexity attention mechanisms. This structural flexibility ensures that Diffolio can be well-scaled and remains practically applicable across diverse portfolio sizes.

}

\section{Conclusion} \label{sec:conclusion}

In this paper, we propose Diffolio, a diffusion-based model for multivariate probabilistic financial time-series forecasting and portfolio construction. Diffolio incorporates a two-stage hierarchical attention architecture with separate asset-level and market-level layers. The first stage, asset-level attention, extracts salient features for each asset from its own historical data and asset-specific covariates, and the second stage, market-level attention, then models the cross-sectional dependencies across all assets and their exposure to systematic covariates. To further improve the learning of these dependencies, we introduce a correlation-guided regularizer that employs a stable target correlation matrix estimated via the Ledoit-Wolf shrinkage method.

{\color{black}
Experiments on 12 industry portfolios demonstrate that Diffolio achieves strong results in both statistical probabilistic forecasting accuracy and forecast calibration, indicating its capability to effectively model the joint distribution and complex dynamics of asset returns.
Notably, Diffolio exhibits substantial economic significance through superior portfolio performance.
In particular, portfolios constructed from Diffolio’s forecasts achieve a Sharpe ratio of 0.7206---over 40\% higher than baseline models and the market---and a certainty equivalent of 0.1611, which implies a utility equivalent to a guaranteed 16.11\% risk-free annual return.
{Moreover, the robust performance of Diffolio under transaction costs and the high consistency of covariate importance with established literature reinforce its economic reliability.}
Overall, these findings demonstrate that the effectiveness of our proposed Diffolio goes beyond statistical forecasting to facilitate practical decision-making in constructing efficient portfolios.

{For future work, we consider the scalability and practical feasibility of Diffolio. While this study focused on 12 industry portfolios at a daily frequency, Diffolio's hierarchical architecture is fundamentally frequency-agnostic, and thus applicable to diverse time scales. Specifically, future research could assess the model's capacity to capture high-frequency dynamics, such as the potentially dominant influence of asset-specific returns over macroeconomic variables. In terms of computational efficiency,} generating 100 next-time-step samples requires approximately 1.2 seconds, and complexity analysis suggests that scaling to a universe of 100 assets would require roughly 24 seconds. This indicates the potential for Diffolio to support intraday strategies requiring rebalancing every few minutes. However, to push the boundaries of Diffolio toward higher dimensions (e.g., universes spanning thousands of assets) and high-frequency trading regimes, future work could concentrate on evolving the framework into a high-throughput architecture by integrating computationally efficient attention variants including linear-complexity attention mechanisms, \textcolor{black}{while simultaneously addressing potential multi-faceted data challenges such as robust data integration \cite{yu2025mgsfformer, yu2025ginar+} and adaptive modeling of non-stationary environments \cite{cheng2025charging, cheng2026metagnsdformer}.}


Furthermore, integrating the probabilistic forecasting capabilities of Diffolio with deep reinforcement learning (DRL) provides a promising avenue for developing an end-to-end portfolio optimization framework. This integration can be realized through two primary methodological directions. First, DRL agents can directly leverage latent features or shared parameters from Diffolio, inheriting the model’s learned representations of market dynamics to refine allocation policies. Second, Diffolio can be utilized within a model-based reinforcement learning (MBRL) framework, where it serves as a high-fidelity environment simulator. By generating a large set of synthetic sample paths to represent realistic market trajectories, Diffolio provides a robust training ground for agents to explore complex scenarios. This allows the DRL agent to learn optimal policies that account for realistic constraints, such as transaction costs and market impact, through extensive simulation before real-world deployment. Such a framework would effectively bridge the gap between high-fidelity density estimation and autonomous, risk-aware investment execution.
}

\section*{CRediT authorship contribution statement}
\textbf{So-Yoon Cho}: Conceptualization, Software, Investigation, Writing - Original Draft. 
\textbf{Jin-Young Kim}: Conceptualization, Methodology, Software.
\textbf{Kayoung Ban}: Validation, Writing - Review \& Editing. 
\textbf{Hyeng Keun Koo}: Conceptualization, Writing - Review \& Editing, Supervision.
\textbf{Hyun-Gyoon Kim}: Conceptualization, Methodology, Writing - Original Draft.

\section*{Declaration of competing interests}
The authors declare that they have no known competing financial interests or personal relationships that could have appeared to influence the work reported in this paper.

\section*{Declaration of generative AI and AI-assisted technologies in the writing process}
During the preparation of this work the authors used Gemini in order to improve the grammatical quality and readability. After using this service, the authors reviewed and edited the content as needed and take full responsibility for the content of the publication.

\section*{Acknowledgement}

We thank anonymous reviewers for their valuable comments and suggestions.
So-Yoon Cho is supported by Hankuk University of Foreign Studies Research Fund (of 2026).
Kayoung Ban is supported by the KIAS Individual Grant. No. PG097601.
Hyun-Gyoon Kim is supported by the National Research Foundation of Korea (NRF) grant funded by the Korea government (MSIT) (RS-2025-00513038), and by Institute of Information \& communications Technology Planning \& Evaluation (IITP) under the Artificial Intelligence Convergence Innovation Human Resources Development (IITP-2025-RS-2023-00255968) grant funded by the Korea government (MSIT). 

\section*{Data availability}
Data will be made available on request.

\bibliographystyle{splncs04}
\bibliography{bibliography}

\appendix
\appendix

\clearpage
\onecolumn

\section{\color{black}Nomenclature} \label{appendix: Nomenclature}

{\color{black}
\begin{longtable}{l l l}
\caption{Nomenclature} \label{tab:nomenclature} \\

\toprule
\textbf{Symbol} & \textbf{Definition} & \textbf{Dimension/Type} \\
\midrule
\endfirsthead

\toprule
\textbf{Symbol} & \textbf{Definition} & \textbf{Dimension/Type} \\
\midrule
\endhead

\bottomrule
\endlastfoot

\multicolumn{3}{l}{\textit{\textbf{Indices and Sets}}} \\
$t$ & Real time index & $\mathbb{Z}$ \\
$\tau$ & Diffusion step index & $\{1, \dots, T\}$ \\
$i$ & Asset index & $\{1, \dots, N\}$ \\
$N$ & Number of assets & $\mathbb{N}$ \\
$M$ & Lookback window size & $\mathbb{N}$ \\
$N_z$ & Number of asset-specific covariates & $\mathbb{N}$ \\
$N_y$ & Number of systematic covariates & $\mathbb{N}$ \\
$D$ & Hidden dimension size & $\mathbb{N}$ \\
$T$ & Total diffusion steps & $\mathbb{N}$ \\
\midrule

\multicolumn{3}{l}{\textit{\textbf{Data and Covariates}}} \\
$\mathbf{r}_{t+1}$ & Excess return vector at the next time step & $\mathbb{R}^{N}$ \\
$\mathbf{r}_{t-(M-1):t}$ & Historical excess returns over a lookback window & $\mathbb{R}^{M \times N}$ \\
$\mathbf{z}_{i, t-(M-1):t}$ & Sequence of asset-specific covariates for asset $i$ & $\mathbb{R}^{M \times N_z}$ \\
$\mathbf{y}_{t-(M-1):t}$ & Sequence of systematic covariates & $\mathbb{R}^{M \times N_y}$ \\
$\mathbf{c}$ & Set of conditioning information tensors & $\{ \mathbb{R}^{M \times N}, \mathbb{R}^{N \times M \times N_z}, \mathbb{R}^{M \times N_y} \}$ \\
\midrule

\multicolumn{3}{l}{\textit{\textbf{Diffusion Model}}} \\
$\mathbf{x}^0$ & Original data sample (Target variable $\mathbf{r}_{t+1}$) & $\mathbb{R}^{N}$ \\
$\mathbf{x}^\tau$ & Noisy data sample at diffusion step $\tau$ & $\mathbb{R}^{N}$ \\
$\beta_\tau$ & Variance schedule parameter & $\mathbb{R}$ \\
$\alpha_\tau$ & $1 - \beta_\tau$ & $\mathbb{R}$ \\
$\bar{\alpha}_\tau$ & $\prod_{i=1}^\tau \alpha_i$ & $\mathbb{R}$ \\
$\boldsymbol{\epsilon}$ & Standard Gaussian noise & $\mathbb{R}^{N}$ \\
$\boldsymbol{\epsilon}_\theta$ & Denoising network (Predicts noise $\boldsymbol{\epsilon}$) & $\mathbb{R}^N \times \mathbb{N} \times \mathbf{c} \to \mathbb{R}^N$ \\
$\eta$ & Hyperparameter for stochasticity in DDIM sampling & $\mathbb{R}$ \\
\midrule

\multicolumn{3}{l}{\textit{\textbf{Diffolio Architecture}}} \\
$\mathbf{q}_i$ & Query vector for asset $i$ (Stage 1) & $\mathbb{R}^{D}$ \\
$\mathbf{K}_i, \mathbf{V}_i$ & Key and Value tensors for asset $i$ (Stage 1) & $\mathbb{R}^{M \times D}$ \\
$\mathbf{h}_i$ & Asset-specific latent vector & $\mathbb{R}^{D}$ \\
$\mathbf{h}$ & Integrated latent tensor (Stage 2) & $\mathbb{R}^{(N+N_y) \times D}$ \\
$\mathbf{W}_{(\cdot)}, \mathbf{U}_{(\cdot)}$ & Learnable weight matrices & $\mathbb{R}^{D \times D}$ \\
\midrule

\multicolumn{3}{l}{\textit{\textbf{Optimization and Regularization}}} \\
$\mathcal{L}_{\text{DDPM-cond}}$ & Conditional DDPM loss & $\mathbb{R}$ \\
$\mathcal{L}_{\text{corr}}$ & Correlation-guided regularizer & $\mathbb{R}$ \\
$\lambda_{\text{corr}}$ & Coefficient for the correlation-guided regularizer & $\mathbb{R}$ \\
$\mathbf{A}$ & Asset-to-asset submatrix of attention probability & $\mathbb{R}^{N \times N}$ \\
$\bm{\Sigma}_t^{\text{target}}$ & Target correlation matrix (Ledoit-Wolf shrinkage) & $\mathbb{R}^{N \times N}$ \\
$\bm{\Sigma}^{\text{train}}$ & Covariance matrix over the entire training period & $\mathbb{R}^{N \times N}$ \\
\midrule

\multicolumn{3}{l}{\textit{\textbf{Evaluation Metrics and Supplementary Diagnostics}}} \\
$\text{CRPS}$ & Continuous Ranked Probability Score & $\mathbb{R}$ \\
$\text{ES}$   & Energy Score & $\mathbb{R}$ \\
$\text{SR}$   & Sharpe Ratio & $\mathbb{R}$ \\
$\text{CE}$   & Certainty Equivalent & $\mathbb{R}$ \\
$\text{CorrScore}$ & Correlation Score & $\mathbb{R}$ \\
$\text{LogDet}$    & Log-Determinant Divergence & $\mathbb{R}$ \\
$\text{PICP}$      & Prediction Interval Coverage Probability & $[0, 1]$ \\
$\text{ACE}$       & Average Coverage Error & $[-1, 1]$ \\
\end{longtable}
}

\section{Data Description} \label{appendix: data description}

\subsection{Industry Portfolios}
The target assets in our study are the 12 industry portfolios, which provide a standard and diversified representation of the U.S. stock market. The portfolios and their compositions are summarized in Table \ref{tab:industry_portfolios}.

\begin{table}[!h]
\centering
\caption{Descriptions on the 12 Industry Portfolios.}
\label{tab:industry_portfolios}
\begin{tabular}{@{}ll@{}}
\toprule
\textbf{Code} & \textbf{Industry Portfolio Description} \\
\midrule
NoDur & Consumer Nondurables (Food, Tobacco, Textiles, Apparel, Leather, Toys) \\
Durbl & Consumer Durables (Cars, TVs, Furniture, Household Appliances) \\
Manuf & Manufacturing (Machinery, Trucks, Planes, Office Furniture, Paper) \\
Enrgy & Oil, Gas, and Coal Extraction and Products \\
Chems & Chemicals and Allied Products \\
BusEq & Business Equipment (Computers, Software, Electronic Equipment) \\
Telcm & Telephone and Television Transmission \\
Utils & Utilities \\
Shops & Wholesale, Retail, and Some Services (Laundries, Repair Shops) \\
Hlth  & Healthcare, Medical Equipment, and Drugs \\
Money & Finance \\
Other & Other (Mines, Construction, Transportation, Hotels, Entertainment) \\
\bottomrule
\end{tabular}
\end{table}

\subsection{Asset Characteristics} 

We construct 10 asset characteristics for each asset using historical daily excess returns as follows.
\begin{itemize}
\item {Momentum (\texttt{mom1m}, \texttt{mom6m}, \texttt{mom12m}, \texttt{mom36m})}: The cumulative return over the past $k$ trading days. We use $k \in \{21,126,252,756\}$, corresponding to 1, 6, 12, and 36 months,
\begin{equation*}
   \texttt{mom}k\texttt{m}_{i, t} = \left( \prod_{s=t-k+1}^{t} (1 + \mathbf{r}_{i,s}) \right) - 1.
\end{equation*}

\item {Change in momentum (\texttt{chmom})}: The difference between the most recent 6-month momentum and the 6-month momentum from the preceding period.
\begin{equation*}
    \texttt{chmom}_{i,t} = \texttt{mom6m}_{i,t} - \texttt{mom6m}_{i,t-126}.
\end{equation*}

\item {Return volatility (\texttt{retvol})}: The standard deviation of daily excess returns over the past 21 trading days.
\begin{equation*}
   \texttt{retvol}_{i,t} = \sqrt{\frac{1}{20} \sum_{s=t-20}^{t} \left(\mathbf{r}_{i,s} - \bar{\mathbf{r}}_{i,t}\right)^2},
\end{equation*}
where $\bar{\mathbf{r}}_{i,t}$ is the mean return over the same period.

\item {Maximum return (\texttt{maxret})}: The maximum daily excess return over the past 21 trading days.
\begin{equation*}
    \texttt{maxret}_{i,t} = \max_{s \in [t-20, t]} \mathbf{r}_{i,s}.
\end{equation*}

\item {Beta (\texttt{beta})} and {Beta squared (\texttt{betasq})}: We estimate the market beta, $\beta_i$, based on the Capital Asset Pricing Model (CAPM) by running a rolling regression of the asset's excess return on the market excess return using data from the past 252 trading days:
\begin{equation*}
   \mathbf{r}_{i,s} = \alpha_i + \beta_i F^{Mkt}_s + e_{i,s}, \quad s \in [t-251, t],
\end{equation*}
where $F^{Mkt}_t$ is the market excess return. The estimated factor loading $\beta_i$ at time $t$ serves as the $\texttt{beta}_{i,t}$ characteristic, and its square, $\beta_i^2$, as the $\texttt{betasq}_{i,t}$ characteristic.

\item {Idiosyncratic Volatility (\texttt{idiovol})}: We obtain idiosyncratic volatility from a rolling Fama-French 3-factor model regression using data from the past 252 trading days:
\begin{equation*}
\mathbf{r}_{i,s} = \alpha_i + \beta_{i,Mkt} F^{Mkt}_s + \beta_{i,SMB} F^{SMB}_s + \beta_{i,HML} F^{HML}_s + e_{i,s}, \quad s \in [t-251, t],
\end{equation*}
where $F^{SMB}_t$ and $F^{HML}_t$ are the size and value factors, respectively. 
The $\texttt{idiovol}_{i,t}$ characteristic is then defined as the standard deviation of the resulting regression residuals, $\{e_{i,s}\}$, over the estimation window.
\end{itemize}

\subsection{Macroeconomic Variables}
We use eight macroeconomic variables presented in \cite{welch2008comprehensive}, which are summarized in Table \ref{tab:macro_variables}.

\begin{table}[!h]
\centering
\caption{Macroeconomic Variables.}
\label{tab:macro_variables}
\begin{tabular}{@{}ll@{}}
\toprule
\textbf{Variable} & \textbf{Description} \\
\midrule
\texttt{tbl} & Treasury Bill Rate (3-Month) \\
\texttt{d/p} & Dividend-to-Price Ratio \\
\texttt{e/p} & Earnings-to-Price Ratio \\
\texttt{b/m} & Book-to-Market Ratio \\
\texttt{tms} & Term Spread (Long-term yield minus T-bill rate) \\
\texttt{dfy} & Default Yield Spread (BAA-rated minus AAA-rated corporate bonds) \\
\texttt{ntis} & Net Equity Issuance \\
\texttt{svar} & Stock Variance (computed as sum of squared daily returns on S\&P 500) \\
\bottomrule
\end{tabular}
\end{table}

\section{Details of Evaluation Metrics} \label{appendix: evaluation metrics}

This appendix provides the detailed formulations for the evaluation metrics, score and statistics used in our experiments. Each metric is chosen to assess a particular aspect of the model's performance, from the accuracy of the predicted probability distributions to their economic significance.

\subsection{Statistical Forecasting Accuracy}
First, we use proper scoring rules to evaluate the quality of the probabilistic forecasts, which assess the entire predictive distribution rather than just a point estimate.

\textbf{Continuous Ranked Probability Score (CRPS).}
The CRPS is a widely-used score that generalizes the mean absolute error to probabilistic forecasts. It measures the integrated squared difference between the cumulative distribution function (CDF) of the forecast, $F_{D_t}$, and the empirical CDF of the observation, which is a step function at the realized value $r_t$. For a univariate prediction at a given time $t$, it is defined as:
\begin{equation*}
CRPS_t :=CRPS(D_t, r_t) = \int_{-\infty}^{\infty} \left( F_{D_t} (x) - H(x-r_t) \right)^2 dx,
\end{equation*}
where $r_t\in \mathbb{R}$ is the ground truth, $F_{D_t}$ is the CDF of the predictive distribution $D_t$ of a generative model for time $t$, and $H(\cdot)$ is the Heaviside step function. 
A lower CRPS value indicates a more accurate forecast. 
The {CRPS} metric reported in this paper is obtained as follows. First, for each of the $N$ assets, we compute its time-averaged CRPS over the test period. Then, we report the mean and standard deviation of these $N$ scores.

\textbf{Energy Score (ES).}
The ES is a multivariate generalization of the CRPS, making it suitable for evaluating the accuracy of the full joint distribution. It is defined as:
\begin{equation*}
ES_t := ES(D_t, \mathbf{r}_t) = \mathbb{E} \left[
\lVert \mathbf{X}_t-\mathbf{r}_t \rVert_2 \right]
 - \frac{1}{2} \mathbb{E} \left[ \lVert \mathbf{X}_t-\mathbf{X}'_t \rVert_2 \right],
\end{equation*}
where $\mathbf{r}_t \in \mathbb{R}^N$ is the $N$-dimensional ground truth, $\mathbf{X}_t$ and $\mathbf{X}'_t$ are independent and identically distributed random vectors drawn from the predictive distribution $D_t$, and $\Vert\cdot\rVert_2$ denotes the Euclidean norm.
The former term measures the distance between the forecast samples and the observation, rewarding forecasts that are close to the outcome. 
The latter term measures the spread within the forecast samples, rewarding diversity. 
The overall \textit{ES} is the average of $ES_t$ over the test period.

\subsection{Portfolio Performance}
Second, we conduct investment experiments to assess the economic significance of the model's forecasts. The performance of these portfolios serves as an integrated measure of the model's predictive capabilities.
At the end of each day $t$, we generate a one-day-ahead forecast to determine the optimal portfolio weights $\mathbf{w}_{t+1} \in \mathbb{R}^N$. The performance is then calculated based on the realized returns on the next day $\boldsymbol{r}_{t+1}$.

\textbf{Mean-Variance Tangency Portfolio (MVP).} The MVP is a classic portfolio optimization strategy that aims to find the portfolio with the highest possible SR. The optimization objective is defined as:
\begin{equation*}
\begin{gathered}
\max_{\mathbf{w}\in \mathbb{R}^N} \frac{\mathbf{w}^\top \boldsymbol{\hat{\mu}}_{t+1}}{\sqrt{\mathbf{w}^\top\boldsymbol{\hat{\Sigma}}_{t+1} \mathbf{w}}} \\
\text{subject to }\sum_{i=1}^N \mathbf{w}_i=1, \quad \text{and}\quad \mathbf{w}_i \geq 0, \text{ for }i=1,\dots, N,
\end{gathered}
\end{equation*}
where $\boldsymbol{\hat{\mu}}_{t+1}$ and $\boldsymbol{\hat{\Sigma}}_{t+1}$ are the predicted mean and covariance matrix of the excess returns for time $t+1$, respectively. 
The performance of this strategy relies on the accuracy of the estimated first two moments of the return distribution.
Since the objective is to maximize the SR, it is considered the primary indicator of performance for this strategy. 

We also report several other summary statistics: Ret, Vol, MDD, and turnover. Let $\textbf{r}_{p, t}=\mathbf{w}_{t}^\top \mathbf{r}_{t}$ be the daily portfolio return. The statistics are defined as
\begin{equation*}
\begin{gathered}
\text{Ret} = \mathbb{E}[\textbf{r}_{p, t}] \times 252, \quad
\text{Vol} = \text{Std}(\textbf{r}_{p, t}) \times \sqrt{252}, \\
\text{SR} = \frac{\text{Ret}}{\text{Vol}}, \quad
\text{MDD} = \max_{t} \frac{P_t - V_t}{P_t}, \quad \text{Turnover} = \frac{1}{2M} \sum_{m=1}^M \sum_{i=1}^N \left| w_{i, m} - \frac{w_{i, m-1}(1+r_{i, m})}{1 + \sum_{j=1}^N w_{j, m-1} r_{j, m}} \right|,
\end{gathered}
\end{equation*}
where $V_t=\Pi_{s=1}^t (1+\mathbf{r}_{p, s})$ is the cumulative portfolio value at time $t$ (with initial value $V_0=1$), and $P_t = \max_{0\leq s \leq t} V_s $ is the peak portfolio value up to that time.
Here, Ret represents the annualized average return (profitability), and Vol measures the annualized volatility (risk) of the portfolio. SR is the risk-adjusted return, which is the primary optimization objective for the MVP strategy. MDD measures the largest peak-to-trough loss observed during the investment horizon, indicating downside risk.
\textcolor{black}{Turnover represents the trading intensity of each strategy, defined here as the monthly one-way turnover calculated by taking half of the turnover metric proposed in Gu et al. \cite{gu2020empirical}.}
{\color{black} Lower turnover is typically favorable in practical settings to minimize transaction costs. On the other hand, in a frictionless environment, the relatively high values anticipated for daily-rebalanced MVP and GOP portfolios are incidental to the primary objective; they reflect the frequency and magnitude of strategic weight adjustments in response to evolving market conditions.}
Generally, higher values are desirable for Ret and SR, while lower values are desirable for Vol and MDD. However, Vol tends to decrease as risk exposure is reduced, which often leads to a corresponding decrease in Ret. Since the goal of the MVP strategy is to maximize the SR, the minimum possible Vol is neither necessarily the optimal outcome nor our objective. 

\textbf{Growth Optimal Portfolio (GOP).} The GOP, also known as the \textit{log utility maximization portfolio}, maximizes the expected logarithmic utility of the portfolio return. This is equivalent to maximizing the long-term growth rate of wealth. The objective of the optimization is given by
\begin{equation*}
\begin{gathered}
\max_{\mathbf{w}\in \mathbb{R}^N} \mathbb{E} \left[\log ( 1 + \mathbf{w}^\top \hat{\mathbf{r}}_{t+1} )\right] \\
\text{subject to }\sum_{i=1}^N \mathbf{w}_i=1, \quad \text{and}\quad \mathbf{w}_i \geq 0, \text{ for }i=1,\dots, N.
\end{gathered}
\end{equation*}
The expectation is calculated empirically using the generated samples for the return $\hat{\mathbf{r}}_{t+1}$. Unlike the mean-variance approach, maximizing the non-linear log utility function implicitly considers the entire predictive distribution, including higher-order moments like skewness and kurtosis.
Given that the objective is to maximize utility, we consider the CE as the primary performance metric. The CE translates the expected utility into an economically meaningful risk-free rate of return that an investor would consider equivalent to the uncertain portfolio return. The expected utility $U$ and the annualized CE are defined as:
\begin{equation*}
U = \mathbb{E}[\log(1 + \mathbf{r}_{p,t})], \quad
\text{CE} = \left( \exp(U) \right)^{252} - 1.
\end{equation*}
In addition to CE, we also report SR, Ret, Vol, MDD, \textcolor{black}{and turnover} for a comprehensive comparison.

{\color{black}
\color{black}
\textbf{Transaction-Cost-Aware Optimization.} \textcolor{black}{For the experiments involving transaction costs described in Table \ref{tab:performance_weekly_final},} the optimization objectives are reformulated to inherently account for trading frictions. We define the transaction cost penalty at each rebalancing step as $\mathcal{C} = 2 \cdot c \cdot \text{Turnover}$, where $c$ is the transaction cost of 10 bps and the factor of 2 scales the one-way turnover defined above to reflect the total trading volume. 

Under these conditions, the MVP objective is modified to maximize the cost-adjusted Sharpe ratio:
\begin{equation*}
\max_{\mathbf{w} \in \mathbb{R}^N} \frac{\mathbf{w}^\top \boldsymbol{\hat{\mu}}_{t+1} - \mathcal{C}}{\sqrt{\mathbf{w}^\top\boldsymbol{\hat{\Sigma}}_{t+1} \mathbf{w}}}
\end{equation*}
Similarly, the GOP optimization maximizes the expected logarithmic utility of net-of-cost returns:
\begin{equation*}
\max_{\mathbf{w} \in \mathbb{R}^N} \mathbb{E} \left[\log ( 1 + \mathbf{w}^\top \hat{\mathbf{r}}_{t+1} - \mathcal{C} )\right]
\end{equation*}
The constraints for these optimizations, including the budget and long-only constraints, remain identical to the original formulations. This cost-aware framework ensures that a portfolio rebalancing is executed only when the expected gain from the updated joint distribution is sufficient to overcome the transaction cost penalty, thereby preventing excessive turnover from eroding investment welfare.
}

\subsection{Supplementary Diagnostics}
We examine the model's ability to replicate the cross-sectional correlation structure of the assets \textcolor{black}{using two complementary scores}.

\textbf{Correlation Score (CorrScore).}
The {CorrScore} quantifies the correspondence of the cross-sectional dependence structure among assets, which is regarded as critical in portfolio optimization and risk management.
Since estimating a time-varying conditional correlation matrix from single daily realizations is infeasible, we evaluate the unconditional correlation structure over the entire test period. The score is defined as the Frobenius norm of the difference between the empirical and synthetic correlation matrices,
\begin{equation*}
\text{CorrScore} = \lVert \mathbf{C}_{\text{real}} - \mathbf{C}_{\text{synth}} \rVert_F,
\end{equation*}
where $\lVert \cdot \rVert_F$ is the Frobenius norm, $\mathbf{C}_{\text{real}}$ is the sample correlation matrix estimated from the ground truth returns in the test set, and $\mathbf{C}_{\text{synth}}$ is the correlation matrix estimated from the mean paths of $K$ synthetic sample paths generated by the model. A lower score indicates that the synthetic data more realistically replicates the correlations observed in the real market data.

{\color{black}
\textbf{Log-Determinant Divergence (LogDet).} The Log-Determinant divergence, also known as Stein’s loss, provides a spectral measure of the discrepancy between two positive definite matrices. Unlike entry-wise norms, LogDet captures the structural divergence between correlation matrices by accounting for their global geometric alignment. For a system of $N$ assets, the LogDet divergence between $\mathbf{C}_{\text{real}}$ and $\mathbf{C}_{\text{synth}}$ is defined using the trace and determinant operators as
$$\text{LogDet}(\mathbf{C}_{\text{real}}, \mathbf{C}_{\text{synth}}) = \text{tr}(\mathbf{C}_{\text{real}} \mathbf{C}_{\text{synth}}^{-1}) - \log \det(\mathbf{C}_{\text{real}} \mathbf{C}_{\text{synth}}^{-1}) - N,$$
where $\mathbf{C}_{\text{real}} \mathbf{C}_{\text{synth}}^{-1}$ represents the relative correlation structure. This measure can be further expressed in terms of the eigenvalues of the product matrix $\mathbf{C}_{\text{real}} \mathbf{C}_{\text{synth}}^{-1}$ to characterize the spectral discrepancy as
$$\text{LogDet}(\mathbf{C}_{\text{real}}, \mathbf{C}_{\text{synth}}) = \sum_{i=1}^{N} (\lambda_i - \log \lambda_i - 1),$$
where $\lambda_1, \dots, \lambda_N$ are the eigenvalues of $\mathbf{C}_{\text{real}} \mathbf{C}_{\text{synth}}^{-1}$. The divergence is non-negative and attains its minimum of zero if and only if $\mathbf{C}_{\text{real}} = \mathbf{C}_{\text{synth}}$. 

While the CorrScore based on the Frobenius norm quantifies the element-wise Euclidean distance between the individual components of the correlation matrices, the LogDet evaluates their alignment from a spectral perspective. Specifically, the Frobenius norm is primarily sensitive to localized deviations in specific correlation pairs. In contrast, the LogDet is sensitive to the overall dispersion of eigenvalues, heavily penalizing near-singular structures where the generated matrix approaches a lower-rank approximation. By employing both metrics, we ensure that the model not only matches the magnitude of individual correlations but also preserves a well-conditioned dependency structure.
}

\section{Hyperparameter Configuration} \label{appendix: hyperparam}

\subsection{Implementation Details of Diffolio}

\subsubsection{Hyperparameters Search Space}

We performed a grid search to identify the optimal hyperparameter configuration for Diffolio. The selection was based on the best-performing setup on the validation set, as measured by the ES. The search space for the hyperparameters was as follows: window size was first explored in $\{1, 5, 21, 63, 252, 756\}$, which correspond to periods of one day, one week, one month, one quarter, one year, and three years of trading days, respectively; hidden dimension in $\{32, 64, 128, 256\}$; number of attention heads in $\{2, 4, 8\}$; the correlation-guided regularizer coefficient $(\lambda_{corr})$ in \{0.1, \textcolor{black}{0.07, }0.05, \textcolor{black}{0.03, }0.01, 0.005, 0.001, 0\}; and the initial learning rate in $\{\text{1e-3, 1e-4, 1e-5}\}$.

{\color{black} Table \ref{tab:val_es} summarizes the validation ES for each $\lambda_{\text{corr}}$ candidate value with all other hyperparameters fixed at the optimal configurations specified in Section \ref{sec:experimental_setup}.}

\begin{table}[!h]
\centering
\caption{Validation ES for various $\lambda_{\text{corr}}$ candidate values.}
\label{tab:val_es}
\resizebox{\linewidth}{!}{%
\begin{tabular}{lcccccccc}
\toprule
$\lambda_{\text{corr}}$ & 0 & 0.001 & 0.005 & 0.01 & 0.03 & \textbf{0.05} & 0.07 & 0.1 \\
\midrule
Validation ES & 0.033334 & 0.033430 & 0.033313 & 0.033184 & 0.033366 & \textbf{0.033125} & 0.033269 & 0.033302 \\
\bottomrule
\end{tabular}%
}
\end{table}










{\color{black}
\subsubsection{Sensitivity Analysis on Diffusion and DDIM Sampling Steps}

Following the standard convention in the diffusion literature established by Ho et al. \cite{ho2020denoising} and Song et al. \cite{song2020denoising}, we set the total number of diffusion training steps to $T=1000$ and the number of DDIM sampling steps during inference to $S=50$. To investigate the impact of these parameters on the model's performance, we conduct a sensitivity analysis by varying $T$ and $S$ while holding all other hyperparameters fixed. The results are summarized in Table \ref{tab:diffusion_sensitivity}.

Our findings indicate that statistical metrics, such as CRPS and ES, remain remarkably stable across different sampling steps $S$. While economic metrics like MVP-SR and GOP-CE show some fluctuations, the $S=50$ configuration yields strong performance, which is consistent with findings in the well-established literature. On the other hand, we observe that the supplementary diagnostics measuring correlation dependency, CorrScore and LogDet, show higher sensitivity to the number of training steps $T$.
Although the default configuration ($T=1000, S=50$) does not necessarily achieve the best results in every individual metric, it provides consistent and reliable performance across various measures. It should be noted that since other hyperparameters were specifically tuned for this default setup, the performance of other configurations (e.g., $T=500$ or $T=2000$) might be further improved if their own hyperparameters were independently optimized. Overall, these observations confirm that our chosen setup of $T=1000$ and $S=50$ serves as a reliable configuration for generating high-quality multi-asset return paths.
}

\begin{table}[!h]
\centering
\caption{\color{black} Performance sensitivity to diffusion training steps and DDIM sampling steps.}
\label{tab:diffusion_sensitivity}
\resizebox{\linewidth}{!}{%
\begin{tabular}{lcccc cc}
\toprule
& \multicolumn{4}{c}{Evaluation Metrics} & \multicolumn{2}{c}{Supplementary Diagnostics} \\
\cmidrule(lr){2-5} \cmidrule(lr){6-7}
\textbf{Configuration} & \textbf{CRPS} & \textbf{ES} & \textbf{MVP-SR} & \textbf{GOP-CE} & \textbf{CorrScore} & \textbf{LogDet} \\
\midrule
\multicolumn{7}{l}{\textit{Sensitivity to DDIM Sampling Steps ($S$) with $T = 1000$}} \\
$S = 10$ & 0.007185 ($\pm$ 0.001600) & 0.028974 & 0.5746 & 0.0885 & 1.5490 & 2.8179 \\
$S = 25$ & 0.007168 ($\pm$ 0.001602) & 0.028929 & 0.5720 & 0.1265 & 0.9787 & 2.0966 \\
$S = 50$ (Default) & 0.007169 ($\pm$ 0.001601) & 0.028960 & 0.7206 & 0.1611 & 1.0808 & 2.1982 \\
$S = 75$ & 0.007167 ($\pm$ 0.001601) & 0.028941 & 0.6316 & 0.1000 & 1.0875 & 2.1931 \\
$S = 100$ & 0.007171 ($\pm$ 0.001597) & 0.028951 & 0.6116 & 0.1080 & 1.1031 & 2.2690 \\
\midrule
\multicolumn{7}{l}{\textit{Sensitivity to Diffusion Training Steps ($T$) with $S = 50$}} \\
$T = 500$ & 0.007138 ($\pm$ 0.001613) & 0.028855 & 0.6192 & 0.1297 & 3.9985 & 5.1684 \\
$T = 1000$ (Default) & 0.007169 ($\pm$ 0.001601) & 0.028960 & 0.7206 & 0.1611 & 1.0808 & 2.1982 \\
$T = 2000$ & 0.008133 ($\pm$ 0.002145) & 0.037647 & 0.3889 & 0.0592 & 6.3083 & 8.6651 \\
\bottomrule
\end{tabular}%
}
\end{table}

\subsection{Hyperparameters of Baseline Models}

To ensure a fair and robust comparison, we conducted a comprehensive hyperparameter search for all baseline models. For hyperparameters that are shared with our proposed Diffolio model (e.g., learning rate scheduler, optimizer, window size, batch size), we adopted the same configurations unless specified otherwise. For model-specific parameters, we followed the methodologies and search spaces outlined in their original papers. The hyperparameter configuration for each baseline was selected based on the best performance on the validation set, as measured by ES.
The model-specific architectures and the corresponding hyperparameter search spaces explored are detailed below for each baseline.

\textbf{TimeGAN}
TimeGAN is distinguished by its training process, which consists of three sequential stages: first, training an autoencoder for embedding and recovery; second, training the generator with a supervised loss; and third, jointly training the generator and discriminator in an adversarial manner. The search space for its hyperparameters in our experiments was defined as follows,
\begin{itemize}
    \item \vspace{-0.7em}{Training steps (per stage) }= $\{100, 300, 500, 1000, 10000\}$
    \item \vspace{-0.7em}{Hidden dimension }= $\{64, 100, 128, 200, 256\}$
    \item \vspace{-0.7em}{Supervised loss coefficient (stage 3) }= $\{0.1, 1, 10\}$
    \item \vspace{-0.7em}{Latent dimension for normal random variable $z$ }= $\{16, 32, 64\}$
\end{itemize}

\textbf{TimeGrad}
The architecture of TimeGrad is based on WaveNet \cite{van2016wavenet}, which is composed of 1D convolutional networks with residual connections. We tuned its hyperparameters as follows,
\begin{itemize}
    \item \vspace{-0.7em}{Hidden dimension }= $\{64, 100, 128, 200, 256\}$
    \item \vspace{-0.7em}{Number of residual channels }= $\{4, 8, 16\}$
    \item \vspace{-0.7em}{Number of layers per residual block }= $\{2, 4, 8\}$
\end{itemize}

\textbf{CSDI}
As CSDI is a Transformer-based model, we tuned its architectural components, such as the number of attention heads and the hidden dimension. The search space was defined as
\begin{itemize}
    \item \vspace{-0.7em}{Training steps }= $\{5\times10^4, 1\times 10^5\}$
    \item \vspace{-0.7em}{Hidden dimension (channels) }= $\{64, 96, 128\}$
    \item \vspace{-0.7em}{Number of attention heads }= $\{4, 8\}$
\end{itemize}

\textbf{Diffusion-TS}
A key feature of Diffusion-TS is its use of classifier-free guidance. Accordingly, we tuned the guidance strength parameter in addition to other hyperparameter settings as
\begin{itemize}
    \item \vspace{-0.7em}{Training steps }= $\{5\times 10^4, 1\times 10^5\}$
    \item \vspace{-0.7em}{Guidance strength }= $\{0.1, 0.05, 0.01\}$
    \item \vspace{-0.7em}{Embedding hidden dimension }= $\{64, 96\}$
\end{itemize}

\textbf{MG-TSD}
The hyperparameter search for MG-TSD centered on the configuration of its residual block structure. The explored values were given as
\begin{itemize}
    \item \vspace{-0.7em}{Training steps }= $\{5\times 10^4, 1\times 10^5\}$
    \item \vspace{-0.7em}{Number of residual layers}: $\{2, 3, 4\}$
    \item \vspace{-0.7em}{Residual block hidden dimension }= $\{64, 96, 128\}$
    \item \vspace{-0.7em}{Number of residual channels }= $\{16, 32\}$
\end{itemize}

\textbf{SigCWGAN}
The tuning for SigCWGAN focused on the dimensions of its hidden and latent spaces along with the depth of the signature transform. The search space included the \textcolor{black}{following},
\begin{itemize}
    \item \vspace{-0.7em}{Training steps}: $\{5\times 10^4, 1\times10^5\}$
    \item \vspace{-0.7em}{Hidden dimension}: $\{64, 96, 128\}$
    \item \vspace{-0.7em}{Signature truncation depth}: $\{2, 3, 4\}$
    \item \vspace{-0.7em}{Latent dimension}: $\{10, 30, 50\}$
\end{itemize}

The adopted hyperparameter configurations for each baseline are summarized in Table \ref{tab:baseline_hyperparams}.

\begin{longtable}{@{}ll@{}}
\caption{Adopted hyperparameter configurations for baseline models.}
\label{tab:baseline_hyperparams} \\

\toprule
\textbf{Model} & \textbf{Adopted Hyperparameter Configuration} \\
\midrule
\endfirsthead

\toprule
\textbf{Model} & \textbf{Adopted Hyperparameter Configuration} \\
\midrule
\endhead

\bottomrule
\endlastfoot

TimeGAN (H) &
\begin{tabular}{@{}l@{}}
Training steps (Stage 1 / 2 / 3) = (100 / 500 / 1000) \\
Hidden dimension = 128 \\
Supervised loss coefficient = 10 \\
Latent dimension = 32
\end{tabular} \\
\addlinespace
TimeGAN (H+C) &
\begin{tabular}{@{}l@{}}
Training steps (Stage 1 / 2 / 3) = (100 / 500 / 1000) \\
Hidden dimension = 128 \\
Supervised loss coefficient = 10 \\
Latent dimension = 32
\end{tabular} \\
\midrule
TimeGrad (H) &
\begin{tabular}{@{}l@{}}
Hidden dimension = 128 \\
Number of residual channels = 8 \\
Number of layers per residual block = 8
\end{tabular} \\
\addlinespace
TimeGrad (H+C) &
\begin{tabular}{@{}l@{}}
Hidden dimension = 200 \\
Number of residual channels = 8 \\
Number of layers per residual block = 16
\end{tabular} \\
\midrule
CSDI (H) &
\begin{tabular}{@{}l@{}}
Training steps = $5\times10^4$ \\
Hidden dimension = 64 \\
Number of attention heads = 4
\end{tabular} \\
\addlinespace
CSDI (H+C) &
\begin{tabular}{@{}l@{}}
Training steps = $1\times10^5$ \\
Hidden dimension = 64 \\
Number of attention heads = 4
\end{tabular} \\
\midrule
Diffusion-TS (H) &
\begin{tabular}{@{}l@{}}
Training steps = $5\times10^4$ \\
Embedding hidden dimension = 64 \\
Guidance strength = 0.1
\end{tabular} \\
\addlinespace
Diffusion-TS (H+C) &
\begin{tabular}{@{}l@{}}
Training steps = $1\times10^5$ \\
Embedding hidden dimension = 96 \\
Guidance strength = 0.01
\end{tabular} \\
\midrule
MG-TSD (H) &
\begin{tabular}{@{}l@{}}
Training steps = $1\times 10^5$ \\
Number of residual layers = $4$ \\
Residual block hidden dimension = $64$ \\
Number of residual channels = $16$
\end{tabular} \\
\addlinespace
MG-TSD (H+C) &
\begin{tabular}{@{}l@{}}
Training steps = $1\times 10^5$ \\
Number of residual layers = $2$ \\
Residual block hidden dimension = $128$ \\
Number of residual channels = $32$
\end{tabular} \\
\midrule
SigCWGAN (H) &
\begin{tabular}{@{}l@{}}
Training steps = $1\times10^5$ \\
Hidden dimension = $128$ \\
Signature truncation depth = $2$ \\
Latent dimension = $30$
\end{tabular} \\
\addlinespace
SigCWGAN (H+C) &
\begin{tabular}{@{}l@{}}
Training steps = $1\times 10^5$ \\
Hidden dimension = $96$ \\
Signature truncation depth = $2$ \\
Latent dimension = $10$
\end{tabular} \\

\end{longtable}

\section{Reliability Diagram} \label{appendix: reliability diagram}
\begin{figure}[!htbp]
    \centering

    \begin{subfigure}[b]{0.24\textwidth}
        \centering
        \includegraphics[width=\textwidth]{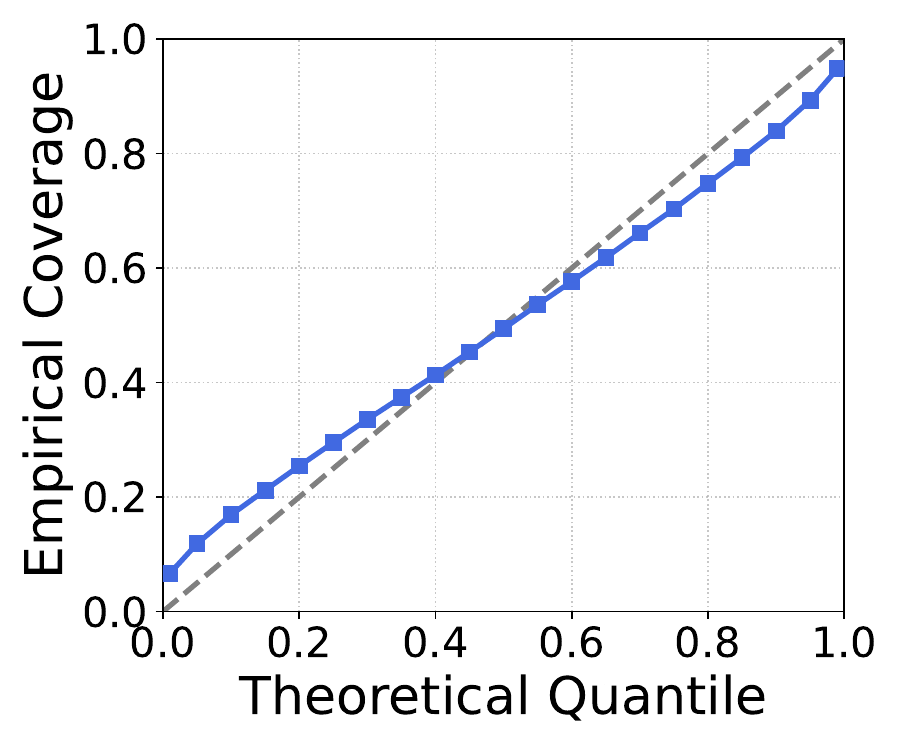}
        \caption{DCC-GARCH (H)}
    \end{subfigure}\hfill
    \begin{subfigure}[b]{0.24\textwidth}
        \centering
        \includegraphics[width=\textwidth]{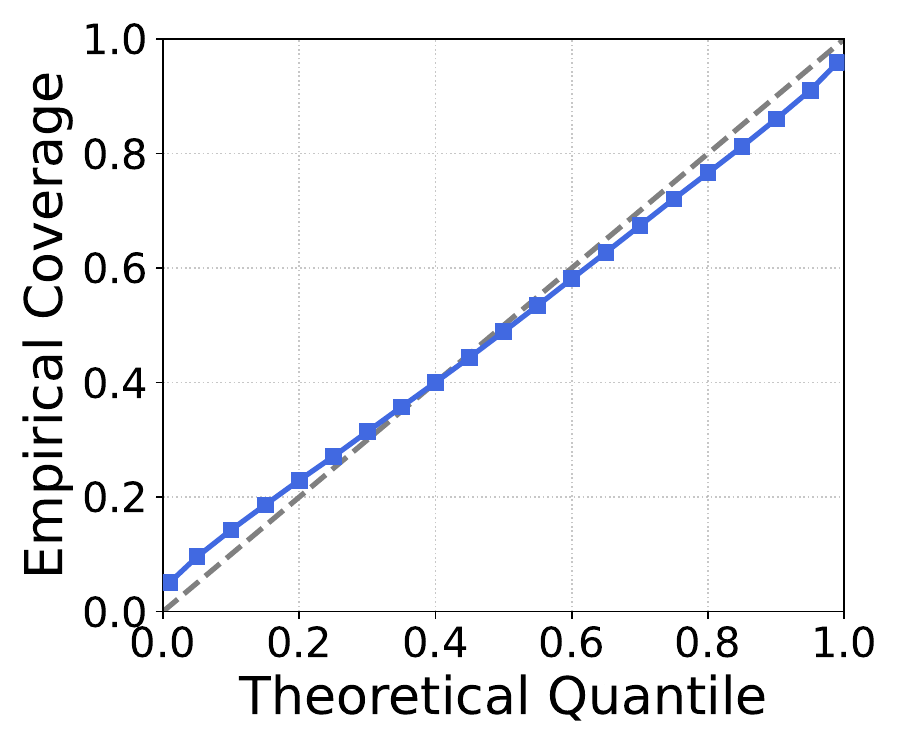}
        \caption{DCC-GARCH (H+C)}
    \end{subfigure}\hfill
    \begin{subfigure}[b]{0.24\textwidth}
        \centering
        \includegraphics[width=\textwidth]{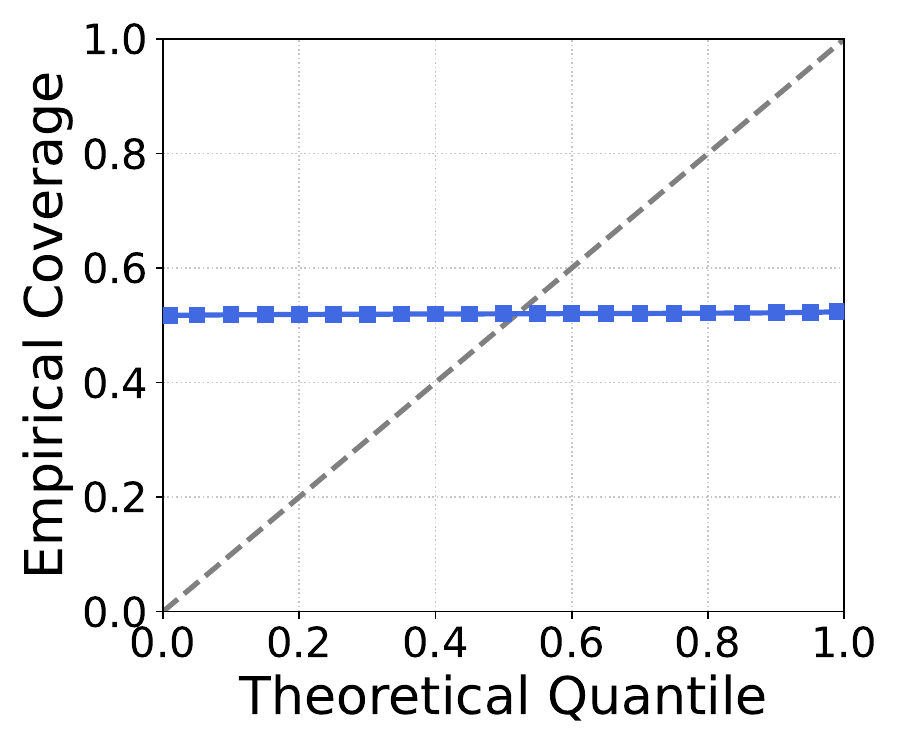}
        \caption{TimeGAN (H)}
    \end{subfigure}\hfill
    \begin{subfigure}[b]{0.24\textwidth}
        \centering
        \includegraphics[width=\textwidth]{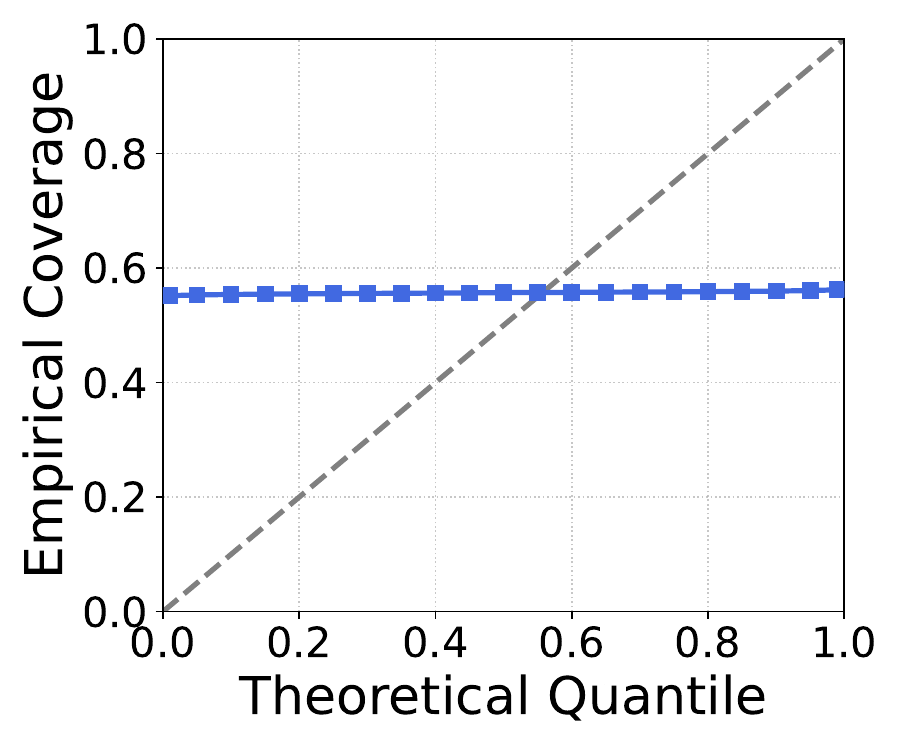}
        \caption{TimeGAN (H+C)}
    \end{subfigure}
    
    \vspace{1em} 
    
    \begin{subfigure}[b]{0.24\textwidth}
        \centering
        \includegraphics[width=\textwidth]{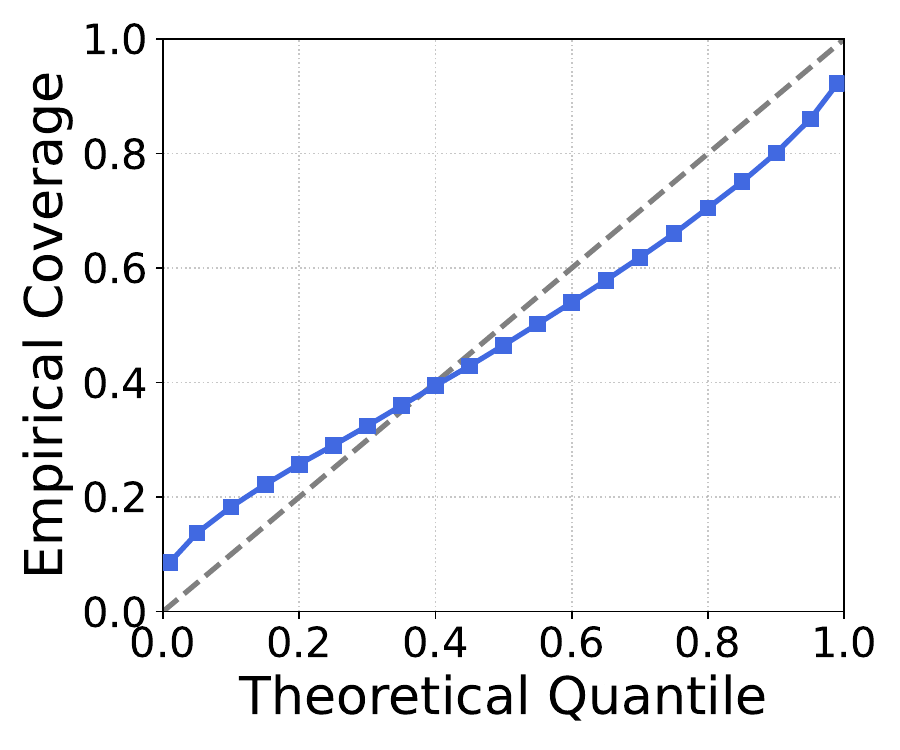}
        \caption{TimeGrad (H)}
    \end{subfigure}\hfill
    \begin{subfigure}[b]{0.24\textwidth}
        \centering
        \includegraphics[width=\textwidth]{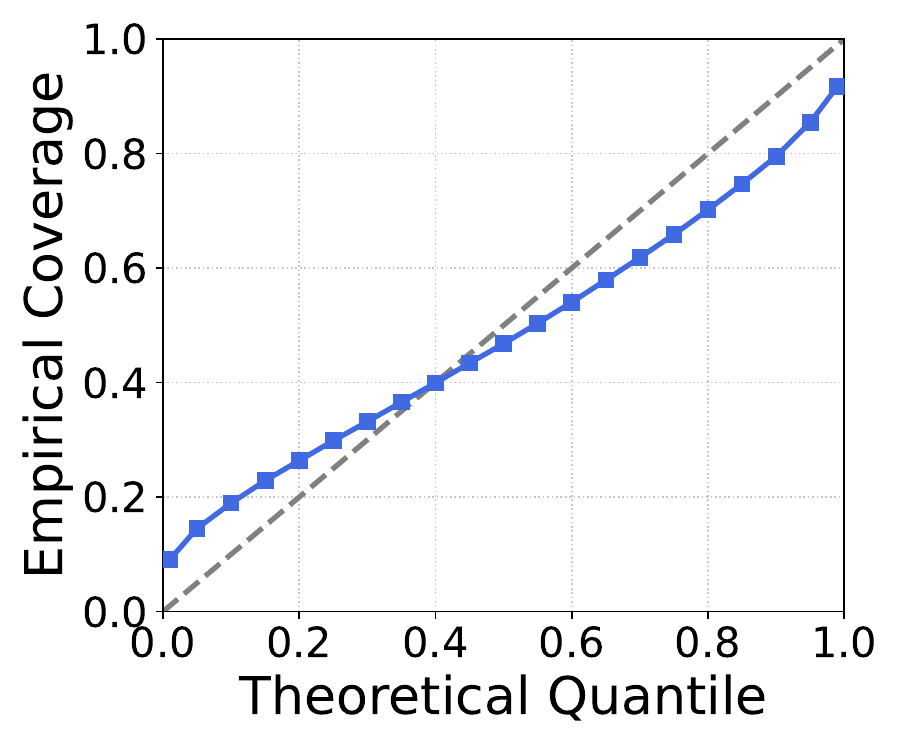}
        \caption{TimeGrad (H+C)}
    \end{subfigure}\hfill
    \begin{subfigure}[b]{0.24\textwidth}
        \centering
        \includegraphics[width=\textwidth]{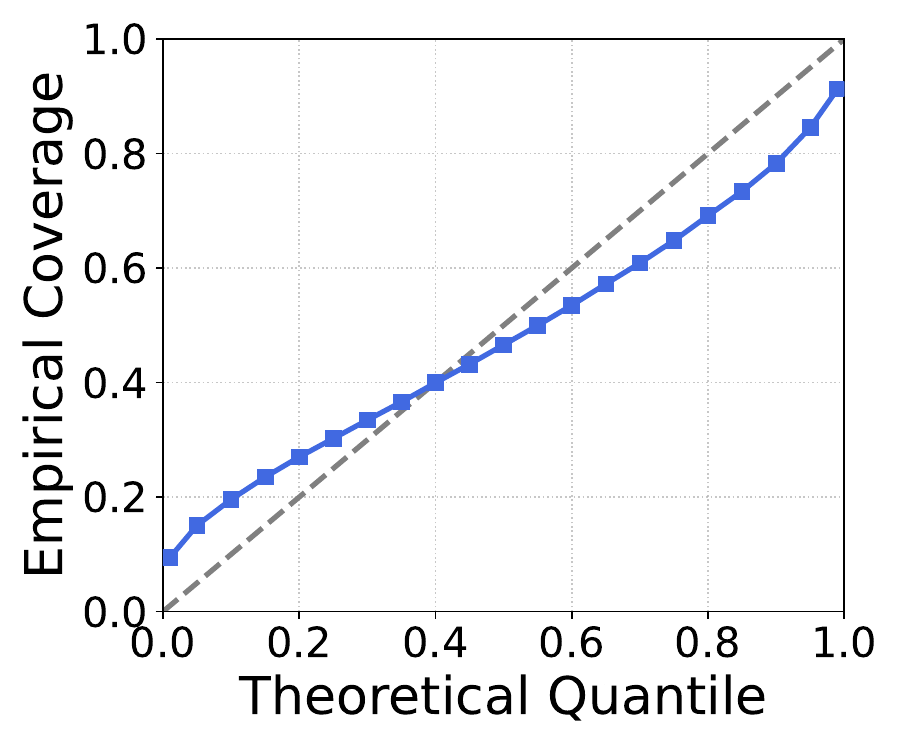}
        \caption{CSDI (H)}
    \end{subfigure}\hfill
    \begin{subfigure}[b]{0.24\textwidth}
        \centering
        \includegraphics[width=\textwidth]{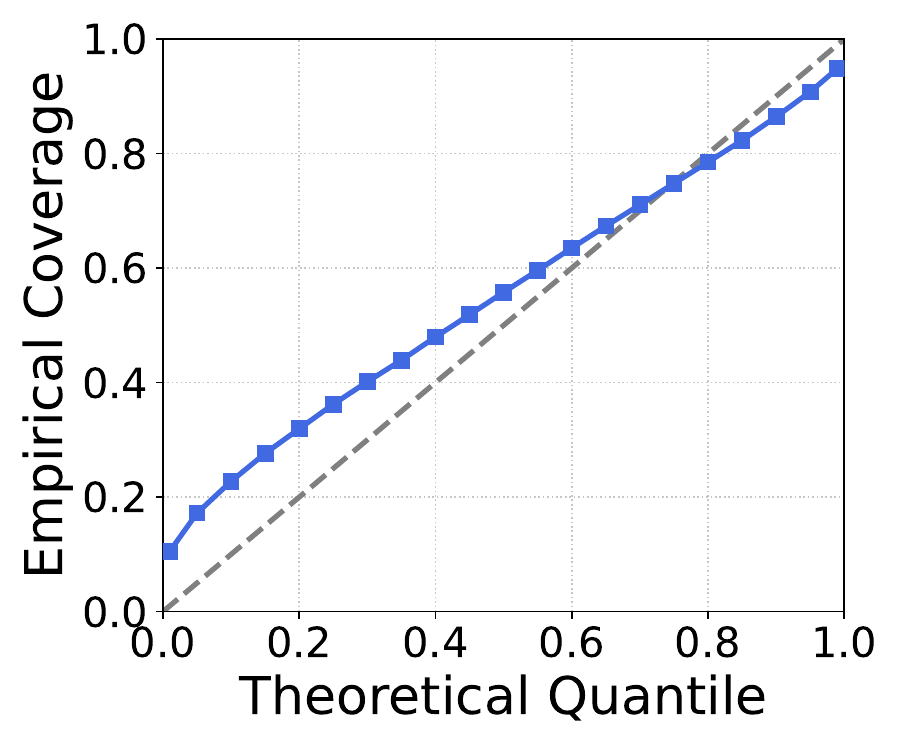}
        \caption{CSDI (H+C)}
    \end{subfigure}
    
    \vspace{1em}

    \begin{subfigure}[b]{0.24\textwidth}
        \centering
        \includegraphics[width=\textwidth]{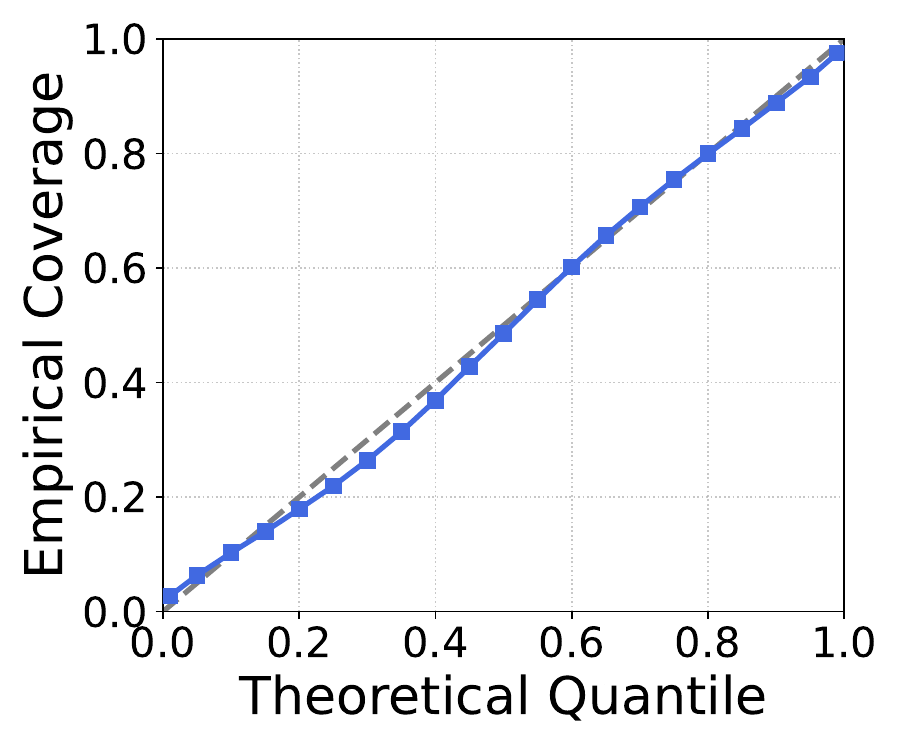}
        \caption{Diffusion-TS (H)}
    \end{subfigure}\hfill
    \begin{subfigure}[b]{0.24\textwidth}
        \centering
        \includegraphics[width=\textwidth]{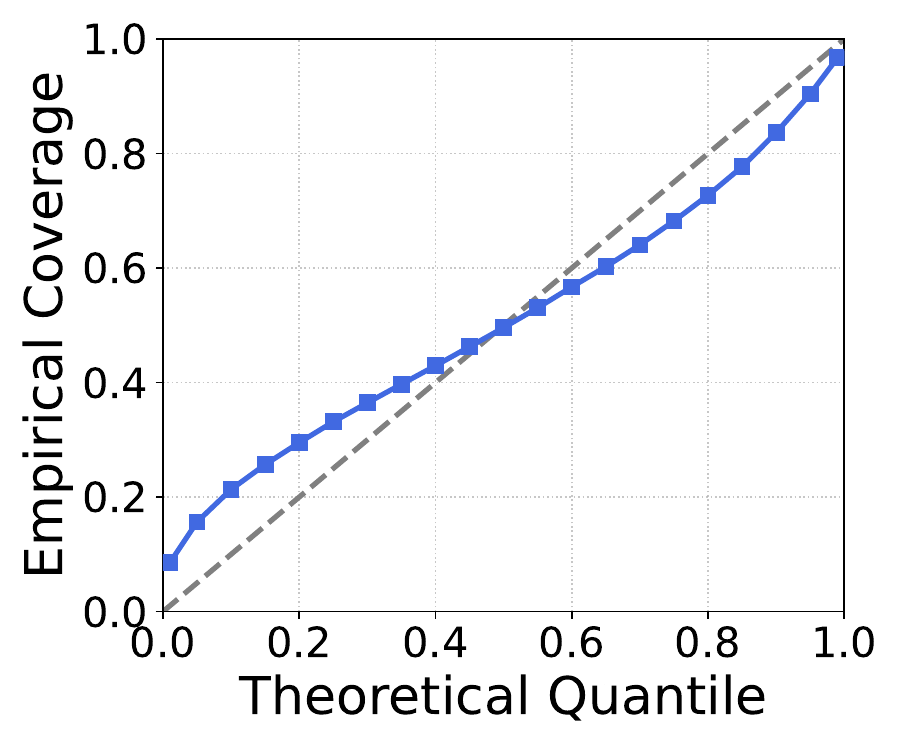}
        \caption{Diffusion-TS (H+C)}
    \end{subfigure}\hfill
    \begin{subfigure}[b]{0.24\textwidth}
        \centering
        \includegraphics[width=\textwidth]{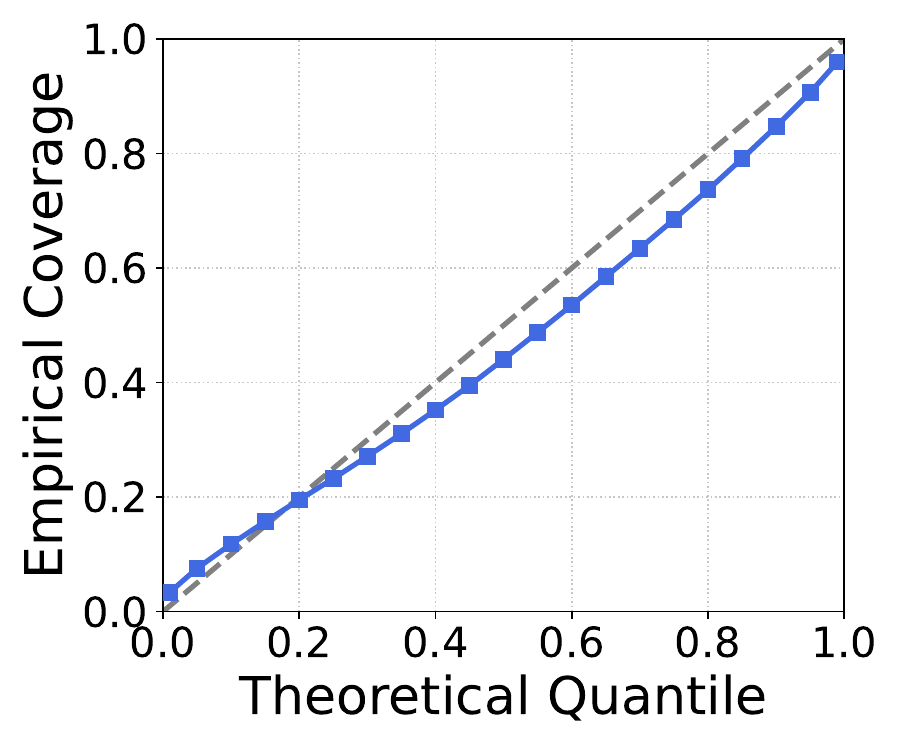}
        \caption{MG-TSD (H)}
    \end{subfigure}\hfill
    \begin{subfigure}[b]{0.24\textwidth}
        \centering
        \includegraphics[width=\textwidth]{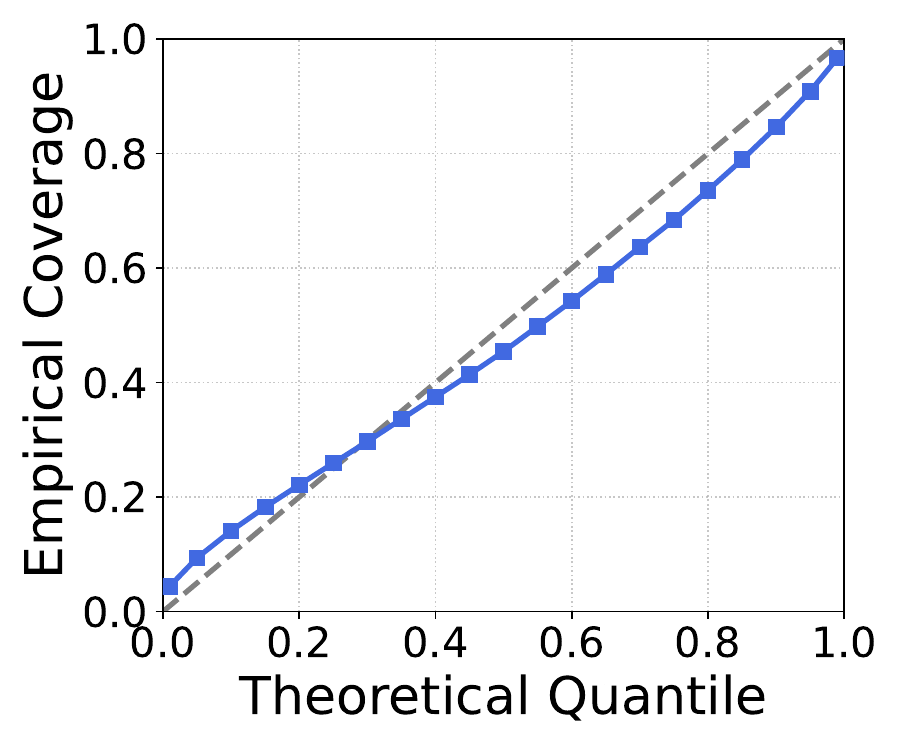}
        \caption{MG-TSD (H+C)}
    \end{subfigure}
    
    \vspace{1em}

    \begin{subfigure}[b]{0.24\textwidth}
        \centering
        \includegraphics[width=\textwidth]{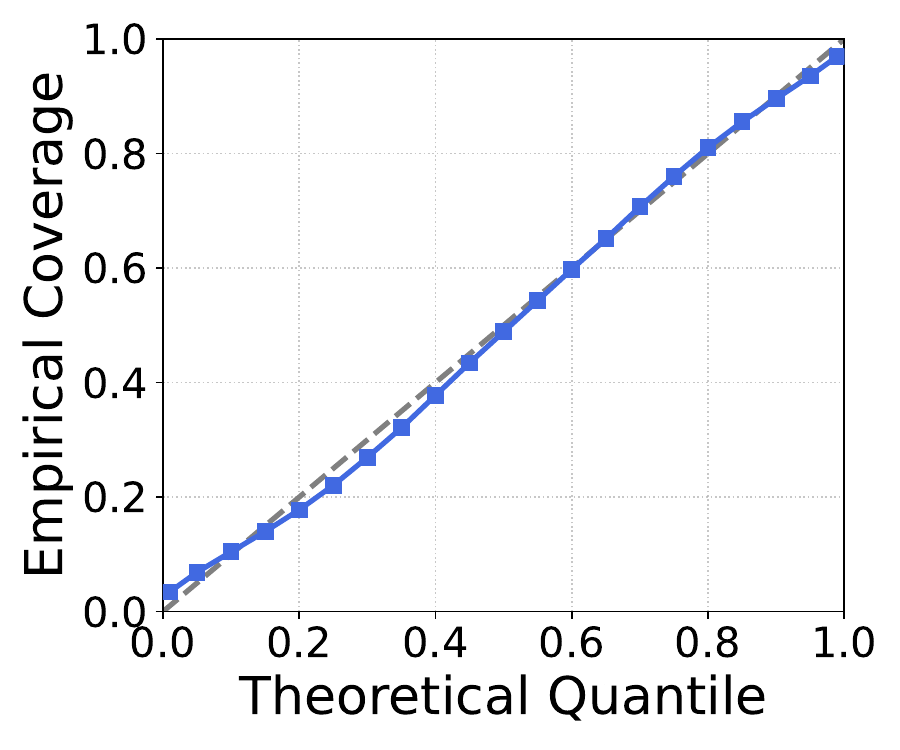}
        \caption{SigCWGAN (H)}
    \end{subfigure}
    \hspace{0.5cm} 
    \begin{subfigure}[b]{0.24\textwidth}
        \centering
        \includegraphics[width=\textwidth]{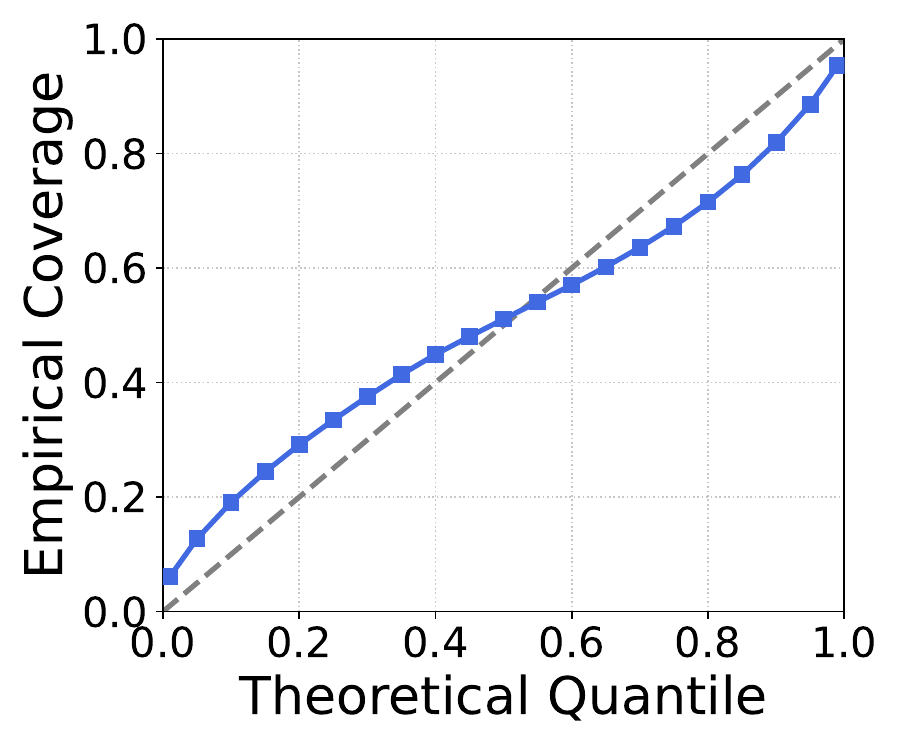}
        \caption{SigCWGAN (H+C)}
    \end{subfigure}
    
    \caption{Reliability diagrams for baseline models.}
    \label{fig:reliability_all_4col}
\end{figure}

\color{black}{
\section{Average Sectoral Betas during the Test Period} \label{appendix: average beta}

This appendix presents the average beta values for the industrial sectors calculated over the test period. These values serve as a reference for the baseline systematic exposure of the assets. We note that several sectors exhibit relatively low betas, which provides context for the high importance assigned to $\texttt{idiovol}$ discussed in Section \ref{sec:variable_importance_analysis}.

\begin{figure}[!h] \centering \includegraphics[width=0.6\linewidth]{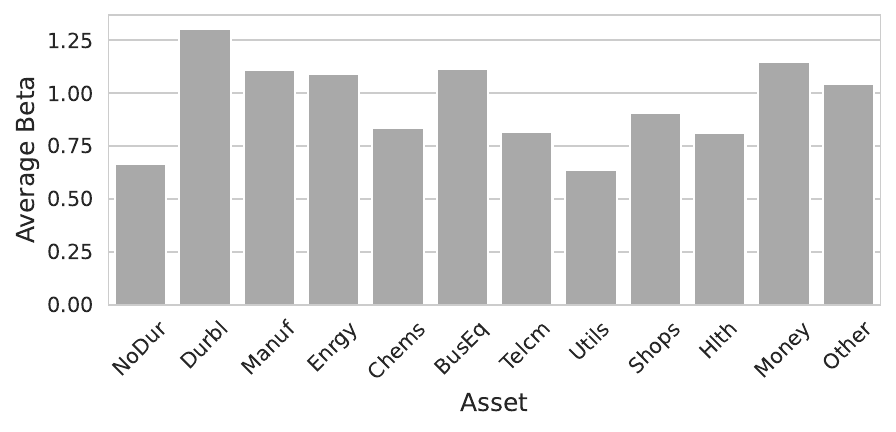} \caption{Average beta of target assets during the test period.} \label{fig:beta_avg_test} \end{figure}
}

\end{document}